\documentclass[a4paper,fleqn,usenatbib]{mnras}

\usepackage{newtxtext,newtxmath}

\usepackage[T1]{fontenc}
\usepackage{scalerel}
\usepackage{tikz}
\usetikzlibrary{svg.path}

\definecolor{orcidlogocol}{HTML}{A6CE39}
\tikzset{
  orcidlogo/.pic={
    \fill[orcidlogocol] svg{M256,128c0,70.7-57.3,128-128,128C57.3,256,0,198.7,0,128C0,57.3,57.3,0,128,0C198.7,0,256,57.3,256,128z};
    \fill[white] svg{M86.3,186.2H70.9V79.1h15.4v48.4V186.2z}
                 svg{M108.9,79.1h41.6c39.6,0,57,28.3,57,53.6c0,27.5-21.5,53.6-56.8,53.6h-41.8V79.1z M124.3,172.4h24.5c34.9,0,42.9-26.5,42.9-39.7c0-21.5-13.7-39.7-43.7-39.7h-23.7V172.4z}
                 svg{M88.7,56.8c0,5.5-4.5,10.1-10.1,10.1c-5.6,0-10.1-4.6-10.1-10.1c0-5.6,4.5-10.1,10.1-10.1C84.2,46.7,88.7,51.3,88.7,56.8z};
  }
}

\newcommand\orcidicon[1]{\href{https://orcid.org/#1}{\mbox{\scalerel*{
\begin{tikzpicture}[yscale=-1,transform shape]
\pic{orcidlogo};
\end{tikzpicture}
}{|}}}}

\usepackage{float}
\usepackage[export]{adjustbox}
\usepackage{graphicx}	

\usepackage{amsmath}	
\usepackage{amssymb}	
\usepackage{multirow}
\usepackage{multicol}
\usepackage{verbatim}   
\usepackage{threeparttable}
\usepackage{upgreek}
\usepackage{rotating}

\usepackage{subcaption}
\usepackage{comment}
\usepackage{longtable,lscape}
\usepackage{caption}
\usepackage{threeparttable}
\newcommand{\lcdm}{$\Lambda$CDM}

\newcommand{\om}{\Omega_{m0}}
\newcommand{\ol}{\Omega_{\Lambda}}
\newcommand{\ok}{\Omega_{k0}}
\newcommand{\FT}[1]{}

\title[Study of QSO X-ray and UV flux data]{Do quasar X-ray and UV flux measurements provide a useful test of cosmological models?}

\author[]{
Narayan Khadka,$^{\orcidicon{0000-0001-5512-2716}1}$\thanks{E-mail: nkhadka@phys.ksu.edu}
Bharat Ratra$^{\orcidicon{0000-0002-7307-0726}1}$\thanks{E-mail: ratra@phys.ksu.edu}\\
$^{1}$Department of Physics, Kansas State University, 116 Cardwell Hall, Manhattan, KS 66502, USA\\
}

\date{Accepted XXX. Received YYY; in original form ZZZ}

\pubyear{2021}

\begin{document}
\label{firstpage}
\pagerange{\pageref{firstpage}--\pageref{lastpage}}
\maketitle

\begin{abstract}
The recent compilation of quasar (QSO) X-ray and UV flux measurements include QSOs that appear to not be standardizable via the X-ray luminosity and UV luminosity ($L_X-L_{UV}$) relation and so should not be used to constrain cosmological model parameters. Here we show that the largest of seven sub-samples in this compilation, the SDSS-4XMM QSOs that contribute about 2/3 of the total QSOs, have $L_X-L_{UV}$ relations that depend on the cosmological model assumed and also on redshift, and is the main cause of the similar problem discovered earlier for the full QSO compilation. The second and third biggest sub-samples, the SDSS-Chandra and XXL QSOs that together contribute about 30\% of the total QSOs, appear standardizable, but provide only weak constraints on cosmological parameters that are not inconsistent with the standard spatially-flat $\Lambda$CDM model or with constraints from better-established cosmological probes.   
\end{abstract}

\begin{keywords}
\textit{(cosmology:)} cosmological parameters -- \textit{(cosmology:)} observations -- \textit{(cosmology:)} dark energy
\end{keywords}



\section{Introduction}
\label{sec:Introduction}
Currently our universe is undergoing accelerated cosmological expansion \citep{eBOSSCollaboration2021, Farooqetal2017, PlanckCollaboration2020, Scolnicetal2018}. General relativistic cosmological models interpret this observational fact as a consequence of dark energy. The simplest cosmological model that is consistent with most observations is the standard spatially-flat $\Lambda$CDM model \citep{Peebles1984}. In this model, at the current epoch, the cosmological constant $(\Lambda)$ contributes about $70\%$ of the cosmological energy budget, non-relativistic cold dark matter (CDM) contributes about $25\%$, and ordinary non-relativistic baryons contributes almost all of the remaining $\sim 5\%$. This model assumes time-independent dark energy density and flat spatial hypersurfaces and is not inconsistent with most observational constraints.\footnote{For recent reviews see \citet{DiValentinoetal2021b} and \citet{PerivolaropoulosSkara2021}.} However, observational data do not rule out a small amount of spatial curvature nor do they rule out mildly dynamical dark energy. So in this paper we also consider spatially non-flat models and dynamical dark energy models.

Cosmological models are largely tested using low and high redshift data. The low redshift data probe the  $0 \leq z \leq 2.3$ part of cosmological redshift space and include baryon acoustic oscillation (BAO), Hubble parameter [$H(z)$], and Type Ia supernova (SNIa) apparent magnitude measurements. The only high cosmological redshift space probe are cosmic microwave background anisotropy data at $z \sim 1100$. In the intermediate redshift range, $2.3 < z < 1100$, cosmological models are poorly tested. There are a handful of astronomical data sets in this range that are starting to be used in cosmology. These data sets include HII starburst galaxy observations reaching to $z \sim 2.4$ \citep{ManiaRatra2012, Chavezetal2014, GonzalezMoran2019, GonzalezMoranetal2021, Caoetal2020, Caoetal2021a, Cao_2021c, Johnsonetal2021}, quasar (QSO) angular size observations reaching to $z \sim 2.7$ \citep{Caoetal2017, Ryanetal2019, Caoetal2020, Caoetal2021b, Zhengetal2021, Lianetal2021}, gamma-ray burst observations reaching to $z \sim 8.2$ \citep{Amati2008, Amati2019, samushia_ratra_2010, Wang_2016, Wangetal2021, Demianski_2019, Dirirsa2019, KhadkaRatra2020c, Khadkaetal2021, Caoetal2021d}, and recently \cite{Vagnozzietal2021b} has used old astrophysical objects in the redshift range $0 \leq z \leq 8$ to constrain cosmological parameters.

Quasar X-ray and UV flux observations reaching to $z \sim 7.5$ \citep{RisalitiLusso2015, RisalitiLusso2019, Lussoetal2020} is another such data set that could be used to constrain cosmological model parameters.\footnote{For recent applications and discussions of these data see \citet{Yangetal2020}, \citet{WeiMelia2020}, \citet{Lindneretal2020}, \citet{KhadkaRatra2020a, KhadkaRatra2020b, KhadkaRatra2021}, \citet{Rezaeietal2020}, \citet{Huetal2020}, \citet{Banerjeeetal2021} \citet{Sperietal2021}, \citet{Zhengetal2021}, \citet{ZhaoXia2021}, \citet{Lietal2021}, \citet{Lianetal2021}, and references therein.} QSO X-ray and UV luminosities are correlated through a non-linear relation, the $L_X-L_{UV}$ relation $L_X = 10^\beta L_{UV}^\gamma$ where $\beta$ and $\gamma$ are parameters. This correlation enables us to standardize and so use these QSO data\footnote{We can employ the $L_X-L_{UV}$ relation to measure the luminosity distance or distance modulus of quasars that obey this correlation. So if the $L_X-L_{UV}$ correlation relation is valid these quasars are standard candles.} to derive cosmological parameter constraints. Basic assumptions underlying the use of the $L_X-L_{UV}$ relation for the standardization of QSO data are: (i) the $L_X-L_{UV}$ relation parameters, intercept $(\beta)$ and slope $(\gamma)$, should not evolve with redshift; and, (ii) these parameters should be independent of the cosmological model used to measure them. While great effort has been made to standardize these QSO data using the $L_X-L_{UV}$ relation the most recent QSO data compilation of \citet{Lussoetal2020} violate both assumptions underpinning QSO standardization via the $L_{X}-L_{UV}$ relation \citep{KhadkaRatra2021}. So at least some of the QSOs in the \citet{Lussoetal2020} compilation are not standardizable through the $L_{X}-L_{UV}$ relation. These QSO data are a possibly heterogeneous compilation of seven different sub-samples that are summarized in Table \ref{tab:QSO1}. In this paper, we study these QSO data more granularly to try to determine whether some of the sub-samples are more responsible for the  redshift-dependent and model-dependent behavior of $\beta$ and $\gamma$ values discovered by \citet{KhadkaRatra2021}.

\begin{table}
	\centering
	\small\addtolength{\tabcolsep}{-2.5pt}
	\small
	\caption{Subsets of the full QSO (2421 QSOs) data compilation.}
	\label{tab:QSO1}
	\begin{threeparttable}
	\begin{tabular}{lccccccccccc} 
		\hline
		Data set & Redshift range & Number of QSOs \\
		\hline
		SDSS-4XMM & $0.131 \leq z \leq 3.412$ & 1644\\
		SDSS-Chandra & $0.489 \leq z \leq 3.981$ & 608\\
		XXL & $0.2436 \leq z \leq 1.9247$ & 106\\
		Local AGN & $0.009 \leq z \leq 0.0866$ & 13\\
		High-$z$ & $4.01 \leq z \leq 7.5413$ & 35\\
		XMM-Newton$\sim 3$ & $3.028 \leq z \leq 3.296$ & 14\\
		XMM-Newton$\sim 4$ & 4.109 & 1\\
		\hline
	\end{tabular}
    \end{threeparttable}
\end{table}

The largest sub-sample in the \citet{Lussoetal2020} QSO compilation are SDSS-4XMM QSOs. Our analyses here show that the $\beta$ and $\gamma$ values measured for this sub-sample depend on the cosmological model used and change with redshift. On the other hand, other large sub-samples, such as SDSS-Chandra QSOs, have model-independent $\beta$ and $\gamma$ values. So the main issues are mostly related with the SDSS-4XMM QSOs and these need to be resolved if these QSOs are to be used for cosmological purposes. We note that there has been a recent claim of a larger than $4\sigma$ tension between these QSO data and the standard spatially-flat $\Lambda$CDM model with $\Omega_{m0}=0.3$ \citep{Lussoetal2020}. We believe that this is more properly interpreted as an indirect indication that these QSOs are not standard candles and so cannot be used for cosmological purposes \citep{KhadkaRatra2021}. While we have discovered an issue with the SDSS-4XMM data, it is at present not known if this was generated during the original compilation process or whether the SDSS-4XMM QSOs have an $L_{X}-L_{UV}$ relation that evolves with $z$ and so are non-standardizable. We also note that other, somewhat lower $z$, reverberation-measured Mg II, radius-luminosity standardized, QSO constraints are consisitent with the standard flat $\Lambda$CDM model \citep{Mary2019,Czerny2021, Michal2021, Yuetal2021, Khadkaetal2021b}.

Our paper is organized as follows. In Sec.\ 2 we describe the cosmological models we study. In Sec.\ 3 we summarize the data sub-sets, and compilations of these, that we analyze. In Sec.\ 4 we describe analysis methods we use. We present our results in Sec.\ 5 and conclude in Sec.\ 6.
\section{Models}
\label{sec:models}

To constrain cosmological model parameters we compare model predictions, that depend on these parameters, to measurements made at known redshifts. So as to understand how model-dependent our results are, we consider six different models, three with non-flat spatial hypersurfaces and three with flat spatial hypersurfaces.\footnote{For spatial curvature observational constraints see \citet{Farooqetal2015}, \citet{Chenetal2016}, \citet{Ranaetal2017}, \citet{Yuetal2018}, \citet{Oobaetal2018a, Oobaetal2018b}, \citet{Wei2018}, \citet{ParkRatra2019a, ParkRatra2019b}, \citet{DESCollaboration2019}, \citet{Handley2019}, \citet{Lietal2020}, \citet{EfstathiouGratton2020}, \citet{DiValentinoetal2021a}, \citet{VelasquezToribioFabris2020}, \citet{Vagnozzietal2020, Vagnozzietal2021}, \citet{KiDSCollaboration2021}, \citet{ArjonaNesseris2021}, \citet{Dhawanetal2021}, and references therein.} For the observables we study in this paper, the model predictions are determined by the Hubble parameter, the cosmological expansion rate, that is a  function of the redshift $z$ and the parameters of the cosmological model.

In these six cosmological models the Hubble parameter is
\begin{equation}
\label{eq:friedLCDM}
    H(z) = H_0\sqrt{\Omega_{m0}(1 + z)^3 + \Omega_{k0}(1 + z)^2 + \Omega_{DE}(z)}.
\end{equation}
Here $H_0$ is the Hubble constant, $\Omega_{m0}$ and $\Omega_{k0}$ are the current values of the non-relativistic matter and curvature energy density parameters, and  $\Omega_{DE}(z)$ is the dark energy density parameter. $\Omega_{k0} = 0$ in the spatially-flat models. We consider three different models of dark energy. In two of the three models we study, the dark energy density evolves as a power of $(1+z)$: $\Omega_{DE}(z) = \Omega_{DE0}(1+z)^{1+\omega_X}$ where $\Omega_{DE0}$ is the current value of the dark energy density parameter and $\omega_X$ is the dark energy equation of state parameter (see below).

In the $\Lambda$CDM model dark energy is the standard cosmological constant with $\omega_X = -1$ and $\Omega_{DE}$ = $\Omega_{DE0}$ = $\Omega_{\Lambda}$. The current values of the three $\Lambda$CDM model energy density parameters are constrained by the energy budget equation $\Omega_{m0} + \Omega_{k0} + \Omega_{\Lambda} = 1$. In the spatially-flat $\Lambda$CDM model $\Omega_{m0}$ is picked to be the free parameter while in the non-flat $\Lambda$CDM model $\Omega_{m0}$ and $\Omega_{k0}$ are the free parameters.

In the XCDM parametrization dark energy is the ideal parametrized $X$-fluid with equation of state parameter $\omega_X$ (the ratio of the $X$-fluid pressure and energy density) and $\Omega_{DE0}$ = $\Omega_{X0}$ is the current value of the dynamical $X$-fluid dark energy density parameter. The current values of the three XCDM parametrization energy density parameters are constrained by the energy budget equation $\Omega_{m0} + \Omega_{k0} + \Omega_{X0} = 1$. When $\omega_X > -1$ the $X$-fluid dark energy density decreases with time. In the spatially-flat XCDM parametrization $\Omega_{m0}$ and $\omega_X$ are the free parameters while in the non-flat XCDM parametrization, $\Omega_{m0}$, $\Omega_{k0}$, and $\omega_X$ are picked to be the free parameters. When $\omega_x = -1$ the XCDM parametrization reduces to the $\Lambda$CDM model.

In the $\phi$CDM model \citep{PeeblesRatra1988, RatraPeebles1988, Pavlovetal2013} dynamical dark energy is a scalar field $\phi$ and the dynamical dark energy density parameter $\Omega_{DE}$ dpends on the scalar field potential energy density.\footnote{For $\phi$CDM model observational constraints see \citet{Avsajanishvilietal2015}, \citet{Ryanetal2018}, \citet{SolaPeracaulaetal2018, SolaPercaulaetal2019}, \citet{Zhaietal2017}, \citet{Oobaetal2018c, Oobaetal2019}, \citet{ParkRatra2018, ParkRatra2019c, ParkRatra2020}, \citet{Sangwanetal2018}, \citet{Singhetal2019}, \citet{UrenaLopezRoy2020}, \citet{SinhaBanerjee2021}, and references therein.} We assume an inverse power law scalar field potential energy density
\begin{equation}
\label{eq:phiCDMV}
    V(\phi) = \frac{1}{2}\kappa m_{p}^2 \phi^{-\alpha}, 
\end{equation}
where $m_{p}$ is the Planck mass, $\alpha$ is a positive parameter [$\Omega_{DE}$ = $\Omega_{\phi}(z, \alpha)$ is the dynamical dark energy density parameter], and the constant $\kappa$ is determined by ensuring that, in the numerical integration, the current energy budget constraint $\Omega_{m0} + \Omega_{k0} + \Omega_{\phi}(z = 0, \alpha) = 1$ is satisfied.

With the potential energy density of eq.\ (\ref{eq:phiCDMV}), the equations of motion for a spatially homogeneous scalar field and metric tensor are
\begin{align}
\label{field}
   & \ddot{\phi} + 3\frac{\dot{a}}{a}\dot\phi - \frac{1}{2}\alpha \kappa m_{p}^2 \phi^{-\alpha - 1} = 0,  \\
\label{friedpCDM}
   & \left(\frac{\dot{a}}{a}\right)^2 = \frac{8 \uppi}{3 m_{p}^2}\left(\rho_m + \rho_{\phi}\right) - \frac{k}{a^2}.
\end{align}
Here an overdot denotes a derivative with respect to time, $a$ is the scale factor, $\rho_m$ is the non-relativistic matter energy density, for open, flat, and closed spatial hypersurfaces the curvature index $k$ is negative, zero, and positive (corresponding to $\Omega_{k0} > 0, =0, \rm and < 0$), and the scalar field energy density
\begin{equation}
    \rho_{\phi} = \frac{m^2_p}{32\pi}[\dot{\phi}^2 + \kappa m^2_p \phi^{-\alpha}].
\end{equation}
By numerically integrating eqs.\ (3) and (4), we compute $\rho_{\phi}$, and then determine  $\Omega_{\phi}(z, \alpha)$ by using
\begin{equation}
    \Omega_{\phi}(z, \alpha) = \frac{8 \uppi \rho_{\phi}}{3 m^2_p H^2_0}.
\end{equation}
In the spatially-flat $\phi$CDM model $\Omega_{m0}$ and $\alpha$ are the free parameters while in the non-flat $\phi$CDM model, $\Omega_{m0}$, $\Omega_{k0}$, and $\alpha$ are picked to be the free parameters. When $\alpha = 0$ the $\phi$CDM model reduces to the $\Lambda$CDM model.
\section{Data}
\label{sec:data}

We use the better 2036 (of 2421) QSOs sample\footnote{For some of the QSOs at $z < 0.7$ in the 2421 QSO sample the 2500 \AA\ UV flux was determined by extrapolation from the optical which is less reliable because of possible host-galaxy contamination \citep{Lussoetal2020} and so these QSOs are excluded from the 2036 QSO sample.} from the \cite{Lussoetal2020} compilation. These better QSO data consist of seven different QSO sub-samples that are listed in Table \ref{tab:QSO2}. In  \cite{KhadkaRatra2021} we discovered that the $L_X-L_{UV}$ relation parameters for the complete better QSO sample are model dependent as well as redshift dependent. This means that  some of the QSOs in the complete sample are not standardizable through the  $L_X-L_{UV}$ relation. In this paper, we analyse the individual QSO sub-samples, as well as some combinations of these sub-samples, in an attempt to determine which QSO sub-samples are more responsible for the problem we discovered in \cite{KhadkaRatra2021}. In addition to the larger individual sub-samples of Table \ref{tab:QSO2}, we also consider combinations of sub-samples that are listed in Table \ref{tab:QSO3}. In all, we study 11 different QSO sub-samples, the eight sub-samples listed in Table \ref{tab:QSO3} and the SDSS-4XMM, SDSS-Chandra, and XXL sub-samples of Table \ref{tab:QSO2}.

In this paper, we also use constraints on cosmological parameters obtained from joint analyses of 11 BAO and 31 $H(z)$ measurements to compare with the cosmological constraints obtained using the QSO sub-groups. These data are described in \cite{KhadkaRatra2021}.

\begin{table}
	\centering
	\small\addtolength{\tabcolsep}{-2.5pt}
	\small
	\caption{Subsets of the better QSO (2036 QSOs) data compilation.}
	\label{tab:QSO2}
	\begin{threeparttable}
	\begin{tabular}{lccccccccccc} 
		\hline
		Data set & Redshift range & Number of QSOs \\
		\hline
		SDSS-4XMM & $0.703 \leq z \leq 3.412$ & 1355\\
		SDSS-Chandra & $0.7031 \leq z \leq 3.981$ & 542\\
		XXL & $0.7061 \leq z \leq 1.9247$ & 76\\
		Local AGN & $0.009 \leq z \leq 0.0866$ & 13\\
		High-$z$ & $4.01 \leq z \leq 7.5413$ & 35\\
		XMM-Newton$\sim 3$ & $3.028 \leq z \leq 3.296$ & 14\\
		XMM-Newton$\sim 4$ & 4.109 & 1\\
		\hline
	\end{tabular}
    \end{threeparttable}
\end{table}

\begin{table}
	\centering
	\small\addtolength{\tabcolsep}{-3.5pt}
	\small
	\caption{Combinations of subsets of better QSO (2036 QSOs) data.}
	\label{tab:QSO3}
	\begin{threeparttable}
	\begin{tabular}{lccccccccccc} 
		\hline
		Data set & Redshift range & Number of QSOs \\
		\hline
		SDSS-4XMM-l\tnote{a} & $0.703 \leq z \leq 1.4615$ & 678\\
		SDSS-4XMM-h\tnote{b} & $1.4655 \leq z \leq 3.412$ & 677\\
		Chandra + Newton-3 & $0.7031 \leq z \leq 3.981$ & 556\\
		High-$z$ + Newton-4 & $4.01 \leq z \leq 7.5413$ & 36\\
		Chandra + XXL & $0.7031 \leq z \leq 3.981$ & 618\\
		Chandra + High-$z$ + Newton-4 & $0.7031 \leq z \leq 7.5413$ & 578\\
		Chandra + XXL + Newton-3 & $0.7031 \leq z \leq 3.981$ & 632\\
		QSO\tnote{c} & $0.7031 \leq z \leq 7.5413$ & 668\\
		\hline
	\end{tabular}
	\begin{tablenotes}
	\item[a] Lower $z$ half of SDSS-4XMM.
	\item[b] Higher $z$ half of SDSS-4XMM.
    \item[c] SDSS-Chandra + XXL + High-$z$ + Newton-3 + Newton-4.
    \end{tablenotes}
    \end{threeparttable}
\end{table}

\section{Methods}
\label{sec:methods}

The non-linear $L_X-L_{UV}$ relation relates the X-ray and UV luminosities of selected QSOs. This relation is
\begin{equation}
\label{eq:xuv}
    \log(L_{X}) = \beta + \gamma \log(L_{UV}) ,
\end{equation}
where $\log$ = $\log_{10}$ and $\gamma$ and $\beta$ are free parameters to be measured from the data.

If we express luminosities in terms of fluxes, eq.\ (\ref{eq:xuv}) becomes
\begin{equation}
\label{eq:fx}
    \log(F_{X}) = \beta +(\gamma - 1)\log(4\pi) + \gamma \log(F_{UV}) + 2(\gamma - 1)\log(D_L).
\end{equation}
Here $F_X$ and $F_{UV}$ are the quasar X-ray and UV fluxes and $D_L(z, p)$ is the luminosity distance. The luminosity distance can be computed in each cosmological model at a given redshift $z$ and for a given set of cosmological model parameters $p$, 
\begin{equation}
\label{eq:DL}
  \frac{H_0\sqrt{\left|\Omega_{k0}\right|}D_L(z, p)}{(1+z)} = 
    \begin{cases}
    {\rm sinh}\left[g(z)\right] & \text{if}\ \Omega_{k0} > 0, \\
    \vspace{1mm}
    g(z) & \text{if}\ \Omega_{k0} = 0,\\
    \vspace{1mm}
    {\rm sin}\left[g(z)\right] & \text{if}\ \Omega_{k0} < 0,
    \end{cases}   
\end{equation}
where
\begin{equation}
\label{eq:gz}
   g(z) = H_0\sqrt{\left|\Omega_{k0}\right|}\int^z_0 \frac{dz'}{H(z')},
\end{equation}
and $H(z)$ is a Hubble parameter described in Sec.\ 2 for each model we study.

Once we compute predicted X-ray fluxes using eqs.\ (\ref{eq:fx}) and (\ref{eq:DL}) in a given model, we compare these predicted fluxes with observations by using the likelihood function \citep{Dago2005}
\begin{equation}
\label{eq:chi2}
    \ln({\rm LF}) = -\frac{1}{2}\sum^{N}_{i = 1} \left[\frac{[\log(F^{\rm obs}_{X,i}) - \log(F^{\rm th}_{X,i})]^2}{s^2_i} + \ln(2\pi s^2_i)\right].
\end{equation}
Here $\ln$ = $\log_e$, $s^2_i = \sigma^2_i + \delta^2$, where $\sigma_i$ and $\delta$ are the measurement error on the observed flux $F^{\rm obs}_{X,i}$ and the global intrinsic dispersion\footnote{The intrinsic dispersion of the $L_X-L_{UV}$ relation quantifies unaccounted errors in the measurements. The larger value of the intrinsic dispersion for current QSO data is one of the reasons they are unable to provide precise cosmological constraints. The source of this dispersion is unknown and could include quasar variability and unknown systematics in the measurements \citep{RisalitiLusso2015}.} respectively, and $F^{\rm th}_{X,i}(p)$ is the predicted flux at redshift $z_i$. In the $L_X-L_{UV}$ relation, $\beta$ and $H_0$ are degenerate so QSO data alone cannot constrain $H_0$. We allow $H_0$ to be a free parameter to determine the complete allowed range of $\beta$. To avoid the QSO data circularity problem we simultaneously determine the $L_X-L_{UV}$ relation parameters and the cosmological model parameters when analyzing data. By studying a number of different cosmological models we are able to determine whether the $L_X-L_{UV}$ relation parameters are independent of the cosmological model in which they were measured. If the parameters of the $L_X-L_{UV}$ relation depend on the cosmological model then the corresponding QSOs are not standardized candles and so cannot be used for cosmological parameter estimation.

The BAO + $H(z)$ data constraints used in this paper are from \cite{KhadkaRatra2021}. A detailed description of the method used for these data is given in Sec.\ 4 of \cite{KhadkaRatra2021}.

We use the Markov chain Monte Carlo (MCMC) method, as implemented in the \textsc{MontePython} code \citep{Brinckmann2019}, to perform the likelihood analyses. For each free parameter, we use the Gelman-Rubin criterion $(R-1 < 0.05)$ to determine the convergence of the MCMC chains. For each of the free parameter we use a flat prior non-zero over the range listed in Table \ref{tab:prior}.

\begin{table}
	\centering
	\small
	\caption{Non-zero flat prior parameter ranges.}
	\label{tab:prior}
	\begin{threeparttable}
	\begin{tabular}{l|c}
	\hline
	Parameter & Prior range \\
	\hline
	$\Omega_bh^2$ & $[0, 1]$ \\
	$\Omega_ch^2$ & $[0, 1]$ \\
    $\Omega_{m0}$ & $[0, 1]$ \\
    $\Omega_{k0}$ & $[-2, 2]$ \\
    $\omega_{X}$ & $[-5, 0.33]$ \\
    $\alpha$ & $[0, 10]$ \\
    $\delta$ & $[0, 10]$ \\
    $\beta$ & $[0, 11]$ \\
    $\gamma$ & $[-5, 5]$ \\
	\hline
	\end{tabular}
    \end{threeparttable}
\end{table}
\section{Results}
\label{sec:QSO}

The BAO + $H(z)$ data results are listed in Table 5. These are from \cite{KhadkaRatra2021} and are discussed in Sec.\ 5.3 of that paper. In this paper we use these BAO + $H(z)$ results to compare with the cosmological constraints obtained using the QSO sub-group samples to see whether the QSO sub-group sample results are consistent or not with the better-established BAO + $H(z)$ ones. These BAO + $H(z)$ results are shown in red in all figures. Results for QSO sub-group samples are given in Table \ref{tab:margr}. The QSO one-dimensional likelihood distributions and two-dimensional likelihood contours are shown in blue or green in Figs.\ 1--10.

The $L_X-L_{UV}$ relation parameters values depend on the QSO data sub-set studied. The intercept $\beta$ ranges from $7.290^{+0.400}_{-0.400}$ to $10.240^{+0.600}_{-0.600}$. The minimum value is obtained using the SDSS-4XMM QSO sample in the flat $\Lambda$CDM case while the maximum value is obtained using the SDSS-4XMM-h sample (the higher redshift half of the SDSS-4XMM sub-set) in the non-flat XCDM case. The difference between maximum and minimum values is 4.1$\sigma$ and statistically significant. The slope $\gamma$ ranges from $0.536^{+0.012}_{-0.019}$ to $0.630^{+0.013}_{-0.013}$. The minimum value is obtained using the SDSS-4XMM-h QSO sample in the non-flat XCDM case while the maximum value is obtained using the SDSS-4XMM QSO sample in the spatially-flat $\Lambda$CDM model. The difference between maximum and minimum values is 5.3$\sigma$ and statistically significant. The intrinsic dispersion ($\delta$) is the third free parameter of the $L_X-L_{UV}$ relation. For all data sub-sets, values of $\delta$ lie in the range $\sim 0.2$--$0.24$. 

The SDSS-4XMM and SDSS-Chandra sub-samples are large sub-sets containing 1355 and 542 QSOs respectively, and when one of these sub-groups is part of the data under analysis it effectively determines the $L_X-L_{UV}$ relation parameter values. 

For the SDSS-4XMM QSOs, comparing $\beta$ and $\gamma$ values listed in the last two columns of Table \ref{tab:margr} for each of the six cosmological models, we see that these are significantly model-dependent. This means that the $L_X-L_{UV}$ relation for the SDSS-4XMM QSOs depends on the cosmological model in which it is estimated. This indicates that current SDSS-4XMM QSOs are not standardizable through the $L_X-L_{UV}$ relation and so cannot be used for cosmological parameter estimation purposes. This is similar to the result found in \cite{KhadkaRatra2021} for the complete QSO sample. Given that the SDSS-4XMM sub-set is by far the largest QSO sub-set, our finding here establishes that the SDSS-4XMM sub-set is a significant driver of the earlier result for the complete QSO compilation \citep{KhadkaRatra2021}. The SDSS-4XMM QSO sample quantitative differences between $\beta$ and $\gamma$ values for different pairs of cosmological model are listed in Table 6. The difference between $\gamma$ values $(\Delta \gamma)$ from model to model ranges over $(0-3.54)\sigma$ which is statistically significant. The difference between $\beta$ values $(\Delta \beta)$ from model to model ranges over $(0-3.82)\sigma$ which is statistically significant.

In an attempt to track down the cause of this effect, we divide SDSS-4XMM QSOs into two sub-sets, the lower redshift half, SDSS-4XMM-l, with QSOs at $z \leq 1.4615$ and the higher redshift half, SDSS-4XMM-h, with QSOs at $z > 1.4615$. For a given cosmological model, these two data sub-sets give different $\beta$ and $\gamma$ values. The differences in $\beta$ and $\gamma$ values for the two data sub-sets, for six different cosmological models, are listed in Table 7. In the six models, the difference between $\gamma$ values $(\Delta \gamma)$ range over $(1.55-2.50)\sigma$ and that between $\beta$ values $(\Delta \beta)$ range over $(1.62-2.14)\sigma$. These are statistically significant and show that in their current form SDSS-4XMM QSOs in different redshift regions follow different $L_X-L_{UV}$ relations. Whether this is because the SDSS-4XMM QSO $L_X-L_{UV}$ relation physically evolves, or some  other effect causes this, remains to be established. Another possibly significant result of our analyses of these lower and higher redshift sub-sets is that both low and high $z$ SDSS-4XMM QSOs follow model-independent (but different) $L_X-L_{UV}$ relations. This can be seen from Tables 8 and 9. This indicates that narrower-redshift bins of SDSS-4XMM QSOs obey model-independent $L_X-L_{UV}$ relations and that possibly the model-dependency and the redshift-dependency of $\beta$ and $\gamma$ values for these are related.
\clearpage
\onecolumn
\begin{landscape}
\addtolength{\tabcolsep}{-1pt}
\begin{longtable}{lcccccccccc}
\caption{Marginalized one-dimensional best-fit parameters with 1$\sigma$ confidence intervals, or 2$\sigma$ limits, for sub-groups of the 2036 better QSOs compilation.}
\label{tab:margr}\\
\hline
Model & Data set\hspace{5mm} & $\om$ & $\ol$\footnotesize{$^c$} & $\ok$ & $\omega_{X}$ & $\alpha$ & $H_0$\footnotesize{$^a$} & $\delta$ & $\beta$ & $\gamma$ \\
\hline
\endfirsthead
\hline
Model & Data set\hspace{5mm} & $\om$ & $\ol$\footnotesize{$^c$} & $\ok$ & $\omega_{X}$ & $\alpha$ & $H_0$\footnotesize{$^a$} & $\delta$ & $\beta$ & $\gamma$ \\
\hline
\endhead
\hline
Flat \lcdm\ & SDSS-4XMM & $>0.772$ & < 0.228 & - & - & - &-& $0.225^{+0.005}_{-0.005}$ & $7.290^{+0.400}_{-0.400}$ & $0.630^{+0.013}_{-0.013}$\\
& SDSS-Chandra & $>0.348$ & < 0.652 & - & - & - &-& $0.238^{+0.008}_{-0.008}$ & $8.190^{+0.540}_{-0.540}$ & $0.602^{+0.018}_{-0.018}$\\
& XXL & --- & --- & - & - & - &-& $0.211^{+0.016}_{-0.020}$ & $7.300^{+1.500}_{-1.500}$ & $0.632^{+0.049}_{-0.049}$\\
& Chandra + Newton-3 & $>0.350$ & < 0.650 & - & - & - &-& $0.236^{+0.008}_{-0.008}$ & $7.960^{+0.500}_{-0.500}$ & $0.609^{+0.017}_{-0.017}$\\
& High-$z$ + Newton-4 & --- & --- & - & - & - &-& $0.198^{+0.034}_{-0.047}$ & $9.250^{+1.700}_{-0.560}$ & $0.575^{+0.015}_{-0.055}$\\
& Chandra + XXL & $> 0.378$ & < 0.622 & - & - & - &-& $0.234^{+0.008}_{-0.008}$ & $8.070^{+0.510}_{-0.510}$ & $0.605^{+0.017}_{-0.017}$\\
& Chandra + High-$z$ + Newton-4 & $> 0.435$ & < 0.565 & - & - & - &-& $0.238^{+0.008}_{-0.008}$ & $7.820^{+0.500}_{-0.500}$ & $0.614^{+0.017}_{-0.017}$\\
& Chandra + XXL + Newton-3 & $> 0.384$ & < 0.616 & - & - & - &-& $0.232^{+0.007}_{-0.007}$ & $7.850^{+0.470}_{-0.470}$ & $0.613^{+0.016}_{-0.016}$\\
& SDSS-4XMM-l & $> 0.360$ & < 0.640 & - & - & - &-& $0.233^{+0.007}_{-0.007}$ & $8.000^{+0.680}_{-0.680}$ & $0.606^{+0.023}_{-0.023}$\\
& SDSS-4XMM-h & $> 0.522$ & < 0.478 & - & - & - &-& $0.207^{+0.006}_{-0.006}$ & $9.500^{+0.570}_{-0.570}$ & $0.559^{+0.019}_{-0.019}$\\
& QSO$^{b}$ & $> 0.462$ & < 0.538 & - & - & - &-& $0.232^{+0.007}_{-0.007}$ & $7.590^{+0.440}_{-0.440}$ & $0.621^{+0.015}_{-0.015}$\\
&  BAO + $H(z)$ & $0.299^{+0.015}_{-0.017}$ & $0.700^{+0.017}_{-0.015}$ & - & - & - &$69.300^{+1.800}_{-1.800}$&-&-&-\\
\hline
Non-flat \lcdm\ & SDSS-4XMM & $> 0.601$ & $1.704^{+0.069}_{-0.059}$ & $-1.530^{+0.204}_{-0.106}$ & - &-&-& $0.219^{+0.005}_{-0.005}$ & $9.470^{+0.470}_{-0.470}$ & $0.559^{+0.016}_{-0.016}$\\
& SDSS-Chandra & $0.264^{+0.426}_{-0.084}$ & $1.210^{+0.280}_{-0.033}$ & $-0.660^{+0.250}_{-0.390}$ & - &-&-& $0.236^{+0.008}_{-0.008}$ & $9.070^{+0.700}_{-0.700}$ & $0.574^{+0.023}_{-0.023}$\\
& XXL & --- & $< 1.700$ & $-0.018^{+1.198}_{-0.582}$ & - &-&-& $0.212^{+0.016}_{-0.020}$ & $7.400^{+1.500}_{-1.500}$ & $0.626^{+0.049}_{-0.049}$\\
& Chandra + Newton-3 & $0.247^{+0.393}_{-0.087}$ & $1.250^{+0.230}_{-0.011}$ & $-0.648^{+0.168}_{-0.322}$ & - &-&-& $0.234^{+0.008}_{-0.008}$ & $9.030^{+0.690}_{-0.690}$ & $0.576^{+0.022}_{-0.022}$\\
& High-$z$ + Newton-4 & $> 0.161$ & $< 1.800$ & $-0.398^{+1.058}_{-0.892}$ & - &-&-& $0.196^{+0.038}_{-0.049}$ & $9.200^{+1.800}_{-0.730}$ & $0.574^{+0.024}_{-0.058}$\\
& Chandra + XXL & $0.287^{+0.373}_{-0.107}$ & $1.260^{+0.260}_{-0.067}$ & $-0.697^{+0.187}_{-0.293}$ & - &-&-& $0.232^{+0.008}_{-0.008}$ & $9.020^{+0.660}_{-0.660}$ & $0.576^{+0.022}_{-0.022}$\\
& Chandra + High-$z$ + Newton-4 & $0.349^{+0.301}_{-0.129}$ & $1.365^{+0.093}_{-0.074}$ & $-0.752^{+0.192}_{-0.308}$ & - &-&-& $0.234^{+0.008}_{-0.008}$ & $9.340^{+0.630}_{-0.630}$ & $0.566^{+0.021}_{-0.021}$\\
& Chandra + XXL + Newton-3 & $0.258^{+0.352}_{-0.088}$ & $1.330^{+0.150}_{-0.050}$ & $-0.670^{+0.160}_{-0.300}$ & - &-&-& $0.230^{+0.007}_{-0.007}$ & $9.020^{+0.630}_{-0.630}$ & $0.576^{+0.020}_{-0.020}$\\
& SDSS-4XMM-l & $> 0.387$ & $< 1.900$ & $-1.175^{+1.235}_{-0.445}$ & - &-&-& $0.232^{+0.007}_{-0.007}$ & $8.180^{+0.720}_{-0.720}$ & $0.600^{+0.024}_{-0.024}$\\
& SDSS-4XMM-h & $> 0.376$ & $1.570^{+0.130}_{-0.110}$ & $-1.448^{+0.348}_{-0.202}$ & - &-&-& $0.203^{+0.006}_{-0.006}$ & $10.230^{+0.630}_{-0.630}$ & $0.535^{+0.010}_{-0.020}$\\
& QSO$^{b}$ & $0.367^{+0.273}_{-0.137}$ & $1.385^{+0.084}_{-0.075}$ & $-0.788^{+0.188}_{-0.292}$ & - &-&-& $0.228^{+0.007}_{-0.007}$ & $9.180^{+0.590}_{-0.590}$ & $0.571^{+0.019}_{-0.019}$\\
& BAO + $H(z)$ & $0.292^{+0.023}_{-0.023}$ & $0.667^{+0.093}_{-0.081}$ & $-0.014^{+0.075}_{-0.075}$ & - & - &$68.700^{+2.300}_{-2.300}$&-&-&-\\
\hline
Flat XCDM & SDSS-4XMM & --- & - & - & $0.105^{+0.555}_{-0.755}$ & - &-& $0.224^{+0.005}_{-0.005}$ & $7.480^{+0.420}_{-0.420}$ & $0.623^{+0.014}_{-0.014}$\\
& SDSS-Chandra & > 0.207 & - & - & $< -0.063$ & - &-& $0.238^{+0.008}_{-0.008}$ & $8.260^{+0.540}_{-0.540}$ & $0.600^{+0.018}_{-0.018}$\\
& XXL & --- & - & - & $-0.212$ & - &-& $0.212^{+0.017}_{-0.021}$ & $7.400^{+1.600}_{-1.500}$ & $0.627^{+0.049}_{-0.049}$\\
& Chandra + Newton-3 & > 0.219 & - & - & $< -0.078$ & - &-& $0.236^{+0.008}_{-0.008}$ & $8.020^{+0.500}_{-0.500}$ & $0.608^{+0.017}_{-0.017}$\\
& High-$z$ + Newton-4 & --- & - & - & $< -0.180$ & - &-& $0.198^{+0.036}_{-0.048}$ & $9.260^{+1.800}_{-0.700}$ & $0.575^{+0.023}_{-0.056}$\\
& Chandra + XXL & > 0.248 & - & - & $< -0.100$ & - &-& $0.234^{+0.008}_{-0.008}$ & $8.130^{+0.500}_{-0.500}$ & $0.604^{+0.017}_{-0.017}$\\
& Chandra + High-$z$ + Newton-4 & > 0.242 & - & - & $< 0.038$ & - &-& $0.238^{+0.008}_{-0.008}$ & $7.890^{+0.500}_{-0.500}$ & $0.612^{+0.017}_{-0.017}$\\
& Chandra + XXL + Newton-3 & > 0.248 & - & - & $< -0.083$ & - &-& $0.232^{+0.008}_{-0.008}$ & $7.930^{+0.470}_{-0.470}$ & $0.611^{+0.015}_{-0.015}$\\
& SDSS-4XMM-l & > 0.234 & - & - & $< 0.003$ & - &-& $0.233^{+0.007}_{-0.007}$ & $8.040^{+0.680}_{-0.680}$ & $0.605^{+0.022}_{-0.022}$\\
& SDSS-4XMM-h & --- & - & - & $< 0.112$ & - &-& $0.207^{+0.007}_{-0.007}$ & $9.490^{+0.680}_{-0.580}$ & $0.560^{+0.019}_{-0.022}$\\
& QSO$^{b}$ & > 0.238 & - & - & $< 0.066$ & - &-& $0.232^{+0.008}_{-0.008}$ & $7.650^{+0.440}_{-0.440}$ & $0.620^{+0.015}_{-0.015}$\\
&BAO + $H(z)$ & $0.282^{+0.021}_{-0.021}$ & - & - & $-0.744^{+0.140}_{-0.097}$ & - &$65.800^{+2.200}_{-2.500}$& - & - & -\\
\hline
Non-flat XCDM & SDSS-4XMM & $> 0.362$ & - & $-0.994^{+0.344}_{-0.216}$ & $< -1.300$ & - &-& $0.219^{+0.005}_{-0.005}$ & $9.650^{+0.480}_{-0.480}$ & $0.557^{+0.016}_{-0.016}$\\
& SDSS-Chandra & $> 0.137$ & - & $-0.581^{+0.511}_{-0.509}$ & $-0.767^{+0.757}_{+1.393}$ & - &-& $0.237^{+0.008}_{-0.009}$ & $8.910^{+0.720}_{-0.720}$ & $0.579^{+0.023}_{-0.023}$\\
& XXL & --- & - & $0.289^{+1.061}_{-0.059}$ & $-0.939^{+0.739}_{2.361}$ & - &-& $0.212^{+0.017}_{-0.017}$ & $7.300^{+1.500}_{-1.500}$ & $0.629^{+0.050}_{-0.050}$\\
& Chandra + Newton-3 & $0.244^{+0.476}_{-0.104}$ & - & $-0.581^{+0.471}_{-0.539}$ & $-0.608^{+0.418}_{+1.382}$ & - &-& $0.235^{+0.007}_{-0.009}$ & $8.840^{+0.720}_{-0.720}$ & $0.582^{+0.024}_{-0.024}$\\
& High-$z$ + Newton-4 & --- & - & $-0.154^{+0.904}_{-0.856}$ & --- & - &-& $0.196^{+0.037}_{-0.048}$ & $9.220^{+1.800}_{-0.710}$ & $0.573^{+0.023}_{-0.056}$\\
& Chandra + XXL & --- & - & $-0.588^{+0.578}_{-0.322}$ & $-0.801^{+0.751}_{-1.389}$ & - &-& $0.233^{+0.007}_{-0.008}$ & $8.870^{+0.710}_{-0.710}$ & $0.581^{+0.025}_{-0.022}$\\
& Chandra + High-$z$ + Newton-4 & $0.324^{+0.356}_{-0.174}$ & - & $-0.643^{+0.423}_{-0.547}$ & $-0.892^{+0.632}_{-1.028}$ & - &-& $0.235^{+0.009}_{-0.009}$ & $9.250^{+0.820}_{-0.650}$ & $0.569^{+0.021}_{-0.027}$\\
& Chandra + XXL + Newton-3 & $0.274^{+0.456}_{-0.134}$ & - & $-0.603^{+0.463}_{-0.507}$ & $-0.837^{+0.707}_{-1.203}$ & - &-& $0.231^{+0.008}_{-0.008}$ & $8.860^{+0.690}_{-0.690}$ & $0.582^{+0.022}_{-0.022}$\\
& SDSS-4XMM-l & $0.207$ & - & $-0.600^{+0.770}_{-0.570}$ & $< 0.300$ & - &-& $0.233^{+0.007}_{-0.007}$ & $8.250^{+0.720}_{-0.720}$ & $0.598^{+0.024}_{-0.024}$\\
& SDSS-4XMM-h & $> 0.358$ & - & $-1.018^{+0.358}_{-0.292}$ & $< -0.500$ & - &-& $0.203^{+0.006}_{-0.006}$ & $10.240^{+0.600}_{-0.600}$ & $0.536^{+0.012}_{-0.019}$\\
& QSO$^{b}$ & $0.324^{+0.325}_{-0.184}$ & - & $-0.705^{+0.465}_{-0.526}$ & $-0.902^{+0.572}_{-0.848}$ & - &-& $0.229^{+0.008}_{-0.008}$ & $9.060^{+0.750}_{-0.640}$ & $0.575^{+0.021}_{-0.024}$\\
& BAO + $H(z)$ & $0.293^{+0.027}_{-0.027}$ & - & $-0.120^{+0.130}_{-0.130}$ & $-0.693^{+0.130}_{-0.077}$ & - &$65.900^{+2.400}_{-2.400}$& - & - & -\\
\hline
Flat $\phi$CDM & SDSS-4XMM & $> 0.710$ & - & - & - & --- &-& $0.225^{+0.005}_{-0.005}$ & $7.290^{+0.390}_{-0.390}$ & $0.630^{+0.013}_{-0.013}$\\
& SDSS-Chandra & $> 0.281$ & - & - & - & --- &-& $0.238^{+0.008}_{-0.008}$ & $8.200^{+0.540}_{-0.540}$ & $0.601^{+0.018}_{-0.018}$\\
& XXL & --- & - & - & - & --- &-& $0.211^{+0.015}_{-0.020}$ & $7.400^{+1.500}_{-1.500}$ & $0.626^{+0.048}_{-0.048}$\\
& Chandra + Newton-3 & $> 0.291$ & - & - & - & --- &-& $0.236^{+0.008}_{-0.008}$ & $7.960^{+0.500}_{-0.500}$ & $0.609^{+0.016}_{-0.016}$\\
& High-$z$ + Newton-4 & --- & - & - & - & --- &-& $0.197^{+0.036}_{-0.048}$ & $9.280^{+1.800}_{-0.690}$ & $0.573^{+0.022}_{-0.056}$\\
& Chandra + XXL & $> 0.311$ & - & - & - & --- &-& $0.234^{+0.007}_{-0.007}$ & $8.080^{+0.510}_{-0.510}$ & $0.605^{+0.017}_{-0.017}$\\
& Chandra + High-$z$ + Newton-4 & $> 0.394$ & - & - & - & --- &-& $0.239^{+0.008}_{-0.008}$ & $7.820^{+0.500}_{-0.500}$ & $0.613^{+0.016}_{-0.016}$\\
& Chandra + XXL + Newton-3 & $> 0.321$ & - & - & - & --- &-& $0.232^{+0.007}_{-0.007}$ & $7.850^{+0.470}_{-0.470}$ & $0.612^{+0.015}_{-0.015}$\\
& SDSS-4XMM-l & $> 0.243$ & - & - & - & --- &-& $0.233^{+0.007}_{-0.007}$ & $7.990^{+0.670}_{-0.670}$ & $0.606^{+0.022}_{-0.022}$\\
& SDSS-4XMM-h & $> 0.471$ & - & - & - & --- &-& $0.207^{+0.006}_{-0.006}$ & $9.470^{+0.580}_{-0.580}$ & $0.560^{+0.019}_{-0.019}$\\
& QSO$^{b}$ & $> 0.420$ & - & - & - & --- &-& $0.232^{+0.007}_{-0.007}$ & $7.590^{+0.440}_{-0.440}$ & $0.621^{+0.014}_{-0.014}$\\
& BAO + $H(z)$ & $0.266^{+0.023}_{-0.023}$ & - & - & - & $1.530^{+0.620}_{-0.850}$ &$65.100^{+2.100}_{-2.100}$& - & - & -\\
\hline
Non-flat $\phi$CDM & SDSS-4XMM & $> 0.816$ & - & $-0.915^{+0.101}_{-0.048}$ & - & --- &-& $0.223^{+0.005}_{-0.005}$ & $7.820^{+0.420}_{-0.420}$ & $0.612^{+0.014}_{-0.014}$\\
& SDSS-Chandra& $> 0.369$ & - & $-0.259^{+0.299}_{+0.361}$ & - & --- & $0.238^{+0.008}_{-0.008}$ & $8.370^{+0.570}_{-0.570}$ & $0.595^{+0.019}_{-0.019}$\\
& XXL & --- & - & $0.008^{+0.412}_{+0.308}$ & - & --- &-& $0.211^{+0.016}_{-0.021}$ & $7.400^{+1.500}_{-1.500}$ & $0.627^{+0.049}_{-0.049}$\\
& Chandra + Newton-3 & $> 0.380$ & - & $-0.288^{+0.248}_{-0.352}$ & - & --- &-& $0.235^{+0.008}_{-0.008}$ & $8.150^{+0.530}_{-0.530}$ & $0.602^{+0.018}_{-0.018}$\\
& High-$z$ + Newton-4 & --- & - & $-0.033^{+0.333}_{-0.367}$ & - & --- &-& $0.197^{+0.036}_{-0.048}$ & $9.290^{+1.800}_{-0.690}$ & $0.572^{+0.022}_{-0.056}$\\
& Chandra + XXL & $> 0.404$ & - & $-0.313^{+0.253}_{-0.347}$ & - & --- &-& $0.234^{+0.008}_{-0.008}$ & $8.270^{+0.540}_{-0.540}$ & $0.598^{+0.018}_{-0.018}$\\
& Chandra + High-$z$ + Newton-4 & $> 0.513$ & - & $-0.723^{+0.373}_{-0.187}$ & - & --- &-& $0.237^{+0.008}_{-0.008}$ & $8.200^{+0.560}_{-0.560}$ & $0.600^{+0.018}_{-0.018}$\\
& Chandra + XXL + Newton-3 & $> 0.420$ & - & $-0.329^{+0.249}_{-0.351}$ & - & --- &-& $0.232^{+0.007}_{-0.007}$ & $8.080^{+0.500}_{-0.500}$ & $0.604^{+0.017}_{-0.017}$\\
& SDSS-4XMM-l & $> 0.265$ & - & $-0.156^{+0.326}_{-0.384}$ & - & --- &-& $0.233^{+0.007}_{-0.007}$ & $8.020^{+0.670}_{-0.670}$ & $0.605^{+0.022}_{-0.022}$\\
& SDSS-4XMM-h & $> 0.598$ & - & $-0.788^{+0.288}_{-0.172}$ & - & --- &-& $0.206^{+0.006}_{-0.006}$ & $9.650^{+0.680}_{-0.560}$ & $0.553^{+0.018}_{-0.022}$\\
& QSO$^b$ & $> 0.545$ & - & $-0.745^{+0.326}_{-0.185}$ & - & --- &-& $0.231^{+0.007}_{-0.007}$ & $8.000^{+0.490}_{-0.490}$ & $0.607^{+0.016}_{-0.016}$\\
& BAO + $H(z)$ & $0.271^{+0.024}_{-0.028}$ & - & $-0.080^{+0.100}_{-0.100}$ & - & $1.660^{+0.670}_{-0.830}$ &$65.500^{+2.500}_{-2.500}$& - & - & -\\
\hline
\end{longtable}
\footnotesize{$\hspace{-0.6cm}^a$ ${\rm km}\hspace{1mm}{\rm s}^{-1}{\rm Mpc}^{-1}$.}\\
\footnotesize{$^b$ Chandra + XXL + High-$z$ + Newton$-3$ + Newton$-4$}\\
\footnotesize{$^c$ Here $\Omega_{\Lambda}$ is a derived parameter and $\Omega_{\Lambda}$ chains are derived using the current energy budget equation $\Omega_{\Lambda}= 1-\Omega_{m0}-\Omega_{k0}$ (where in the flat $\Lambda$CDM model $\Omega_{k0}=0$). We use the \textsc{python} package \textsc{getdist} \citep{Lewis_2019} to determine best-fit values and uncertainties for $\Omega_{\Lambda}$ from these chains. We also use this \textsc{python} package to compute the best-fit values and uncertainties of the free parameters and plot the likelihoods.}
\end{landscape}
\clearpage
\twocolumn 

\begin{figure*}
\begin{multicols}{2}
    \includegraphics[width=\linewidth,height=7cm]{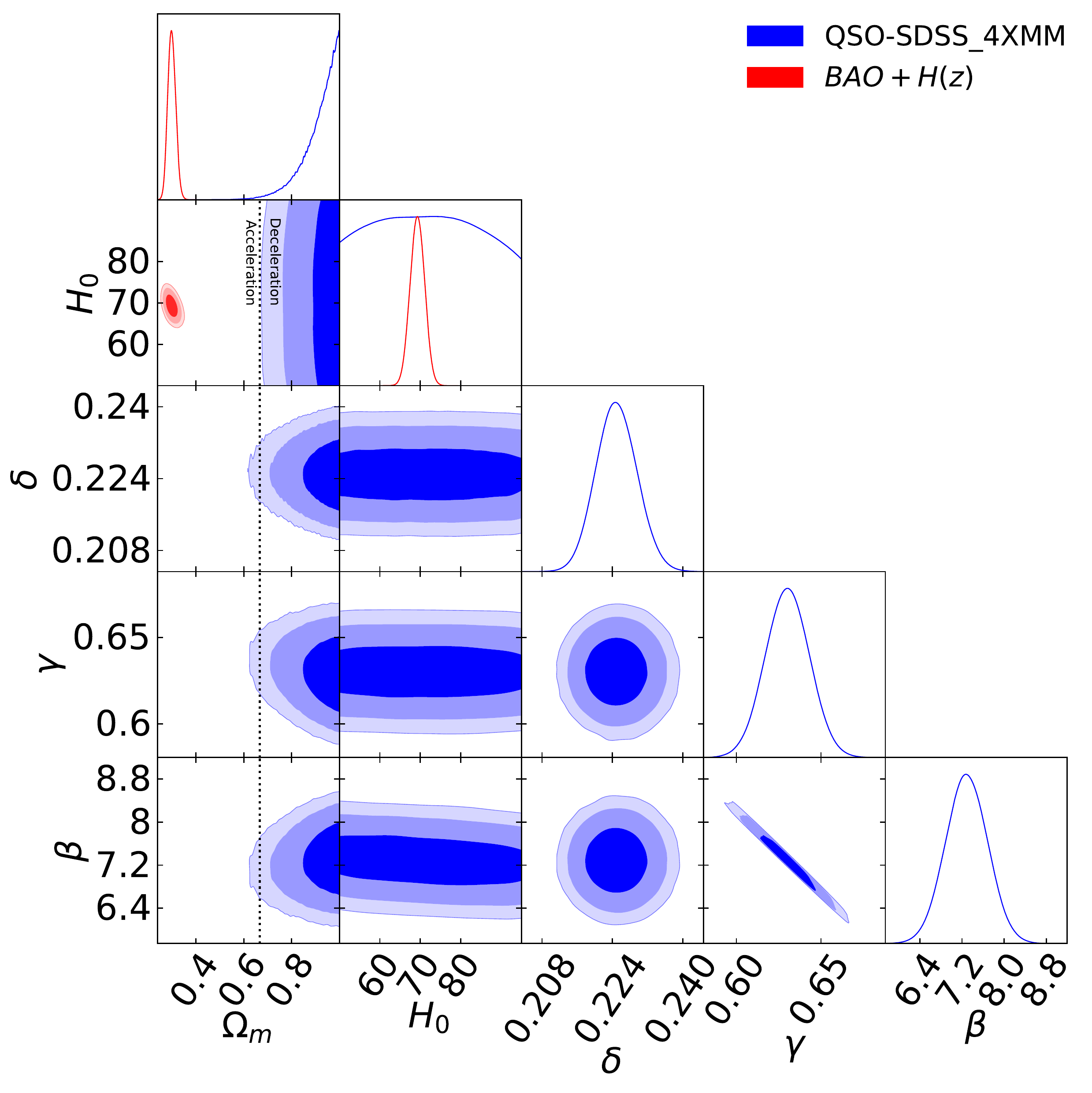}\par
    \includegraphics[width=\linewidth,height=7cm]{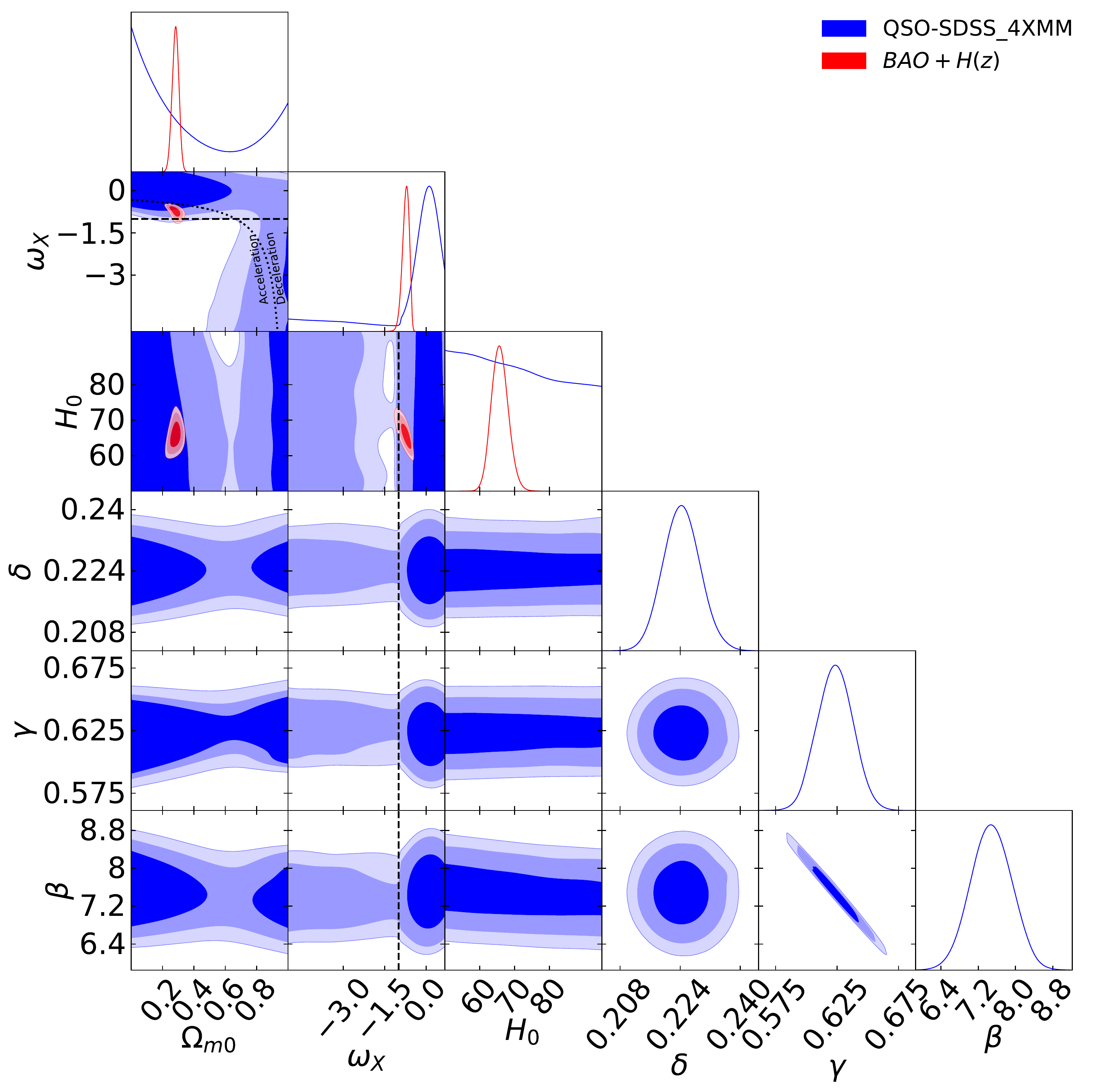}\par
    \includegraphics[width=\linewidth,height=7cm]{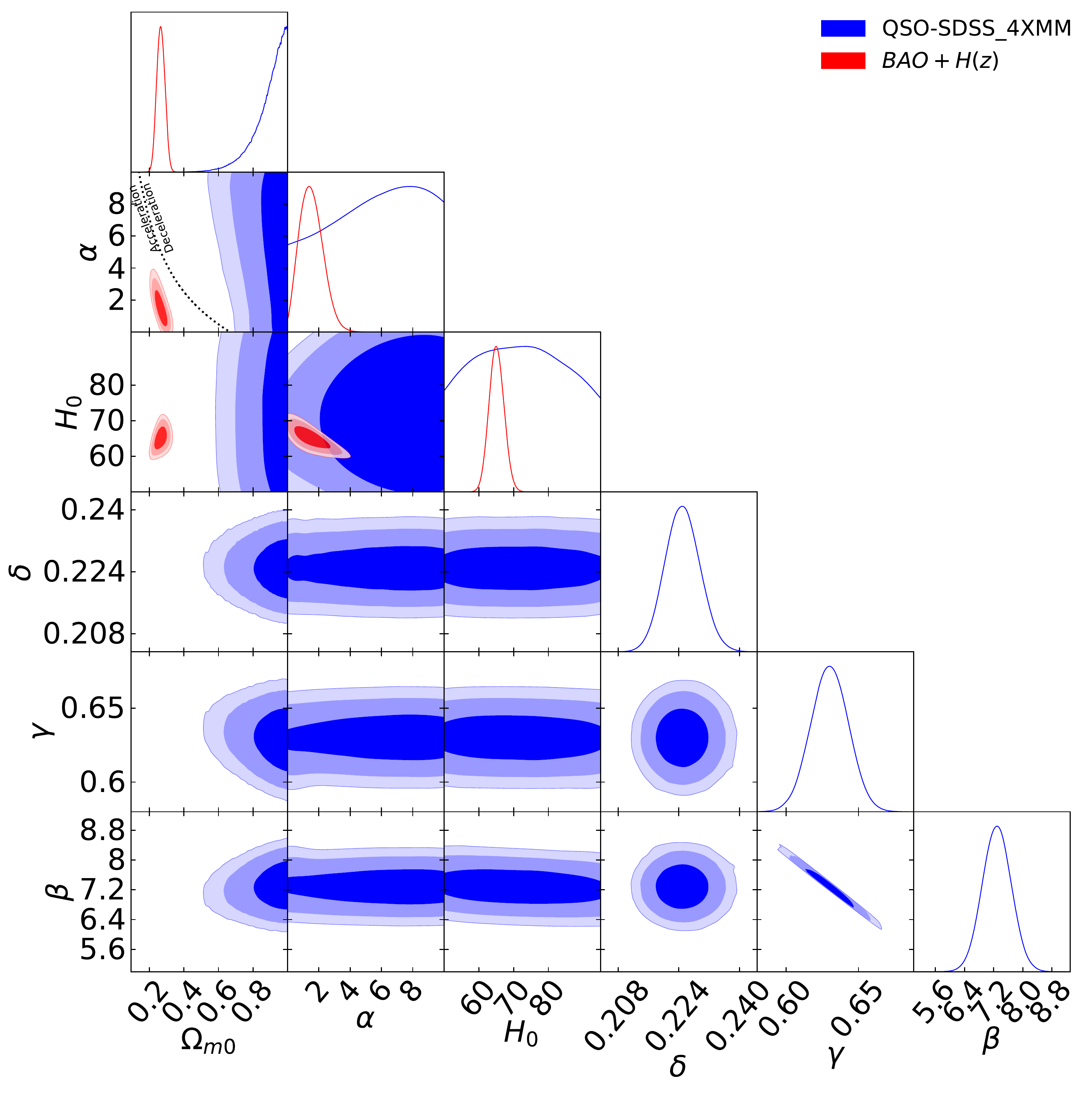}\par
    \includegraphics[width=\linewidth,height=7cm]{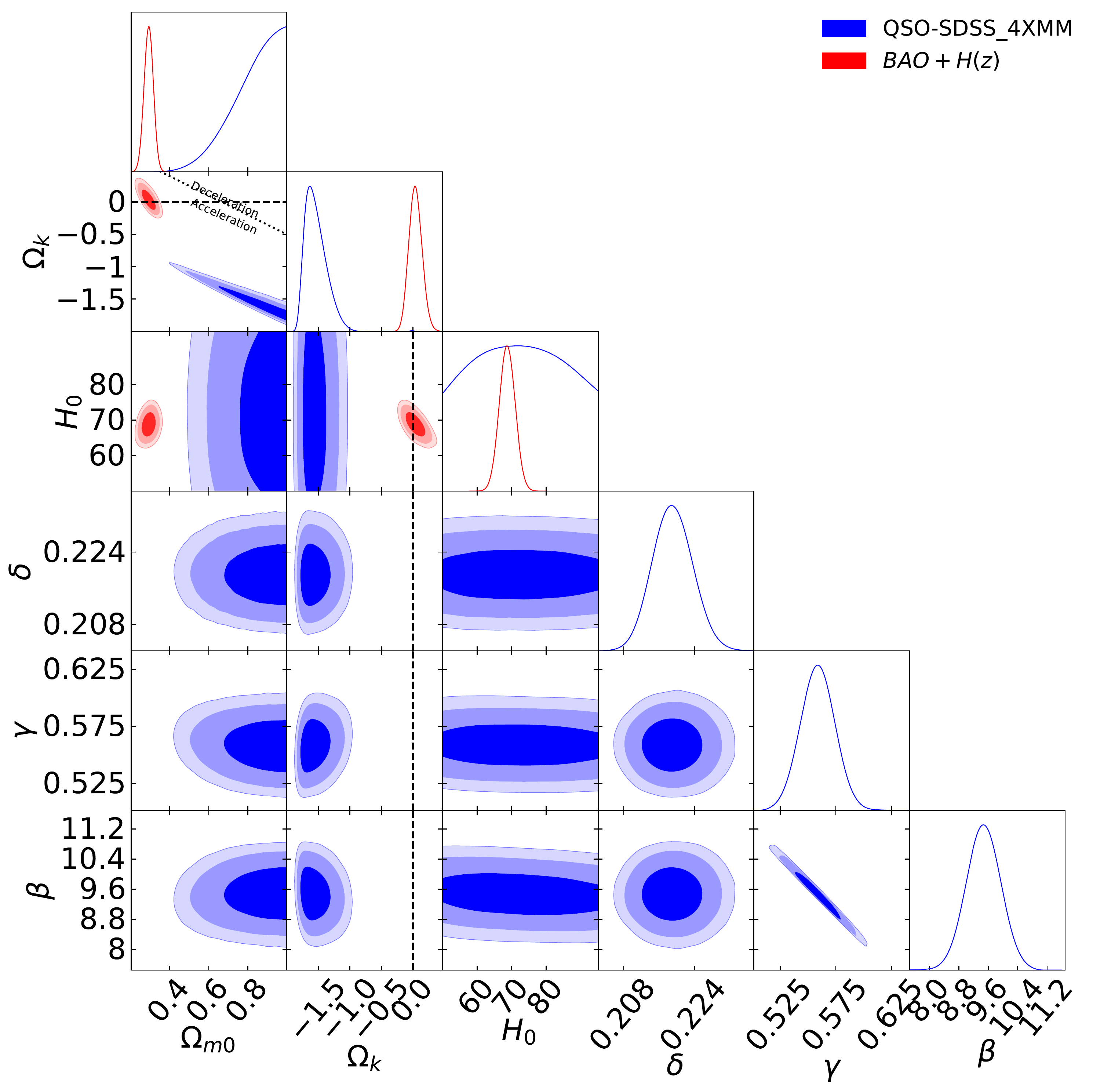}\par
    \includegraphics[width=\linewidth,height=7cm]{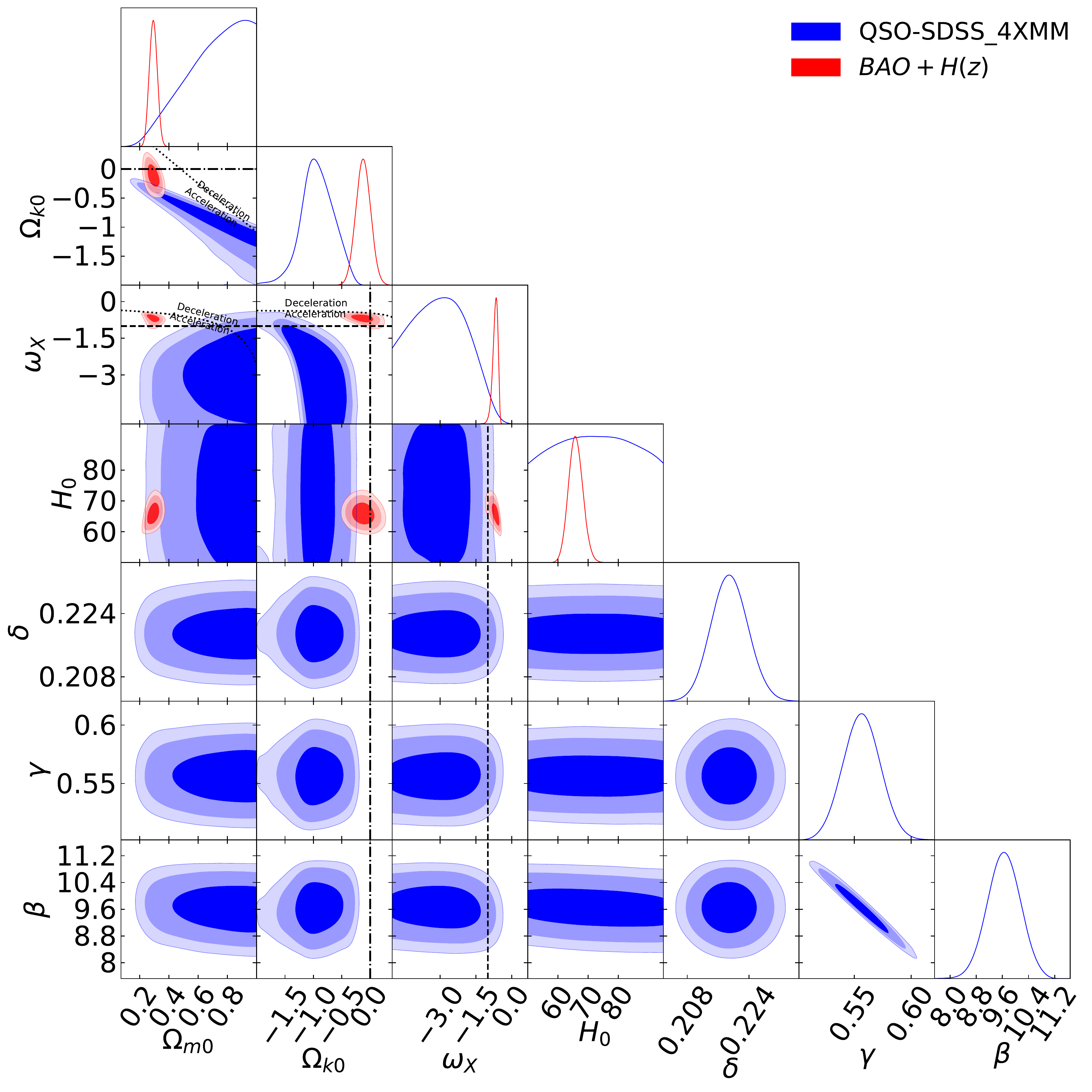}\par
    \includegraphics[width=\linewidth,height=7cm]{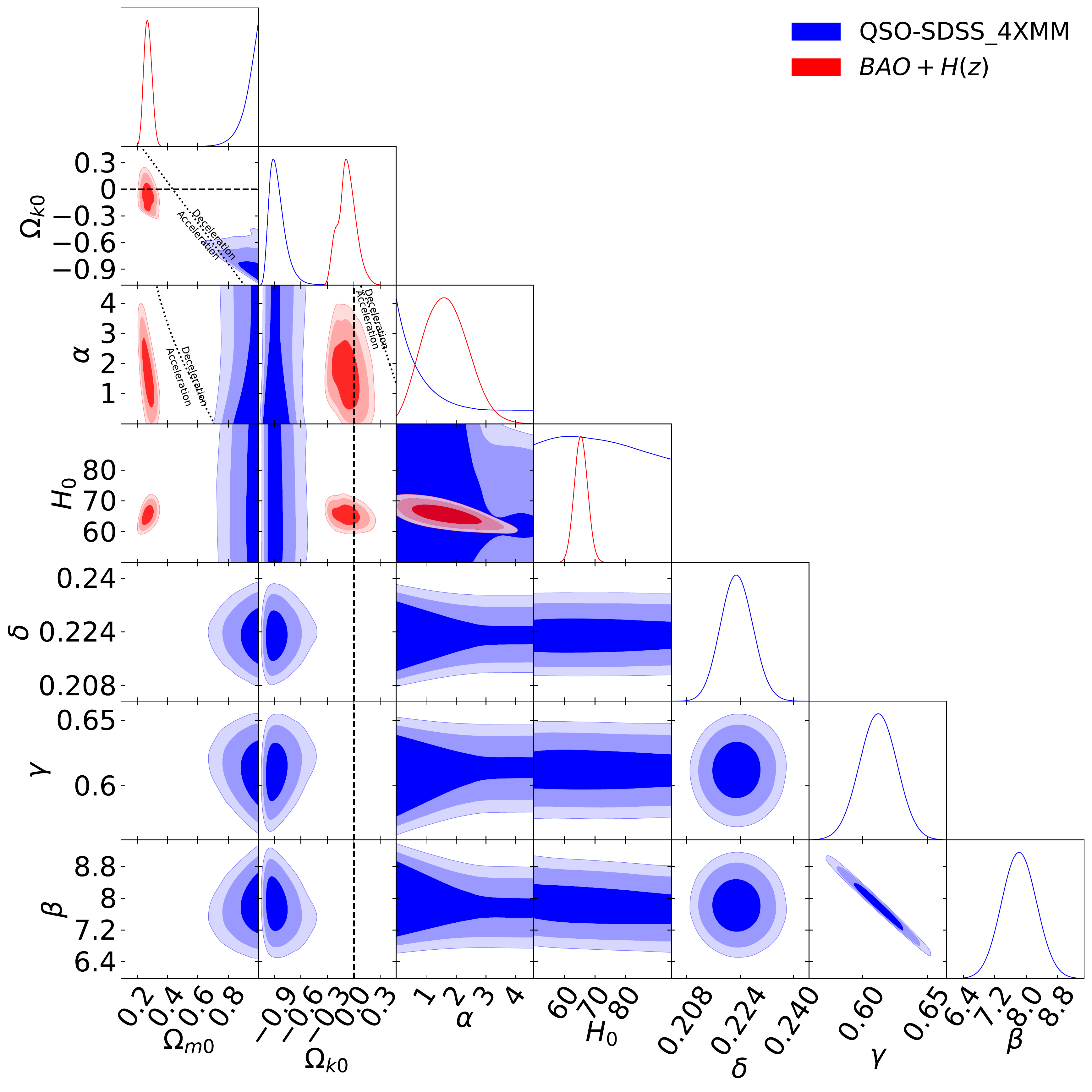}\par
\end{multicols}
\caption{One-dimensional likelihood distributions and two-dimensional likelihood contours at 1$\sigma$, 2$\sigma$, and 3$\sigma$ confidence levels using SDSS-4XMM (blue) and BAO + $H(z)$ (red) data for all free parameters. Left column shows the flat $\Lambda$CDM model, flat XCDM parametrization, and flat $\phi$CDM model respectively. The black dotted lines in all plots are the zero acceleration lines. The black dashed lines in the flat XCDM parametrization plots are the $\omega_X=-1$ lines. Right column shows the non-flat $\Lambda$CDM model, non-flat XCDM parametrization, and non-flat $\phi$CDM model respectively. Black dotted lines in all plots are the zero acceleration lines. Black dashed lines in the non-flat $\Lambda$CDM and $\phi$CDM model plots and black dotted-dashed lines in the non-flat XCDM parametrization plots correspond to $\Omega_{k0} = 0$. The black dashed lines in the non-flat XCDM parametrization plots are the $\omega_X=-1$ lines.}
\label{fig:Eiso-Ep}
\end{figure*}

\begin{figure*}
\begin{multicols}{2}
    \includegraphics[width=\linewidth,height=7cm]{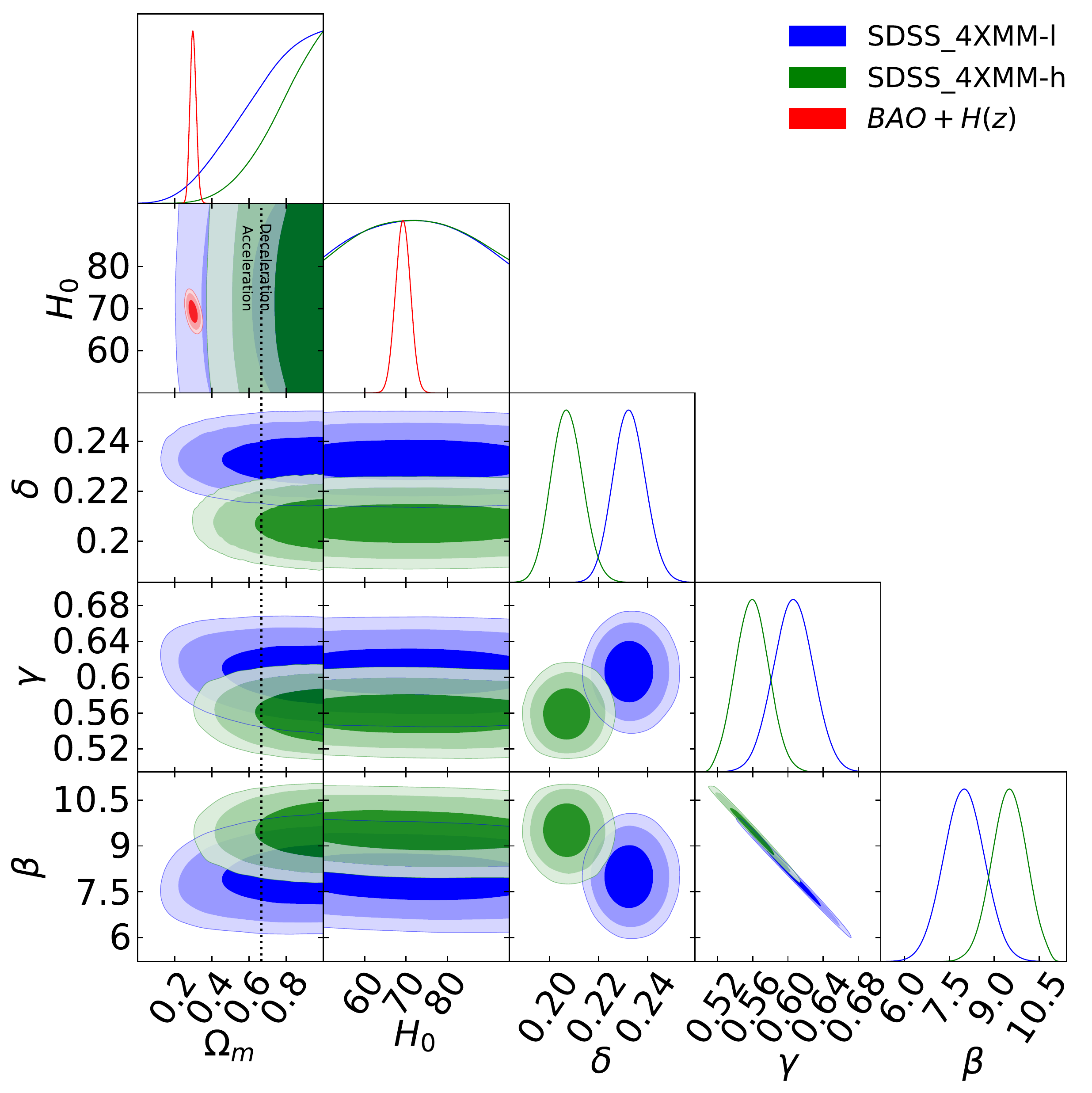}\par
    \includegraphics[width=\linewidth,height=7cm]{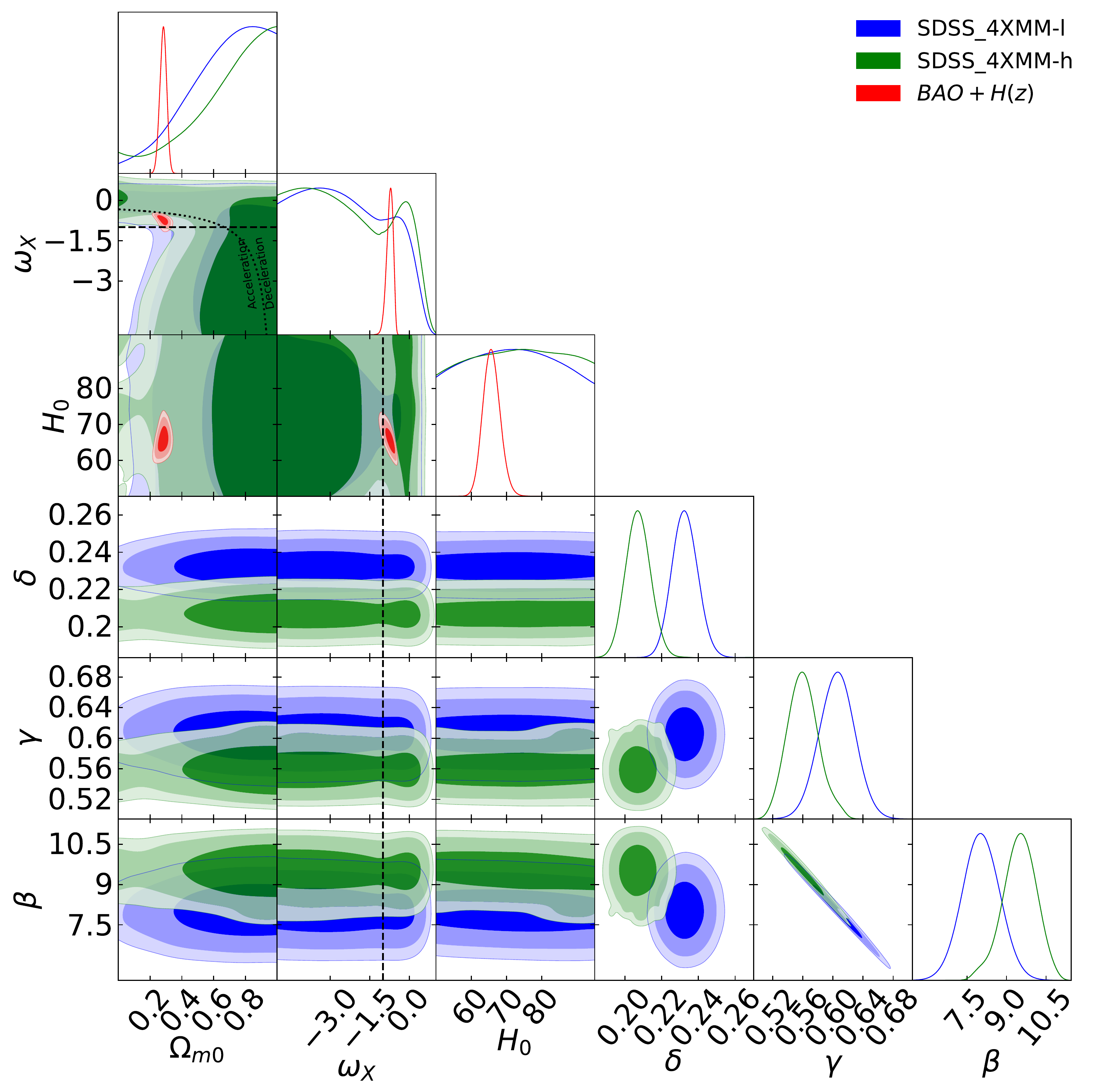}\par
    \includegraphics[width=\linewidth,height=7cm]{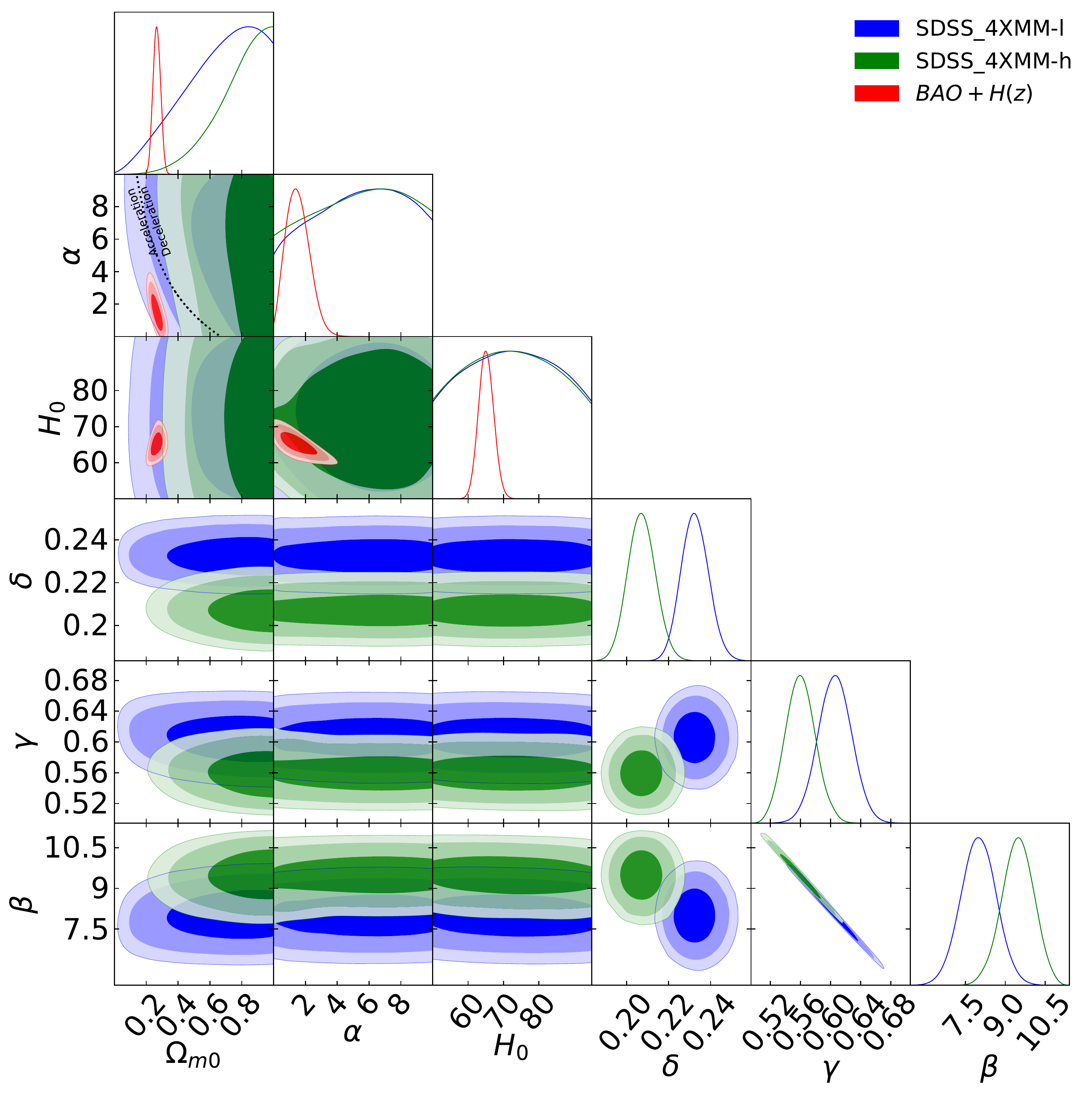}\par
    \includegraphics[width=\linewidth,height=7cm]{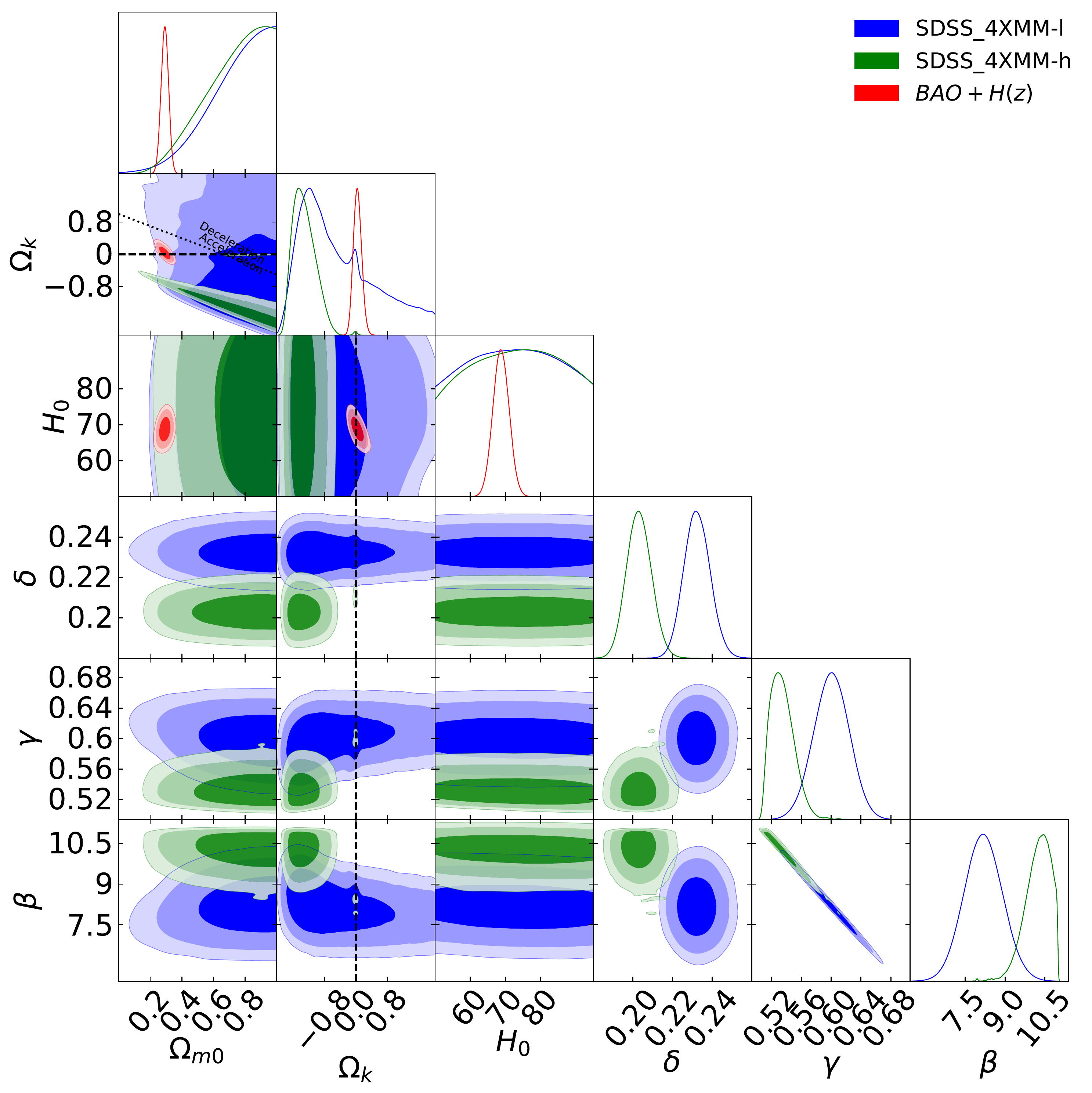}\par
    \includegraphics[width=\linewidth,height=7cm]{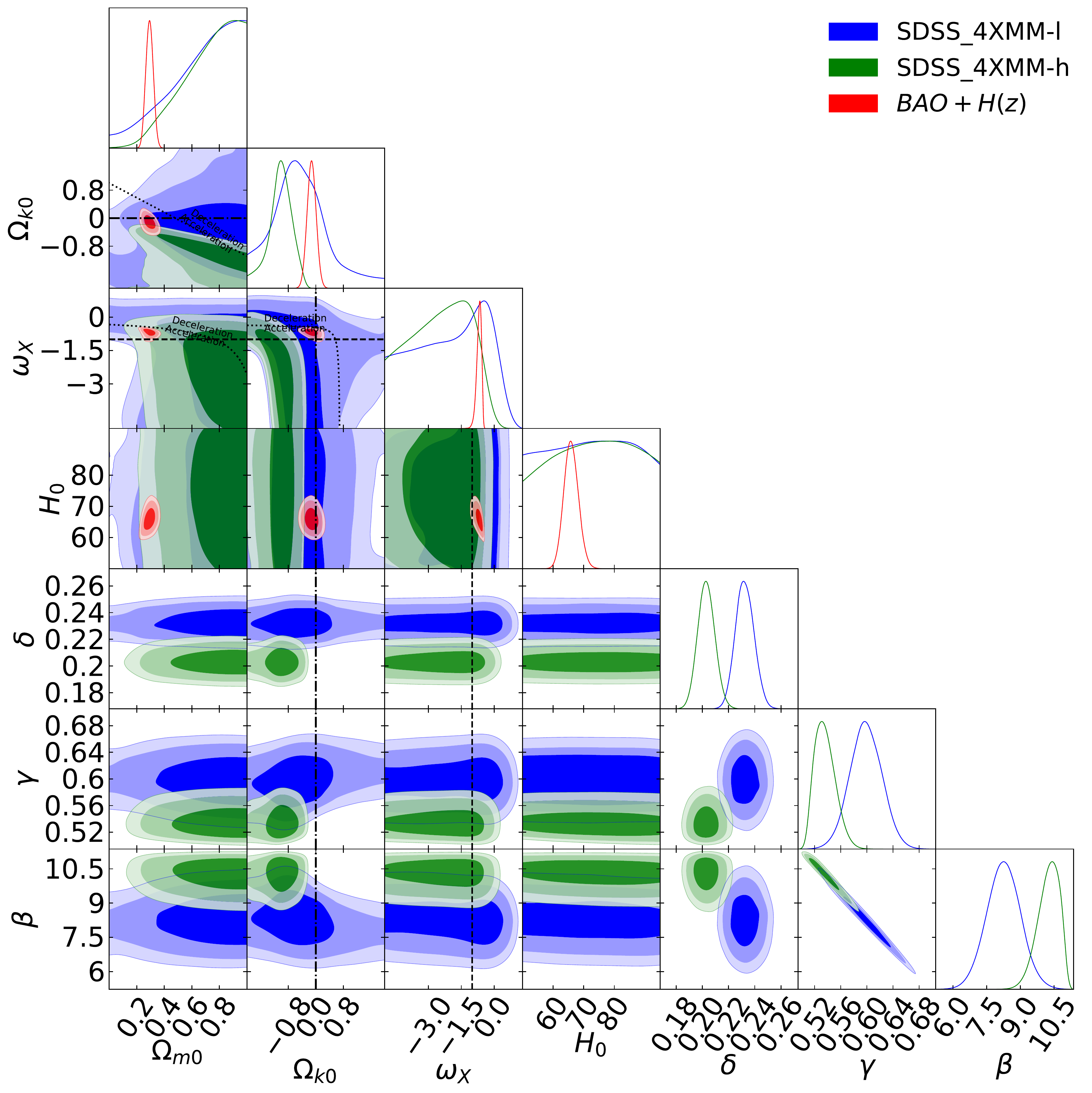}\par
    \includegraphics[width=\linewidth,height=7cm]{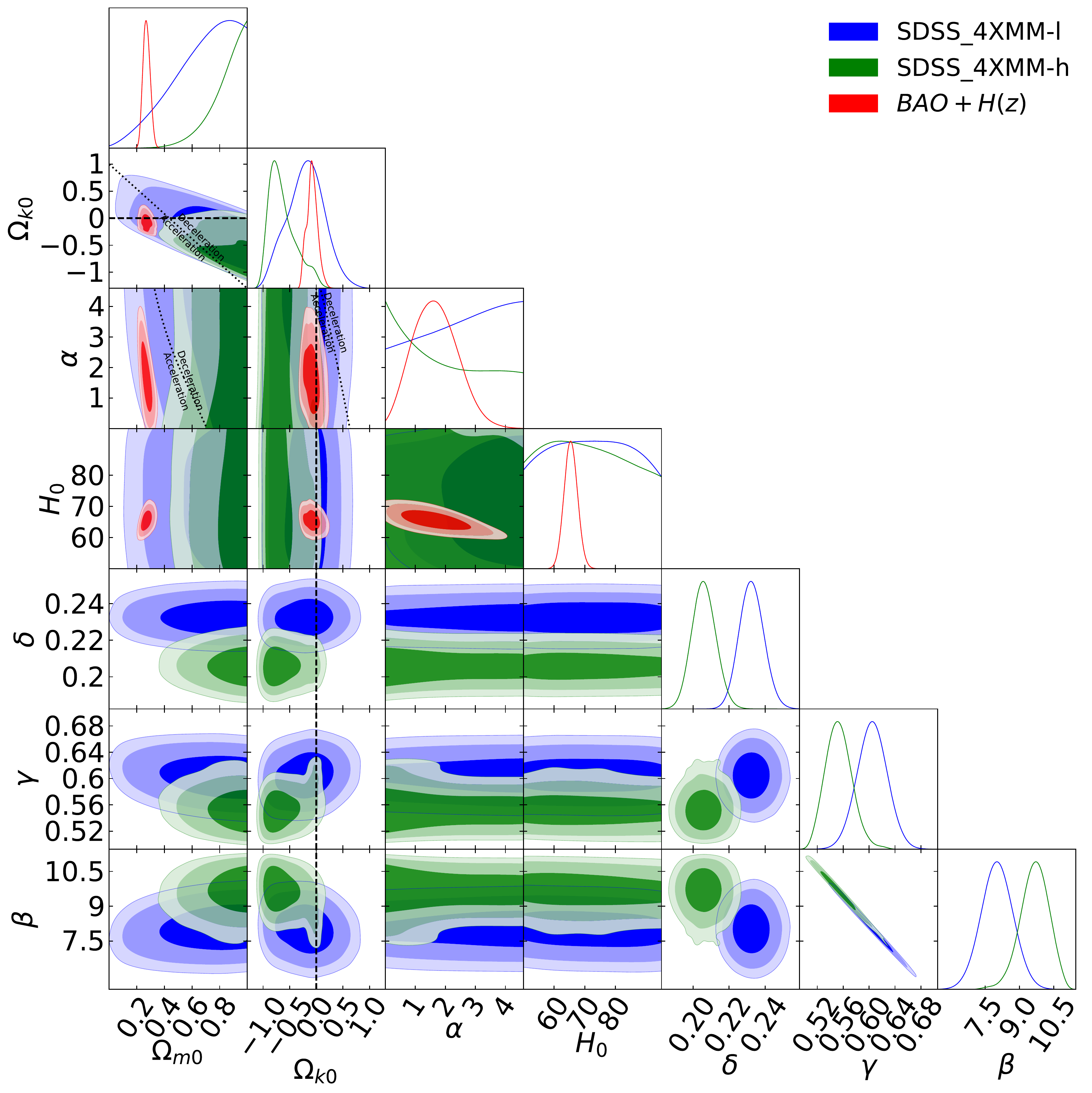}\par
\end{multicols}
\caption{One-dimensional likelihood distributions and two-dimensional likelihood contours at 1$\sigma$, 2$\sigma$, and 3$\sigma$ confidence levels using SDSS-4XMM-l (blue), SDSS-4XMM-h (green), and BAO + $H(z)$ (red) data for all free parameters. Left column shows the flat $\Lambda$CDM model, flat XCDM parametrization, and flat $\phi$CDM model respectively. The black dotted lines in all plots are the zero acceleration lines. The black dashed lines in the flat XCDM parametrization plots are the $\omega_X=-1$ lines. Right column shows the non-flat $\Lambda$CDM model, non-flat XCDM parametrization, and non-flat $\phi$CDM model respectively. Black dotted lines in all plots are the zero acceleration lines. Black dashed lines in the non-flat $\Lambda$CDM and $\phi$CDM model plots and black dotted-dashed lines in the non-flat XCDM parametrization plots correspond to $\Omega_{k0} = 0$. The black dashed lines in the non-flat XCDM parametrization plots are the $\omega_X=-1$ lines.}
\label{fig:Eiso-Ep}
\end{figure*}

\begin{figure*}
\begin{multicols}{2}
    \includegraphics[width=\linewidth,height=7cm]{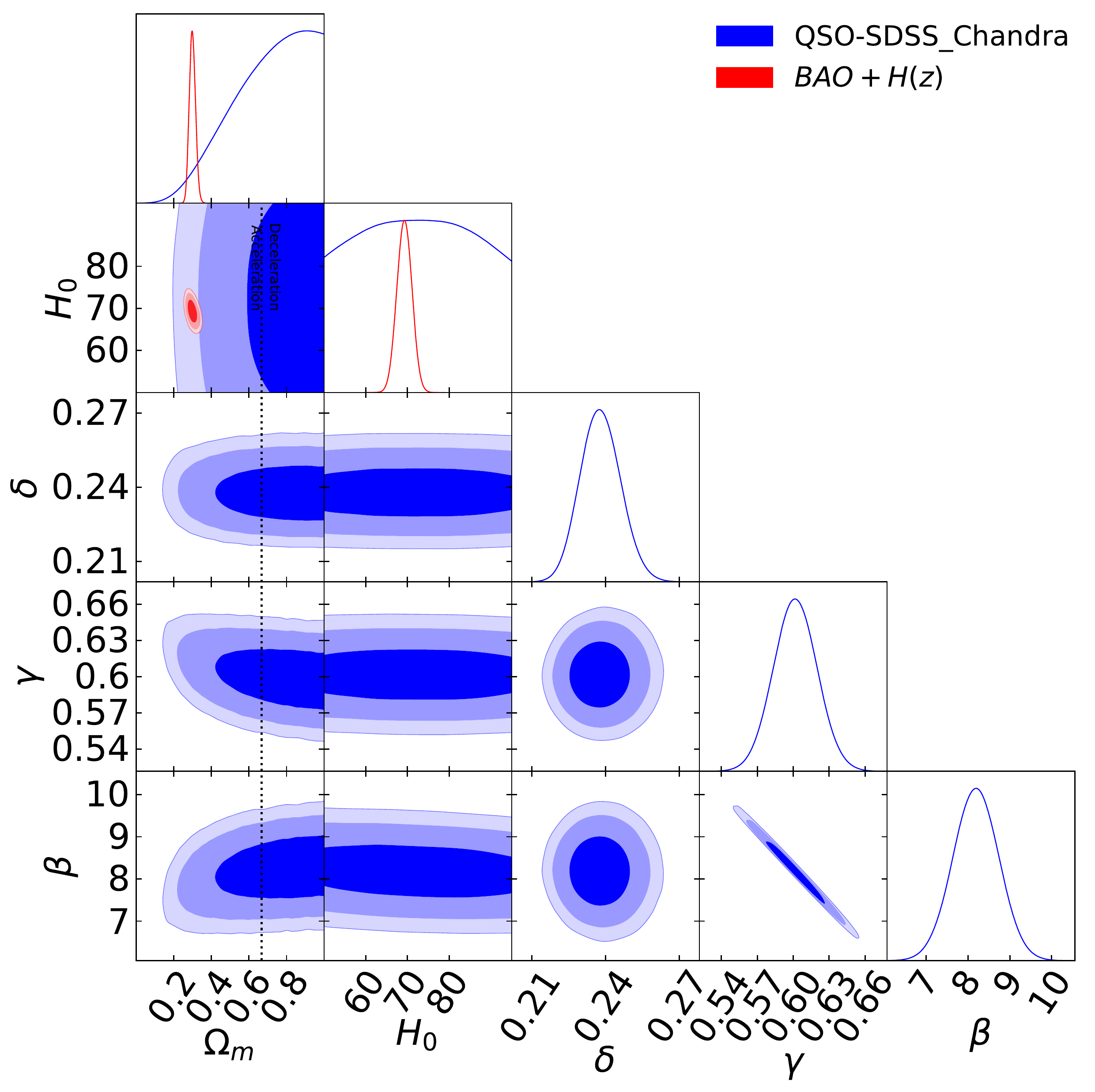}\par
    \includegraphics[width=\linewidth,height=7cm]{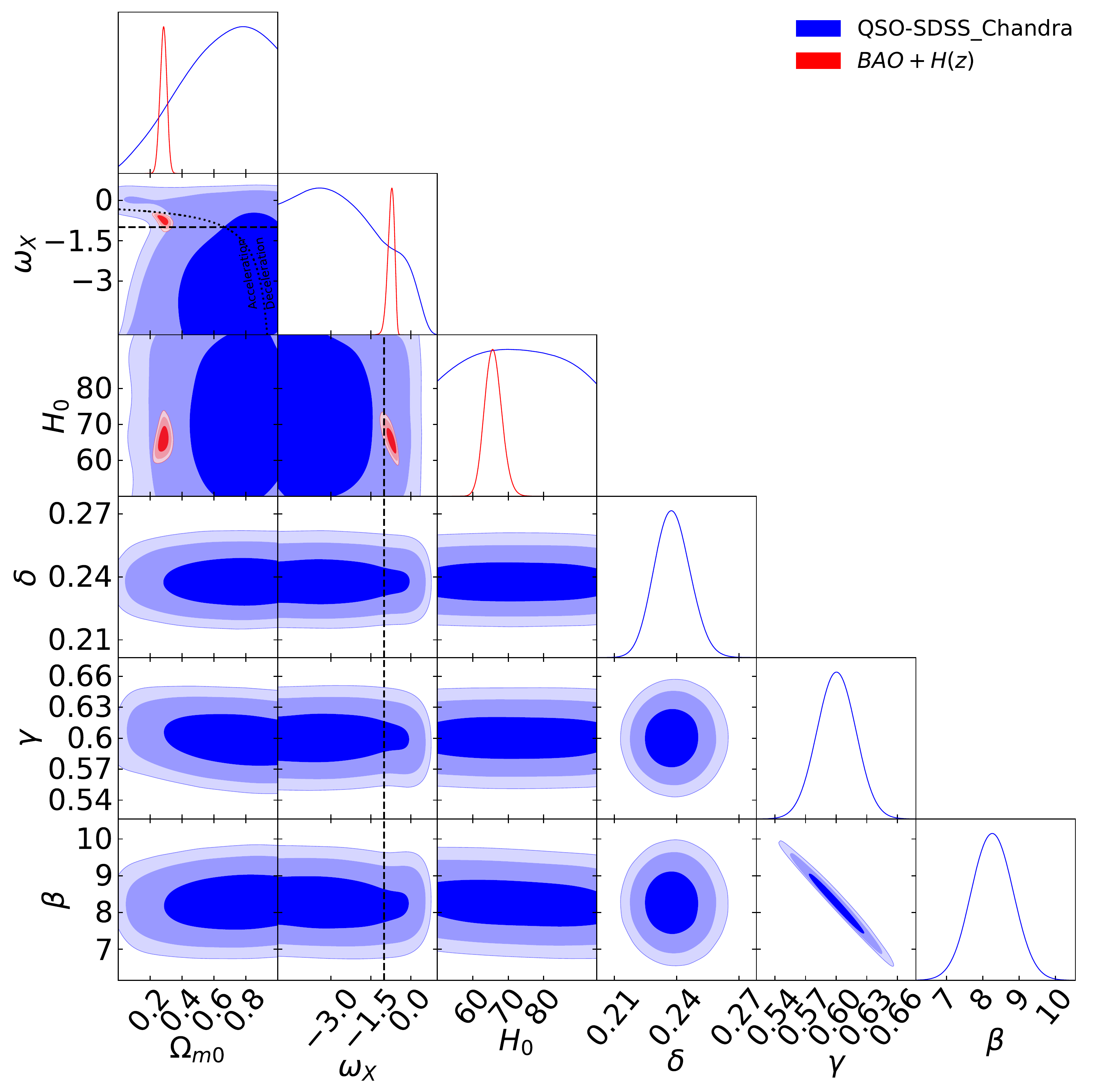}\par
    \includegraphics[width=\linewidth,height=7cm]{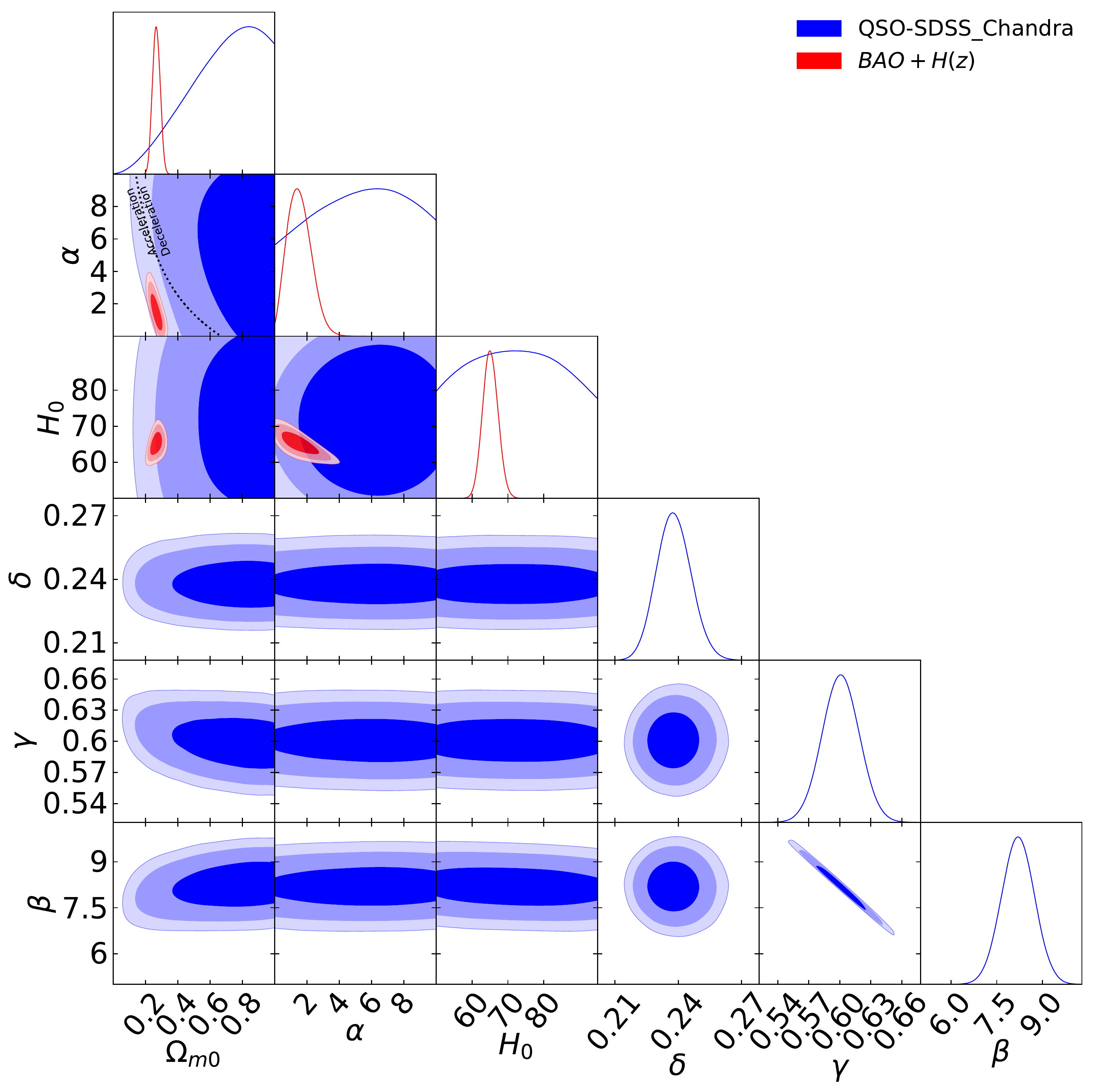}\par
    \includegraphics[width=\linewidth,height=7cm]{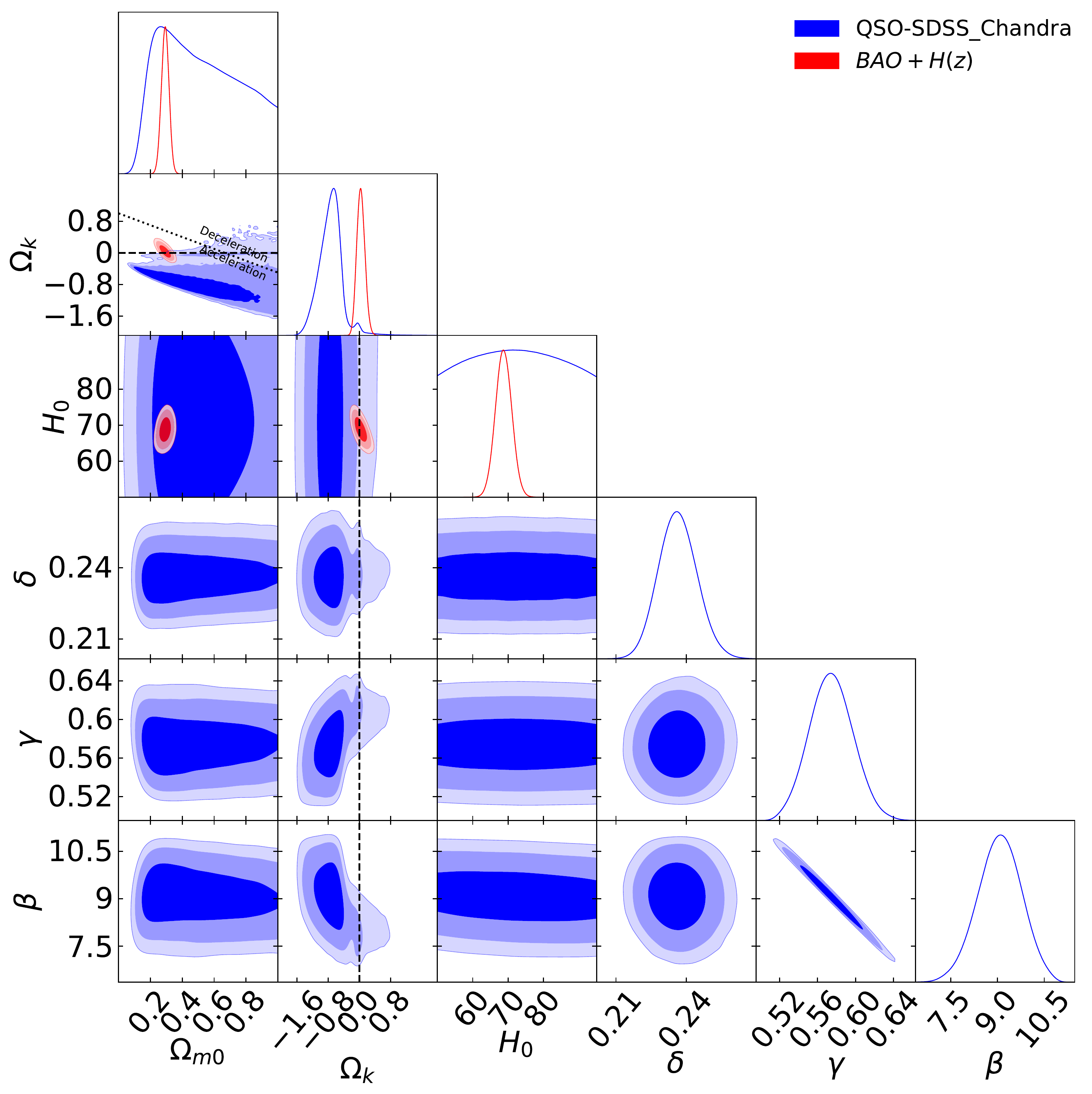}\par
    \includegraphics[width=\linewidth,height=7cm]{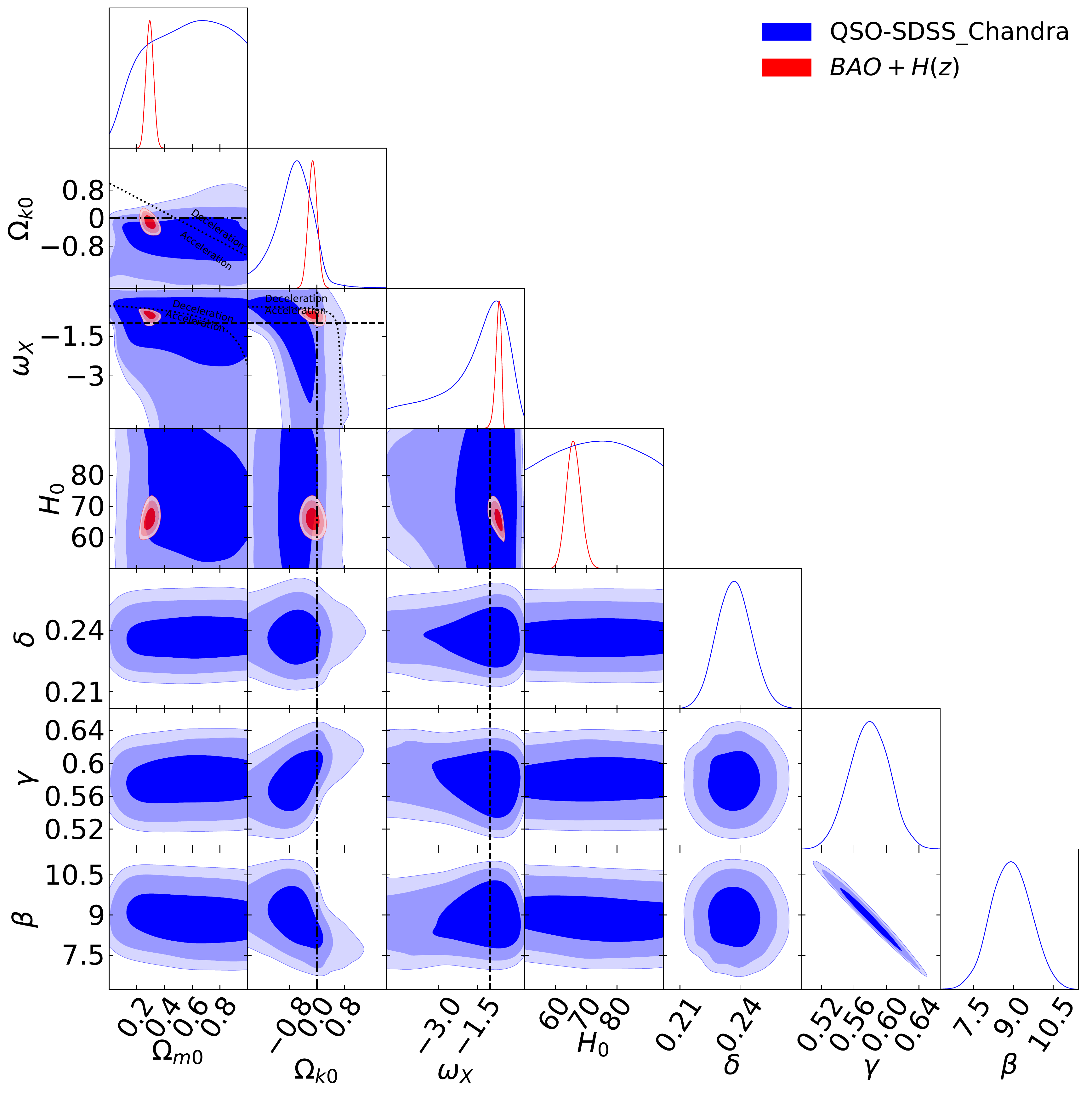}\par
    \includegraphics[width=\linewidth,height=7cm]{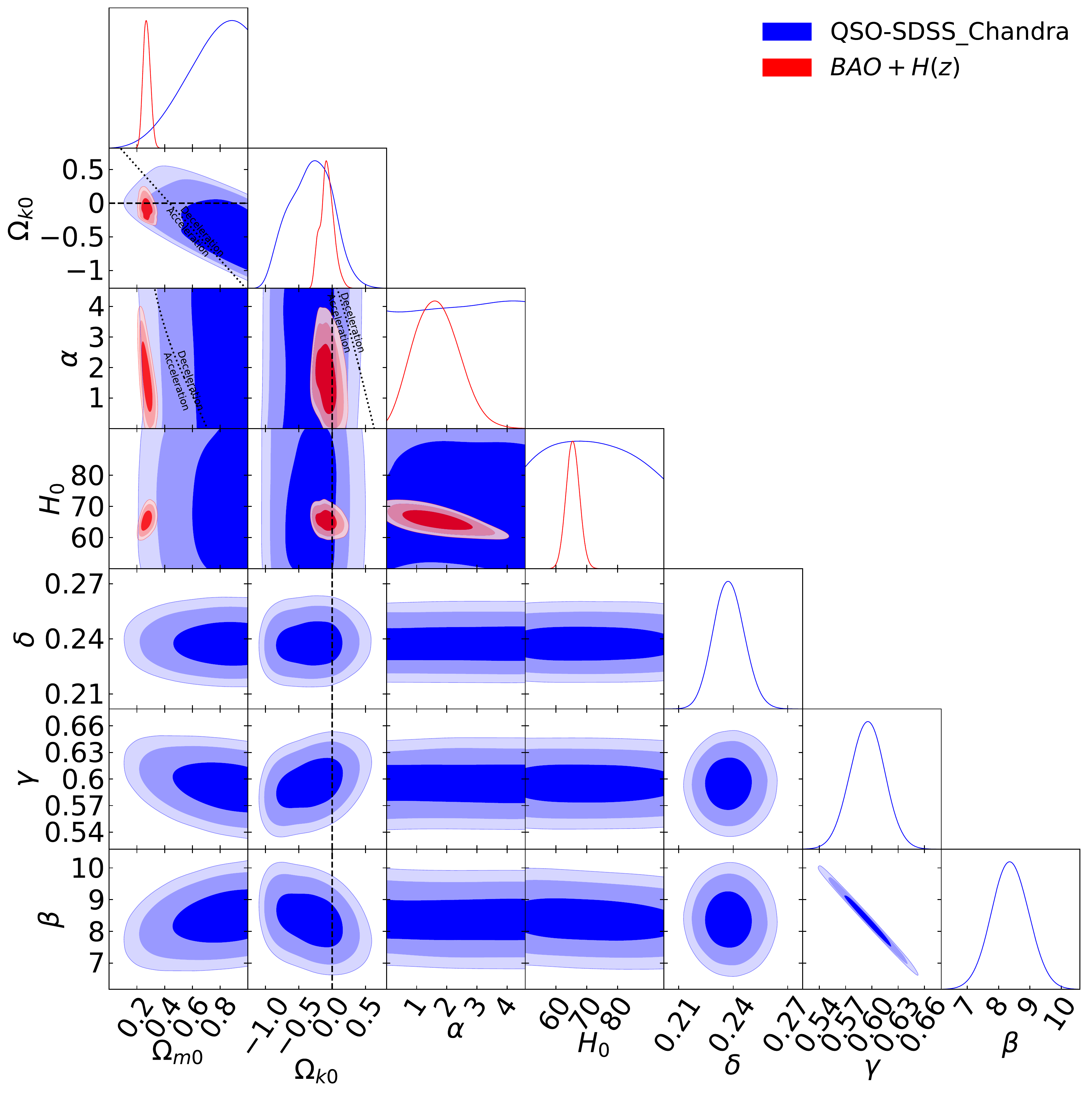}\par
\end{multicols}
\caption{One-dimensional likelihood distributions and two-dimensional likelihood contours at 1$\sigma$, 2$\sigma$, and 3$\sigma$ confidence levels using SDSS-Chandra (blue) and BAO + $H(z)$ (red) data for all free parameters. Left column shows the flat $\Lambda$CDM model, flat XCDM parametrization, and flat $\phi$CDM model respectively. The black dotted lines in all plots are the zero acceleration lines. The black dashed lines in the flat XCDM parametrization plots are the $\omega_X=-1$ lines. Right column shows the non-flat $\Lambda$CDM model, non-flat XCDM parametrization, and non-flat $\phi$CDM model respectively. Black dotted lines in all plots are the zero acceleration lines. Black dashed lines in the non-flat $\Lambda$CDM and $\phi$CDM model plots and black dotted-dashed lines in the non-flat XCDM parametrization plots correspond to $\Omega_{k0} = 0$. The black dashed lines in the non-flat XCDM parametrization plots are the $\omega_X=-1$ lines.}
\label{fig:Eiso-Ep}
\end{figure*}

\begin{figure*}
\begin{multicols}{2}
    \includegraphics[width=\linewidth,height=7cm]{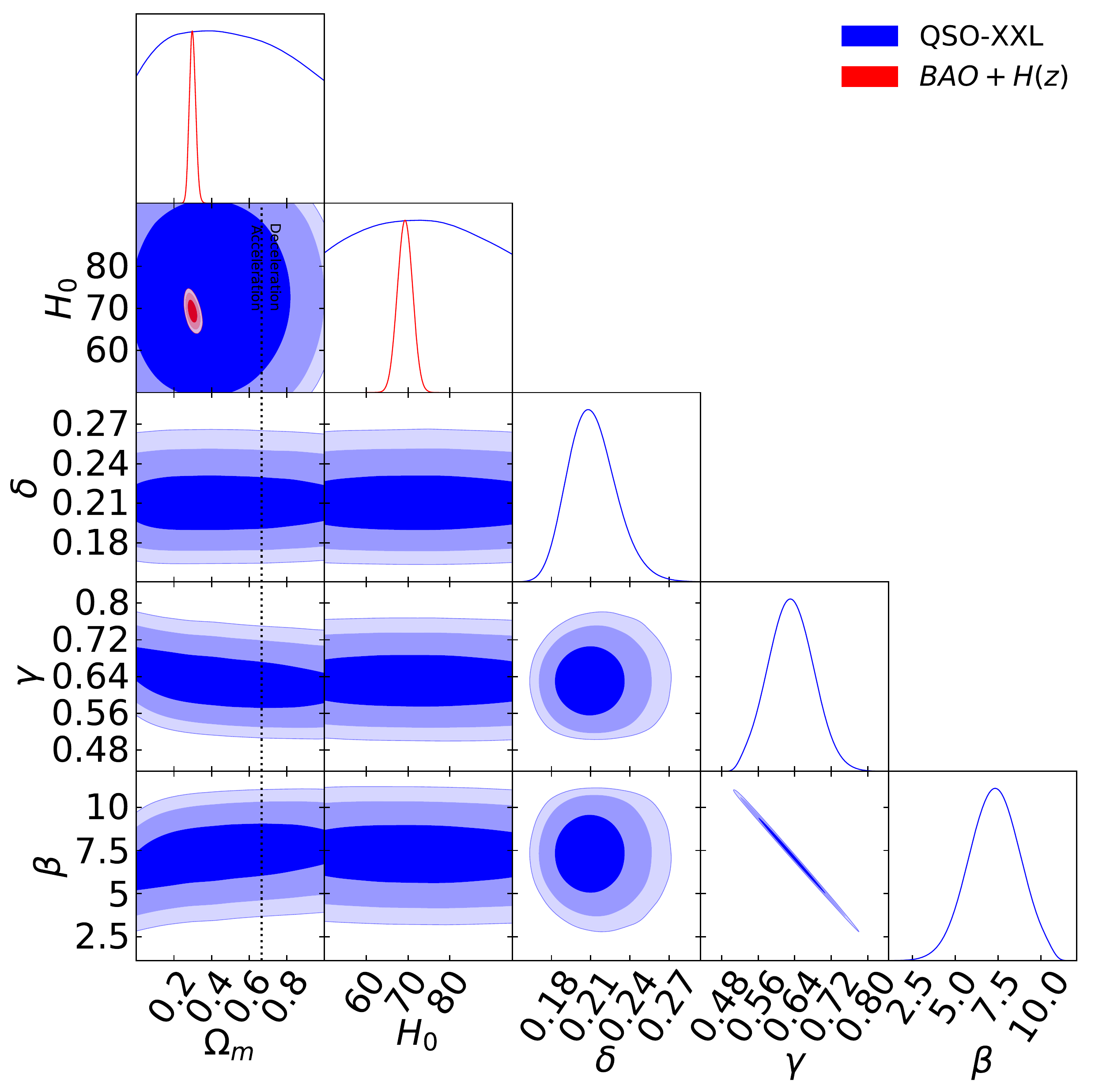}\par
    \includegraphics[width=\linewidth,height=7cm]{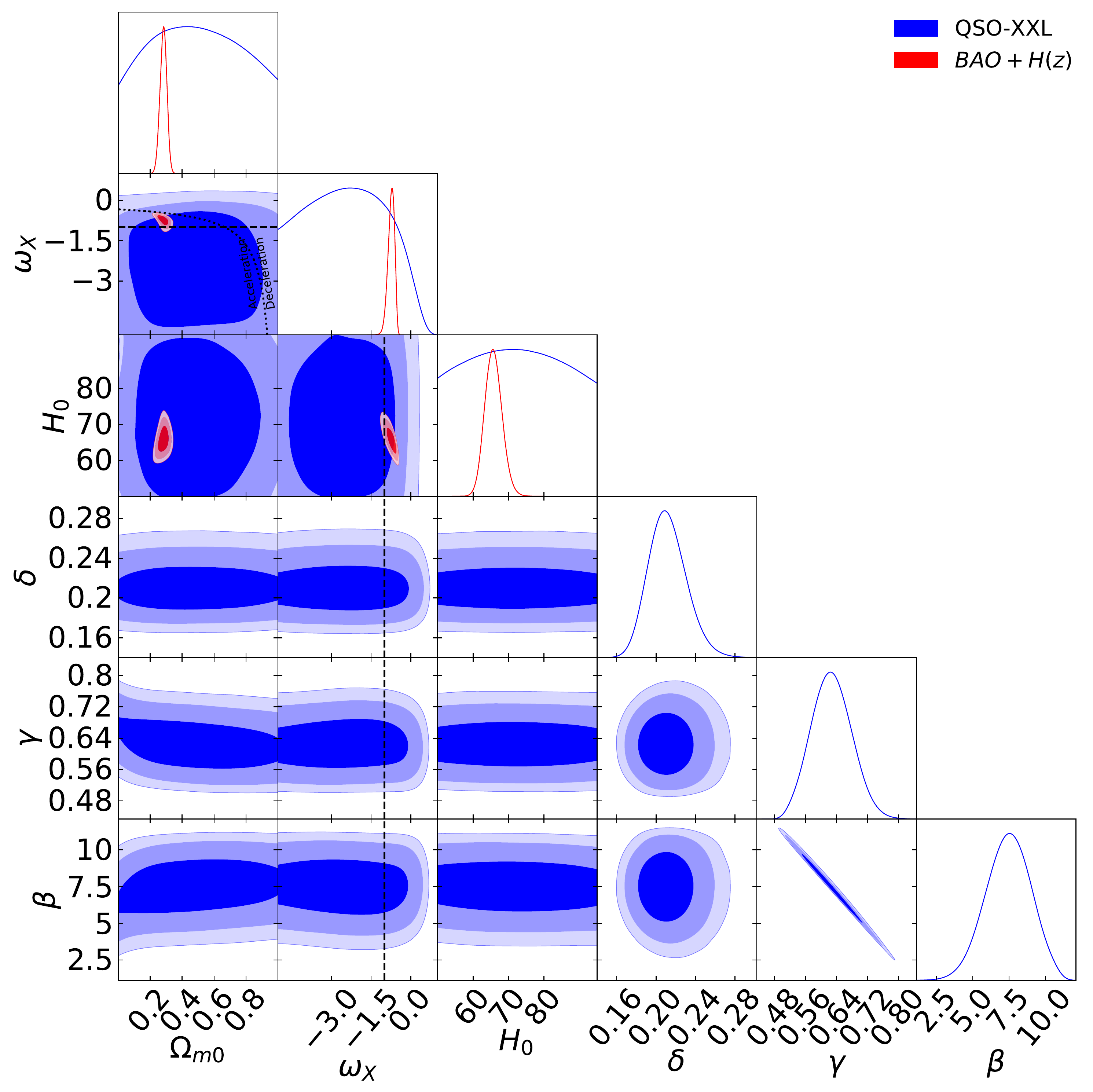}\par
    \includegraphics[width=\linewidth,height=7cm]{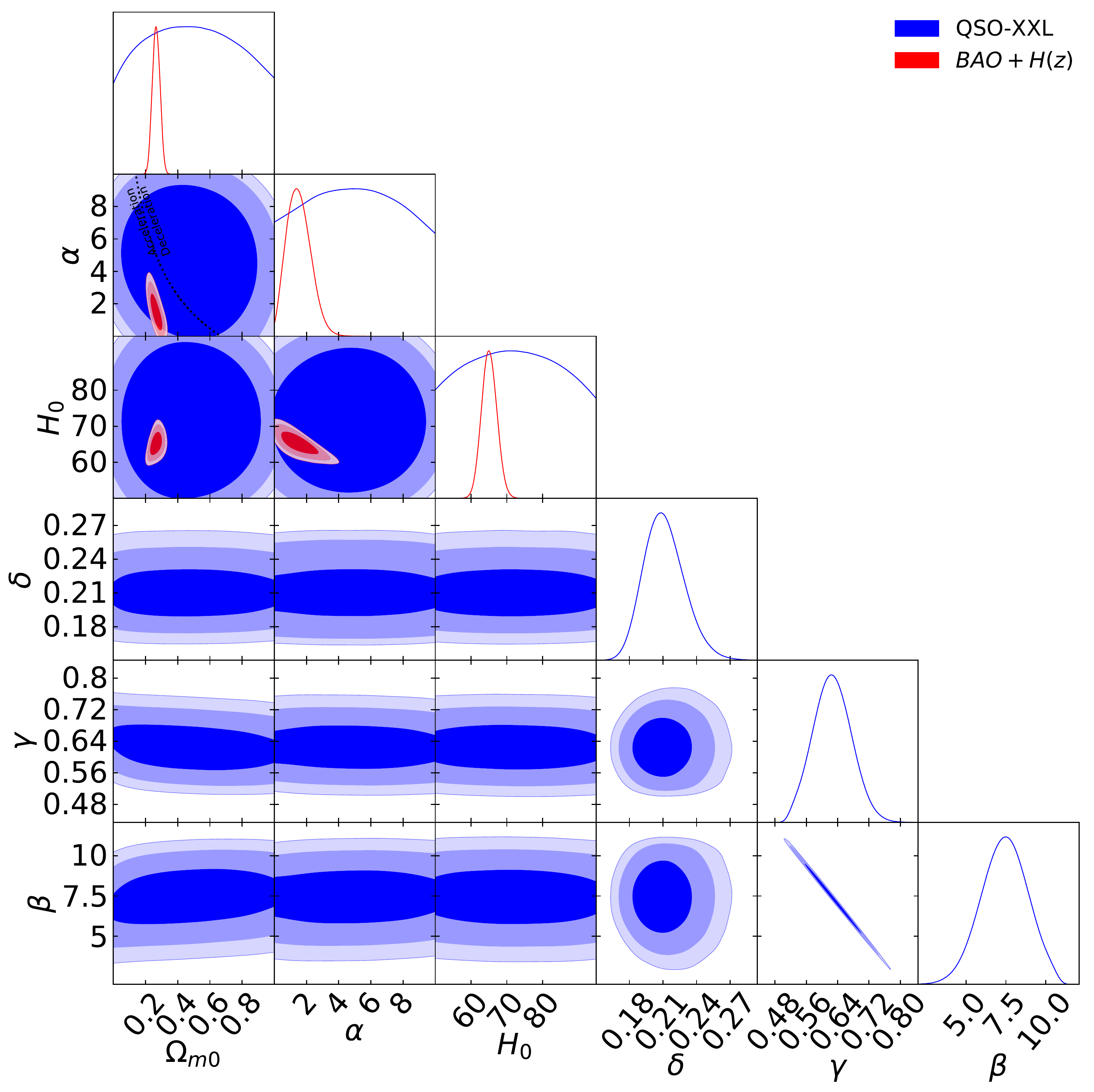}\par
    \includegraphics[width=\linewidth,height=7cm]{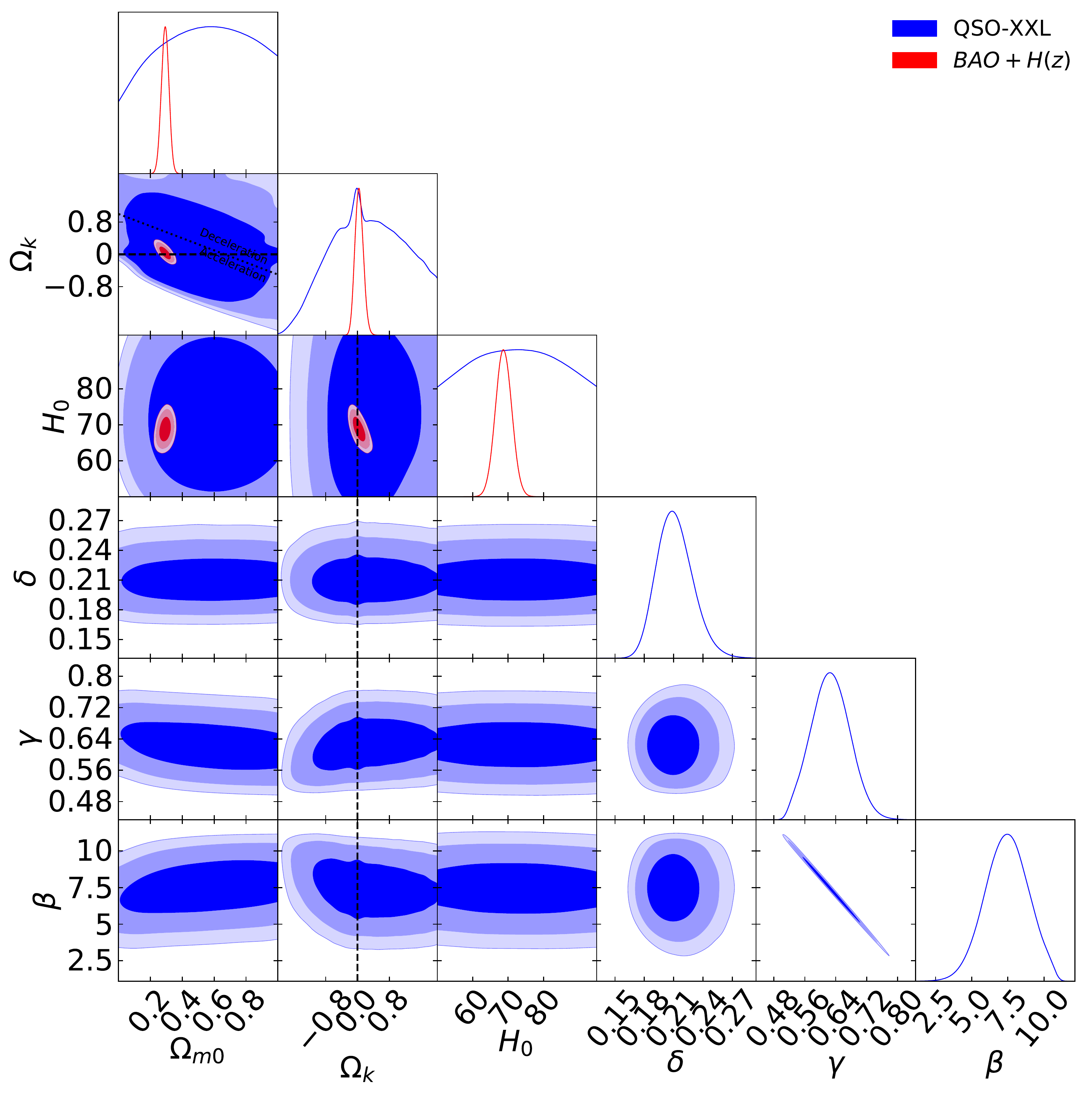}\par
    \includegraphics[width=\linewidth,height=7cm]{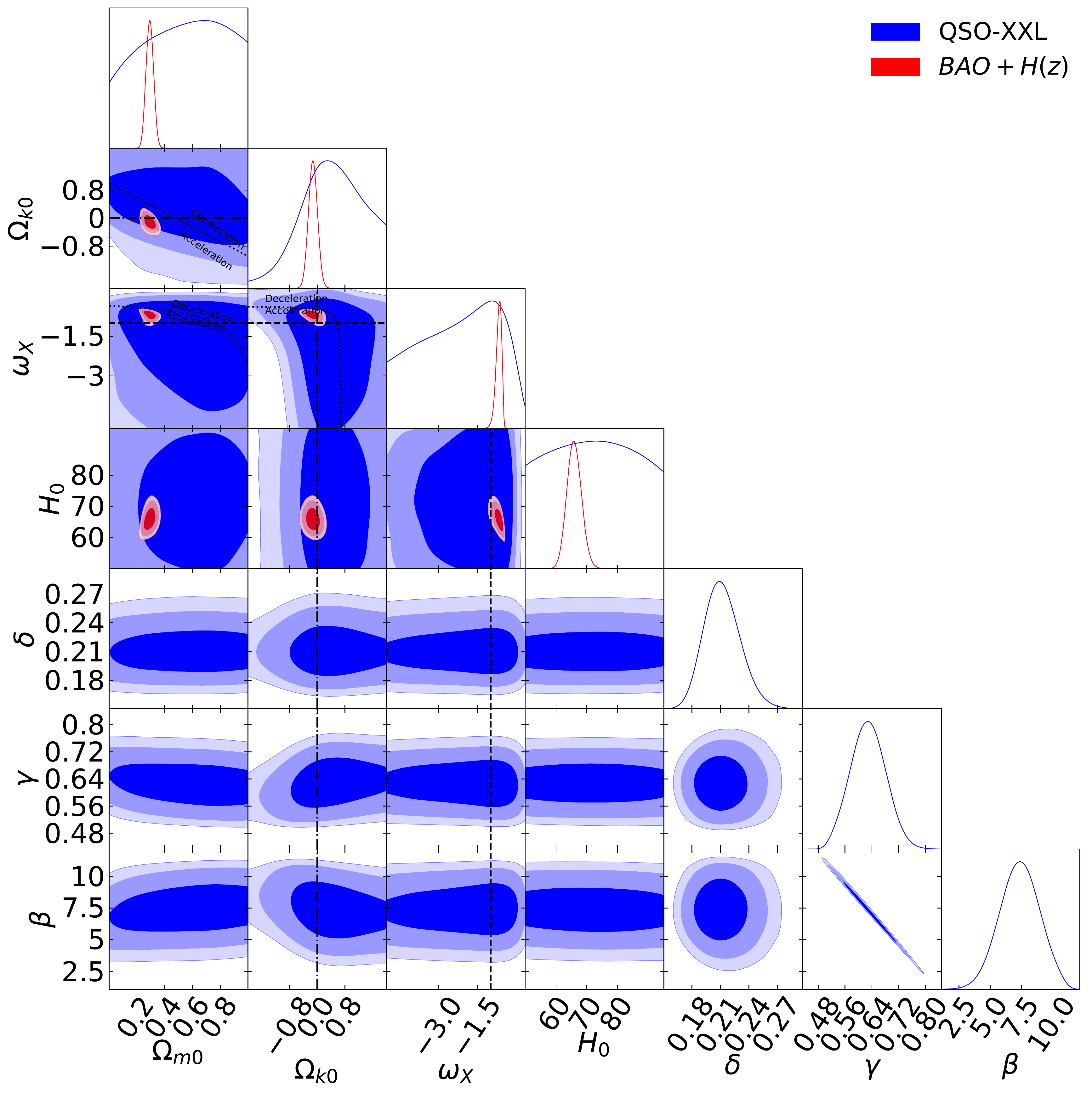}\par
    \includegraphics[width=\linewidth,height=7cm]{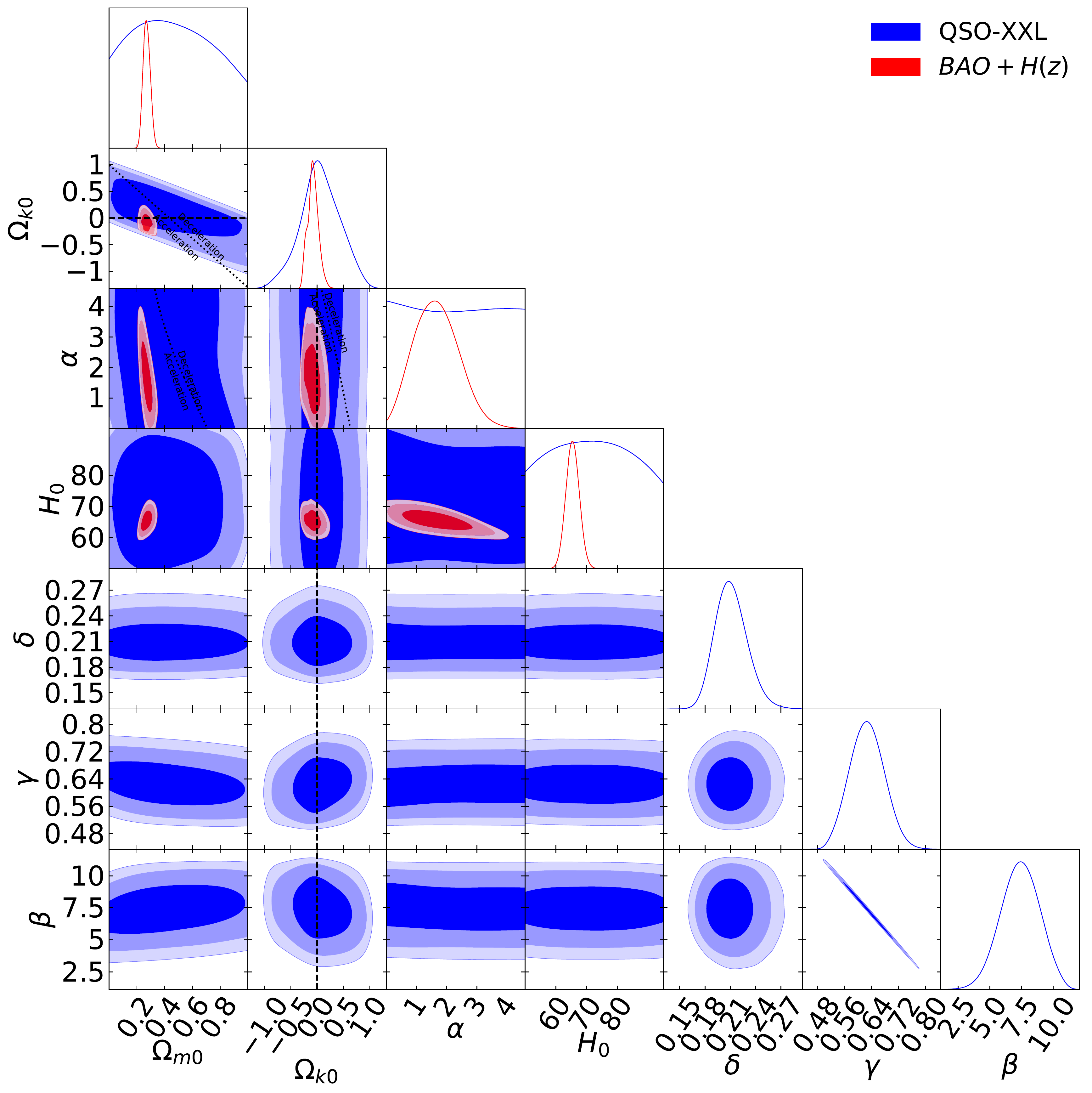}\par
\end{multicols}
\caption{One-dimensional likelihood distributions and two-dimensional likelihood contours at 1$\sigma$, 2$\sigma$, and 3$\sigma$ confidence levels using XXL (blue) and BAO + $H(z)$ (red) data for all free parameters. Left column shows the flat $\Lambda$CDM model, flat XCDM parametrization, and flat $\phi$CDM model respectively. The black dotted lines in all plots are the zero acceleration lines. The black dashed lines in the flat XCDM parametrization plots are the $\omega_X=-1$ lines. Right column shows the non-flat $\Lambda$CDM model, non-flat XCDM parametrization, and non-flat $\phi$CDM model respectively. Black dotted lines in all plots are the zero acceleration lines. Black dashed lines in the non-flat $\Lambda$CDM and $\phi$CDM model plots and black dotted-dashed lines in the non-flat XCDM parametrization plots correspond to $\Omega_{k0} = 0$. The black dashed lines in the non-flat XCDM parametrization plots are the $\omega_X=-1$ lines.}
\label{fig:Eiso-Ep}
\end{figure*}

\begin{figure*}
\begin{multicols}{2}
    \includegraphics[width=\linewidth,height=7cm]{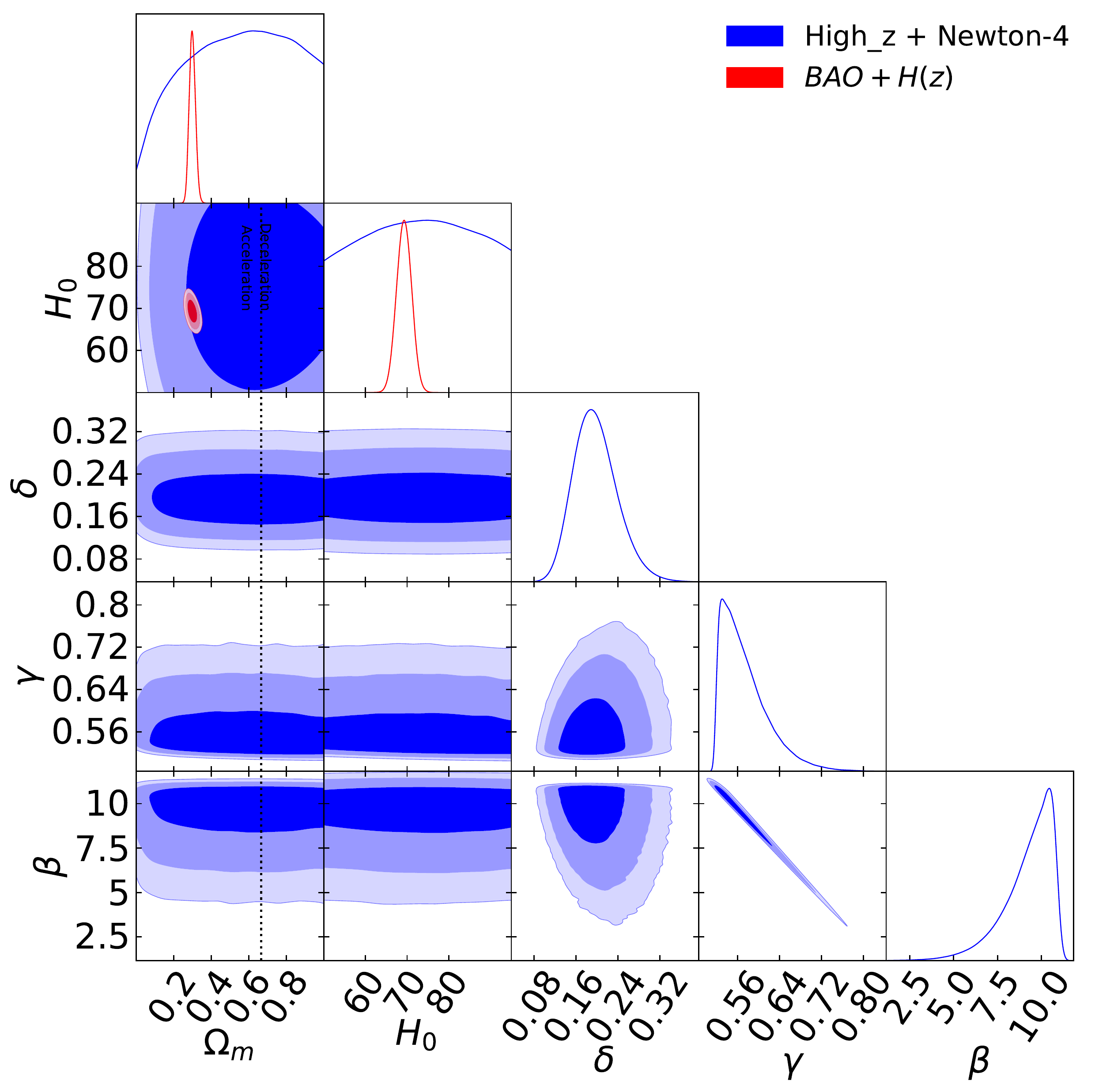}\par
    \includegraphics[width=\linewidth,height=7cm]{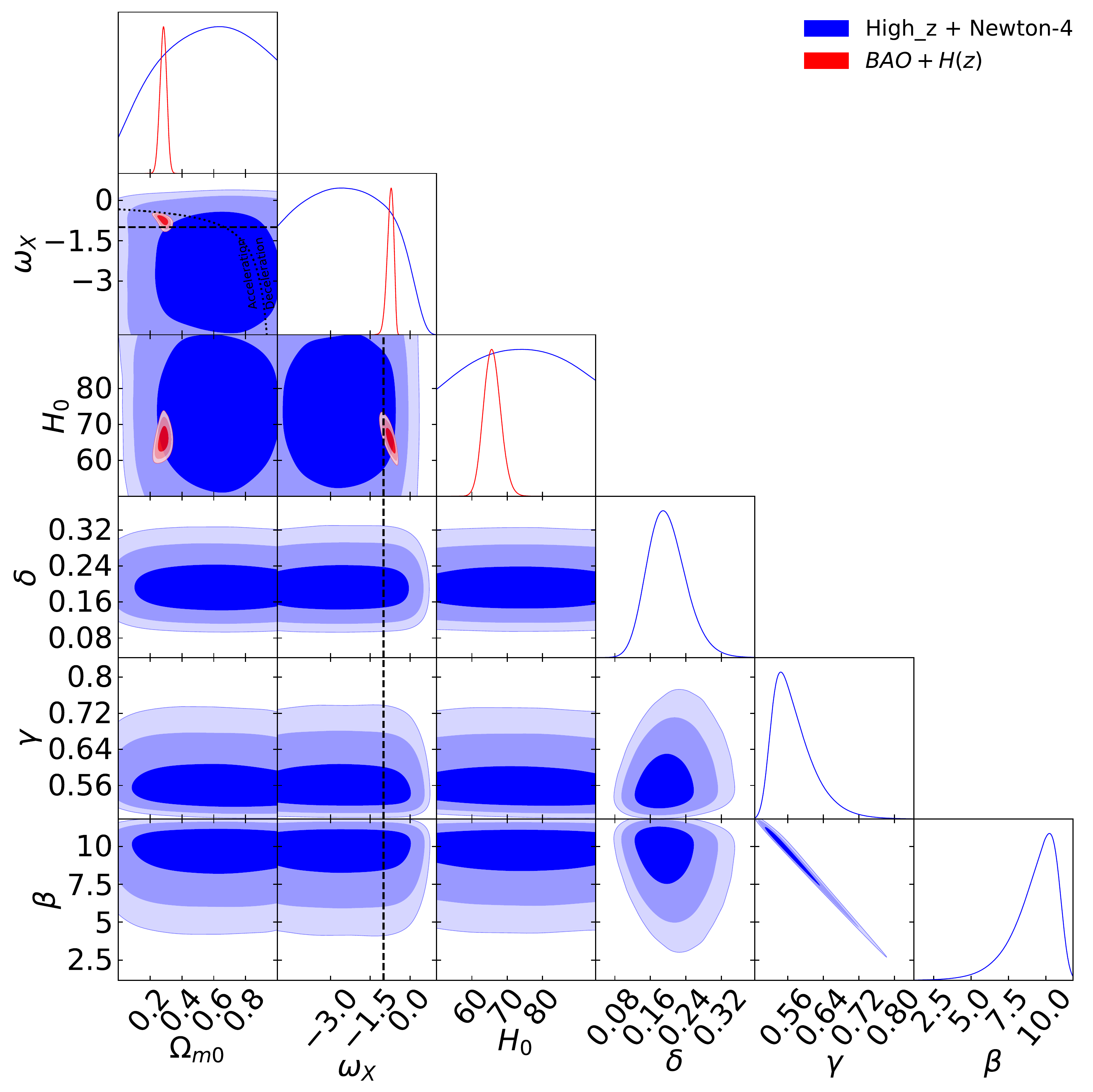}\par
    \includegraphics[width=\linewidth,height=7cm]{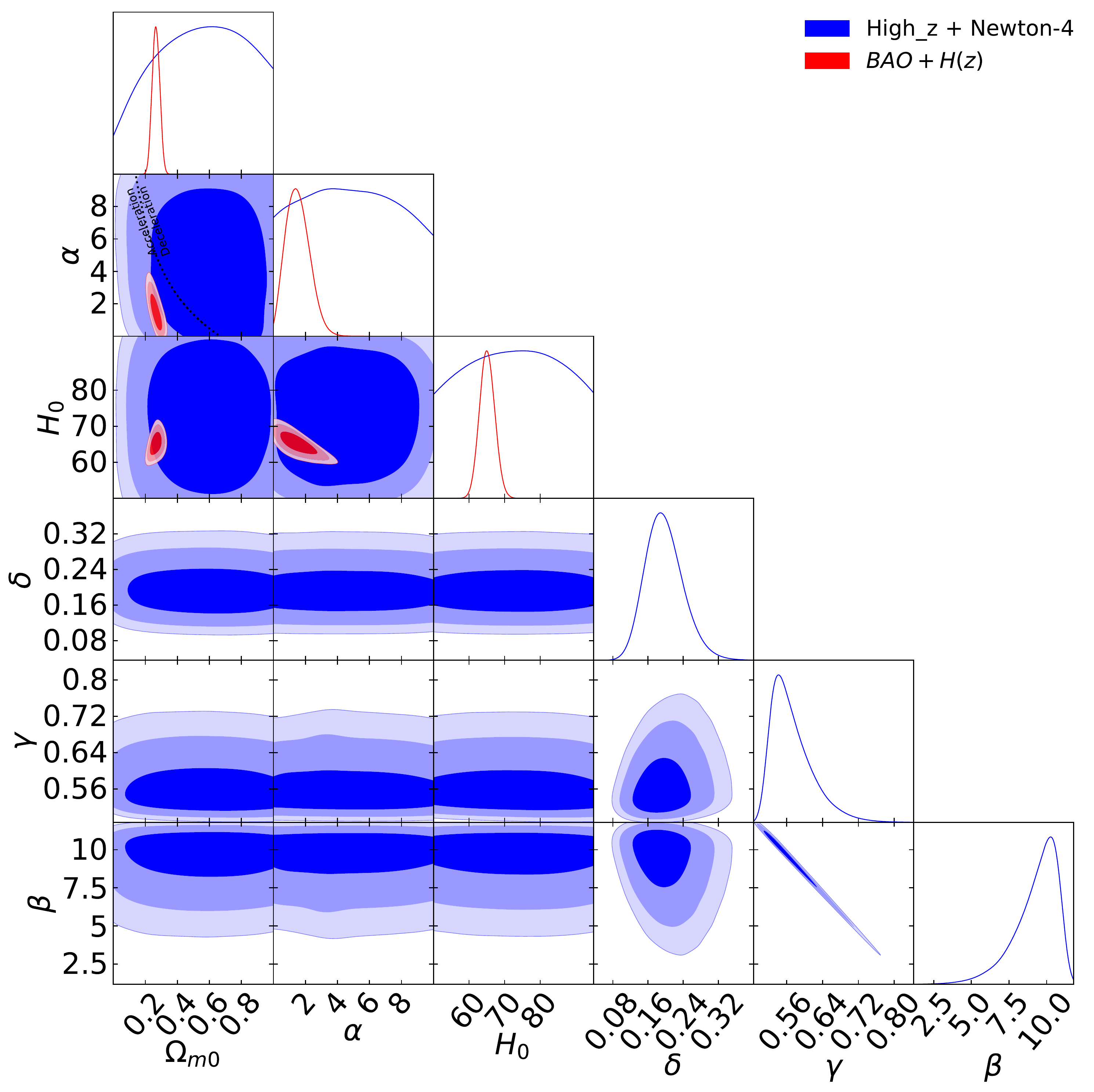}\par
    \includegraphics[width=\linewidth,height=7cm]{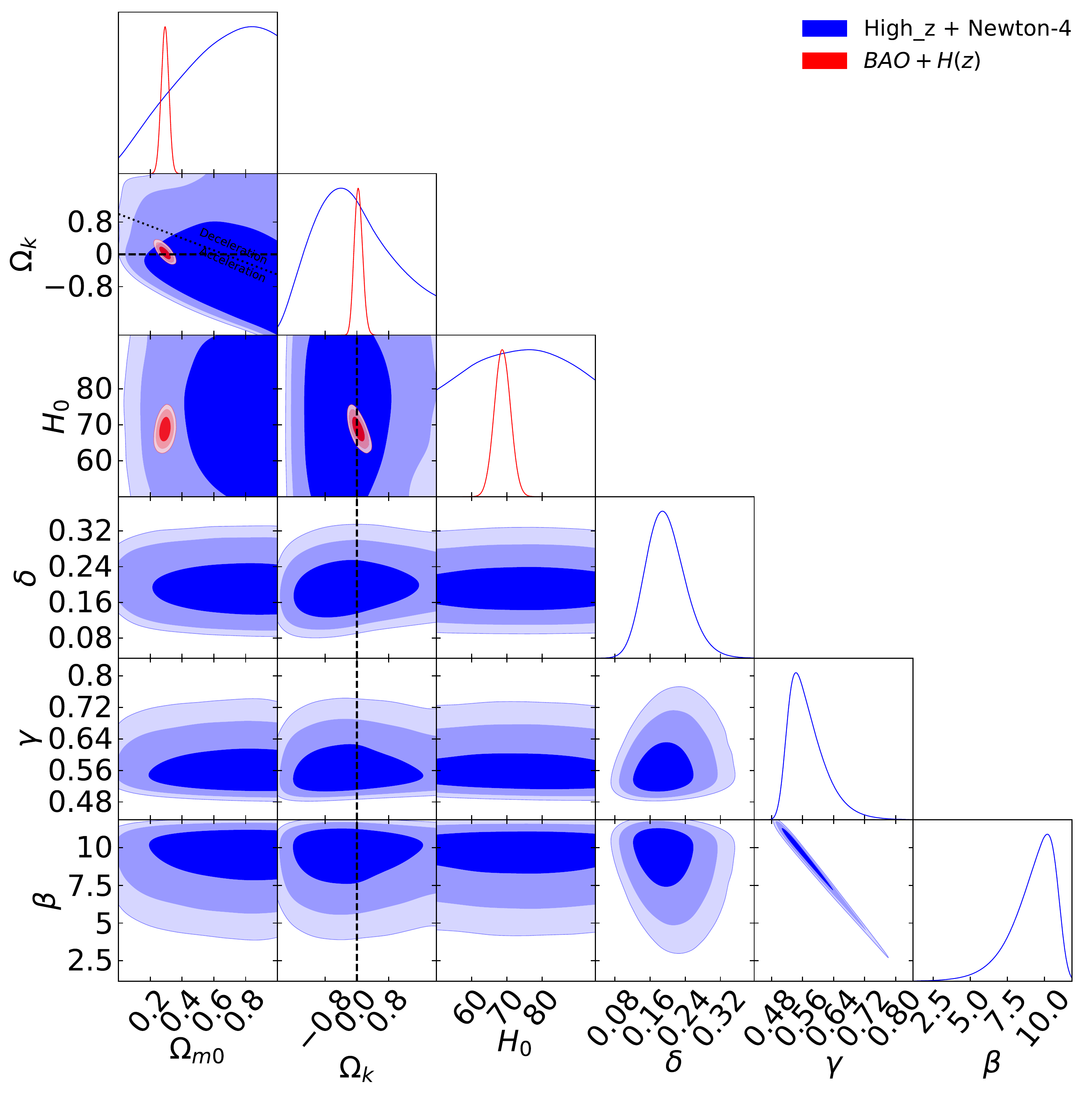}\par
    \includegraphics[width=\linewidth,height=7cm]{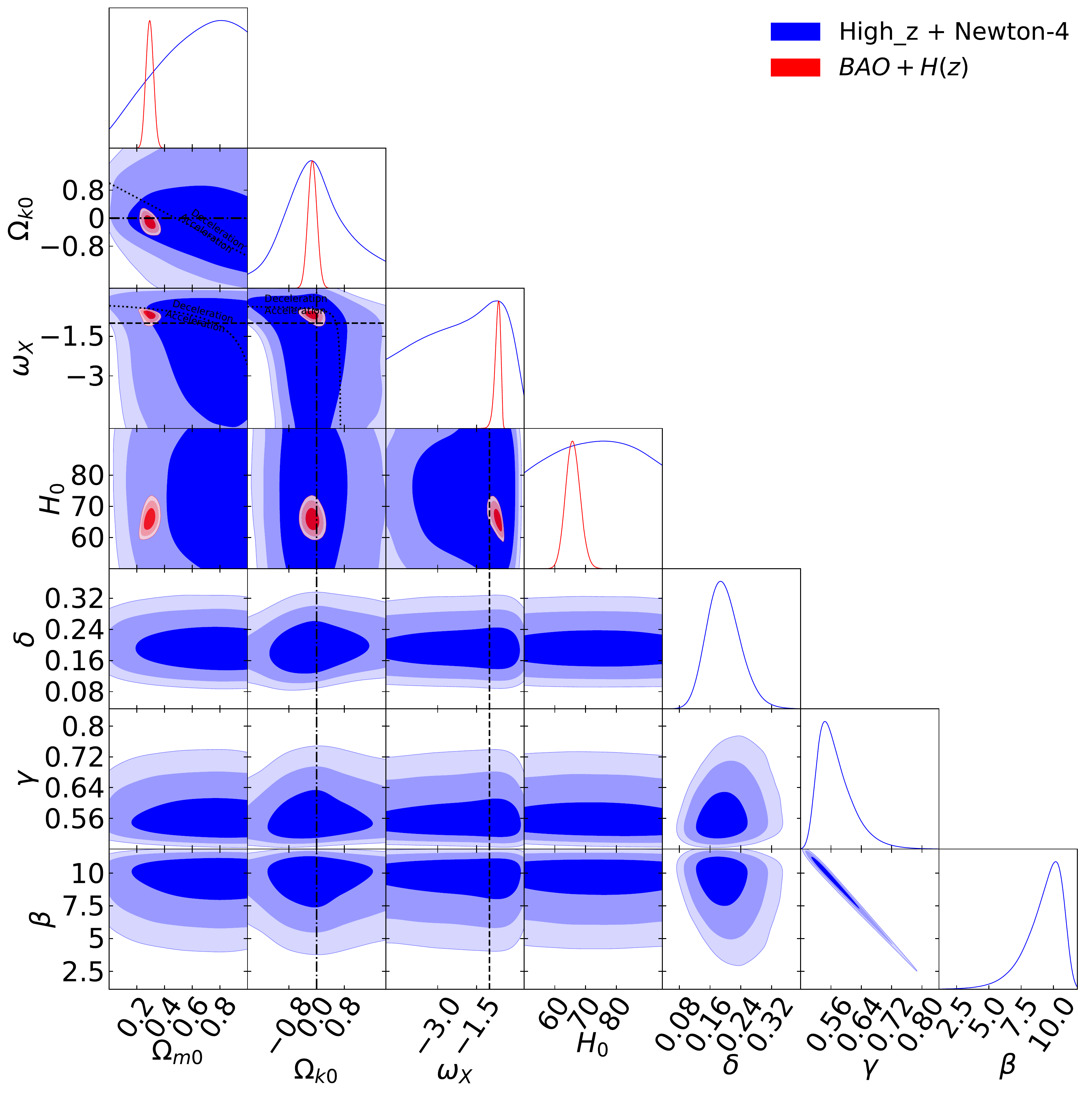}\par
    \includegraphics[width=\linewidth,height=7cm]{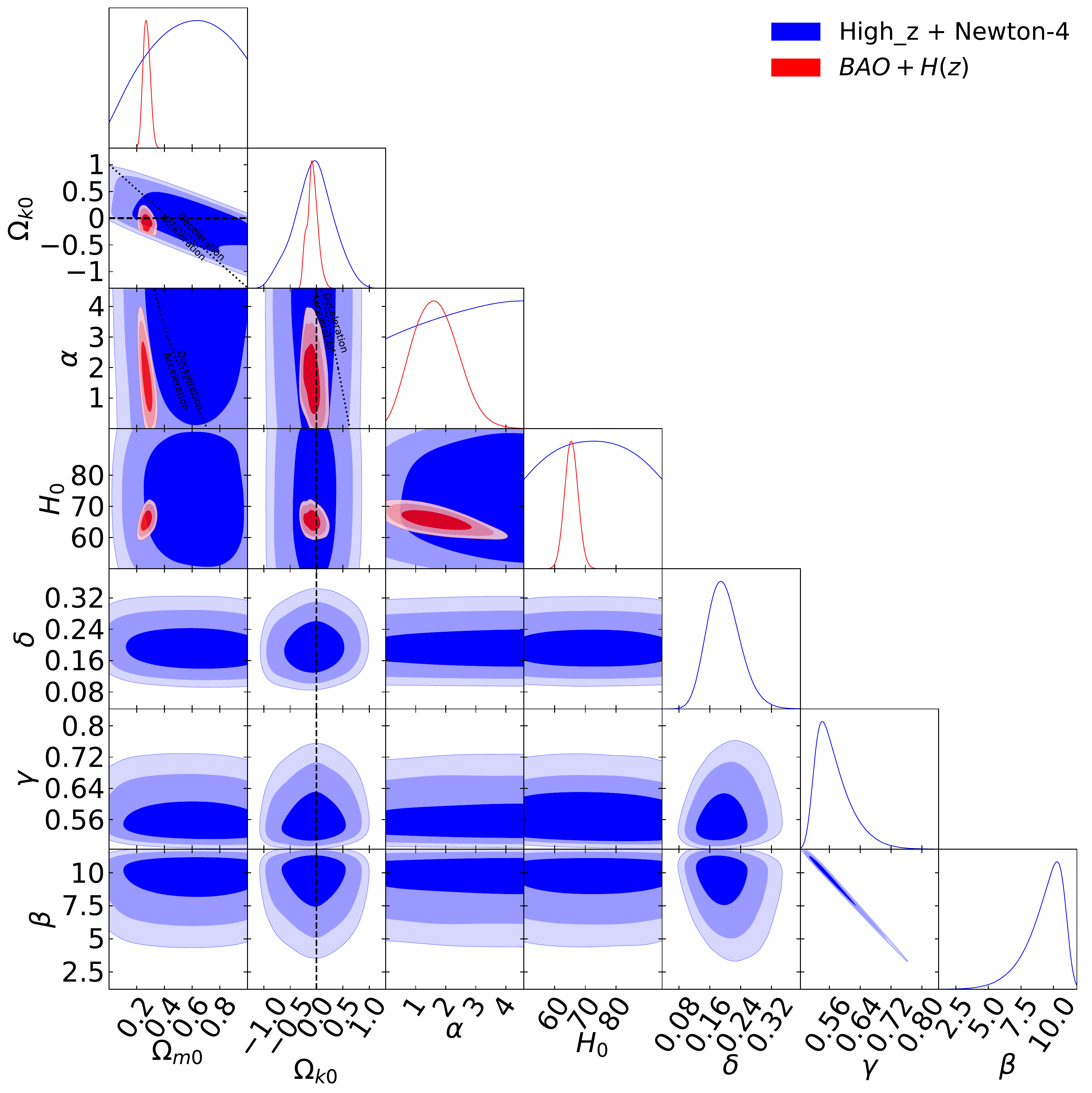}\par
\end{multicols}
\caption{One-dimensional likelihood distributions and two-dimensional likelihood contours at 1$\sigma$, 2$\sigma$, and 3$\sigma$ confidence levels using High-$z$ + Newton-4 (blue) and BAO + $H(z)$ (red) data for all free parameters. Left column shows the flat $\Lambda$CDM model, flat XCDM parametrization, and flat $\phi$CDM model respectively. The black dotted lines in all plots are the zero acceleration lines. The black dashed lines in the flat XCDM parametrization plots are the $\omega_X=-1$ lines. Right column shows the non-flat $\Lambda$CDM model, non-flat XCDM parametrization, and non-flat $\phi$CDM model respectively. Black dotted lines in all plots are the zero acceleration lines. Black dashed lines in the non-flat $\Lambda$CDM and $\phi$CDM model plots and black dotted-dashed lines in the non-flat XCDM parametrization plots correspond to $\Omega_{k0} = 0$. The black dashed lines in the non-flat XCDM parametrization plots are the $\omega_X=-1$ lines.}
\label{fig:Eiso-Ep}
\end{figure*}

\begin{figure*}
\begin{multicols}{2}
    \includegraphics[width=\linewidth,height=7cm]{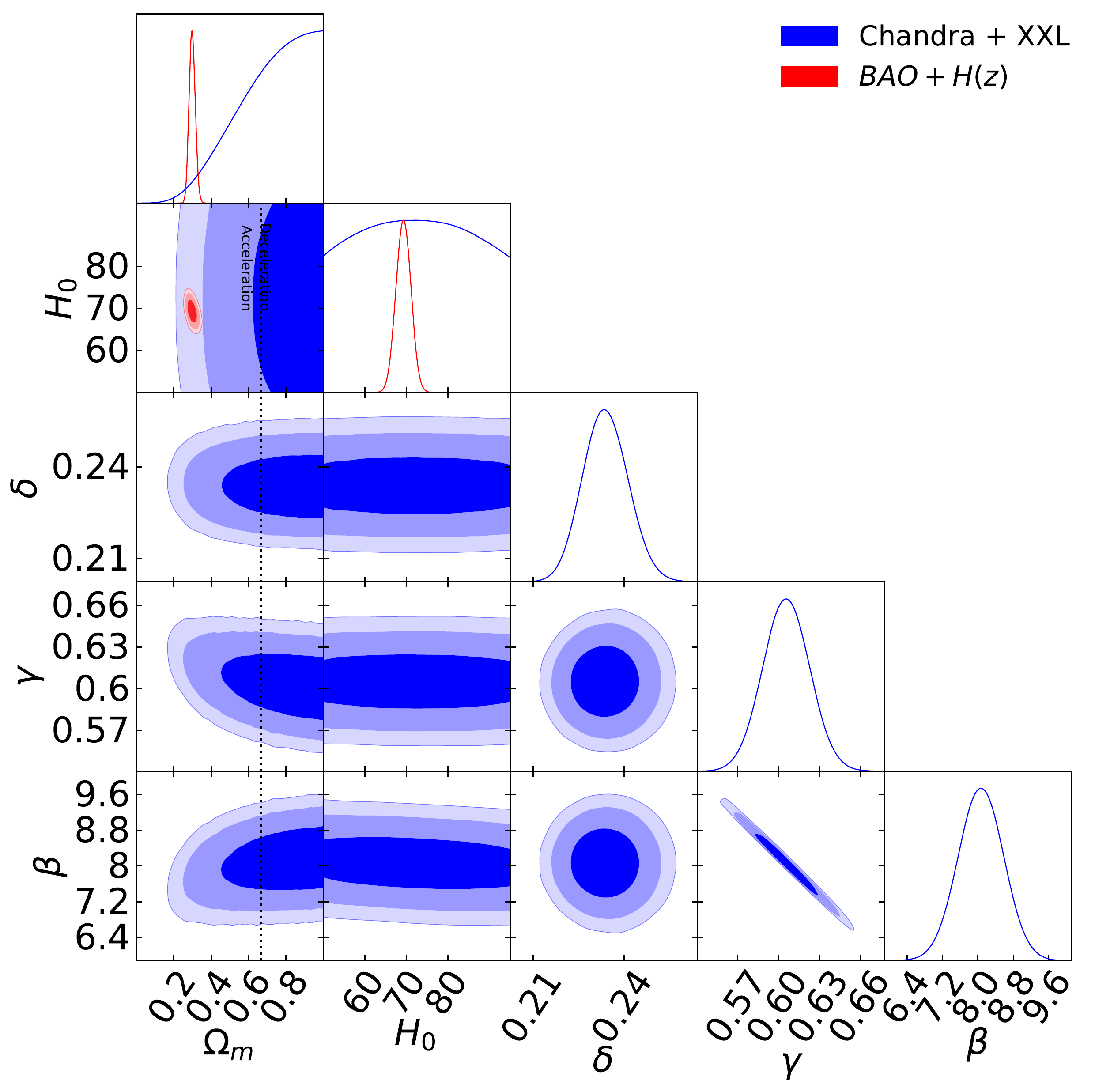}\par
    \includegraphics[width=\linewidth,height=7cm]{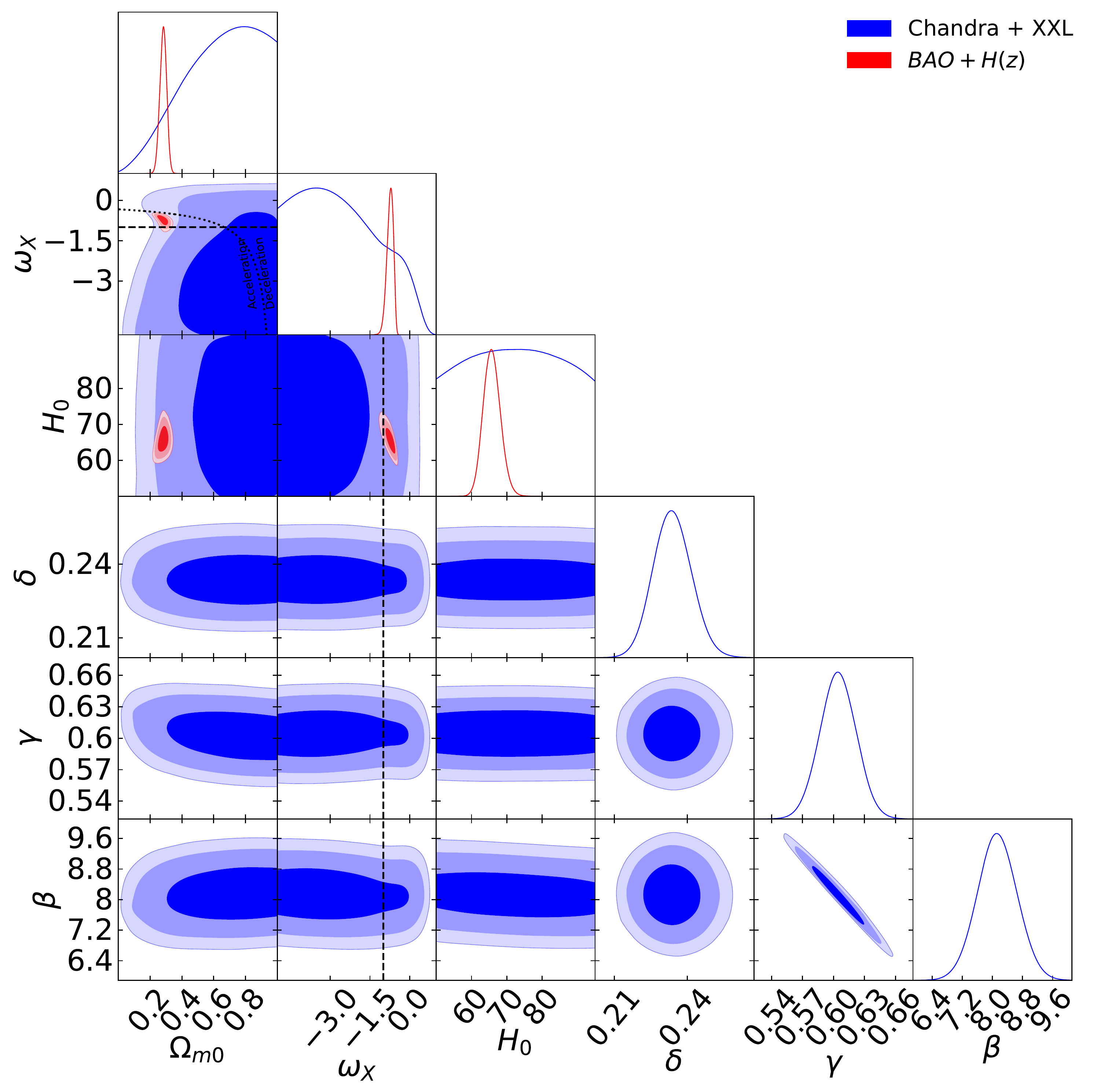}\par
    \includegraphics[width=\linewidth,height=7cm]{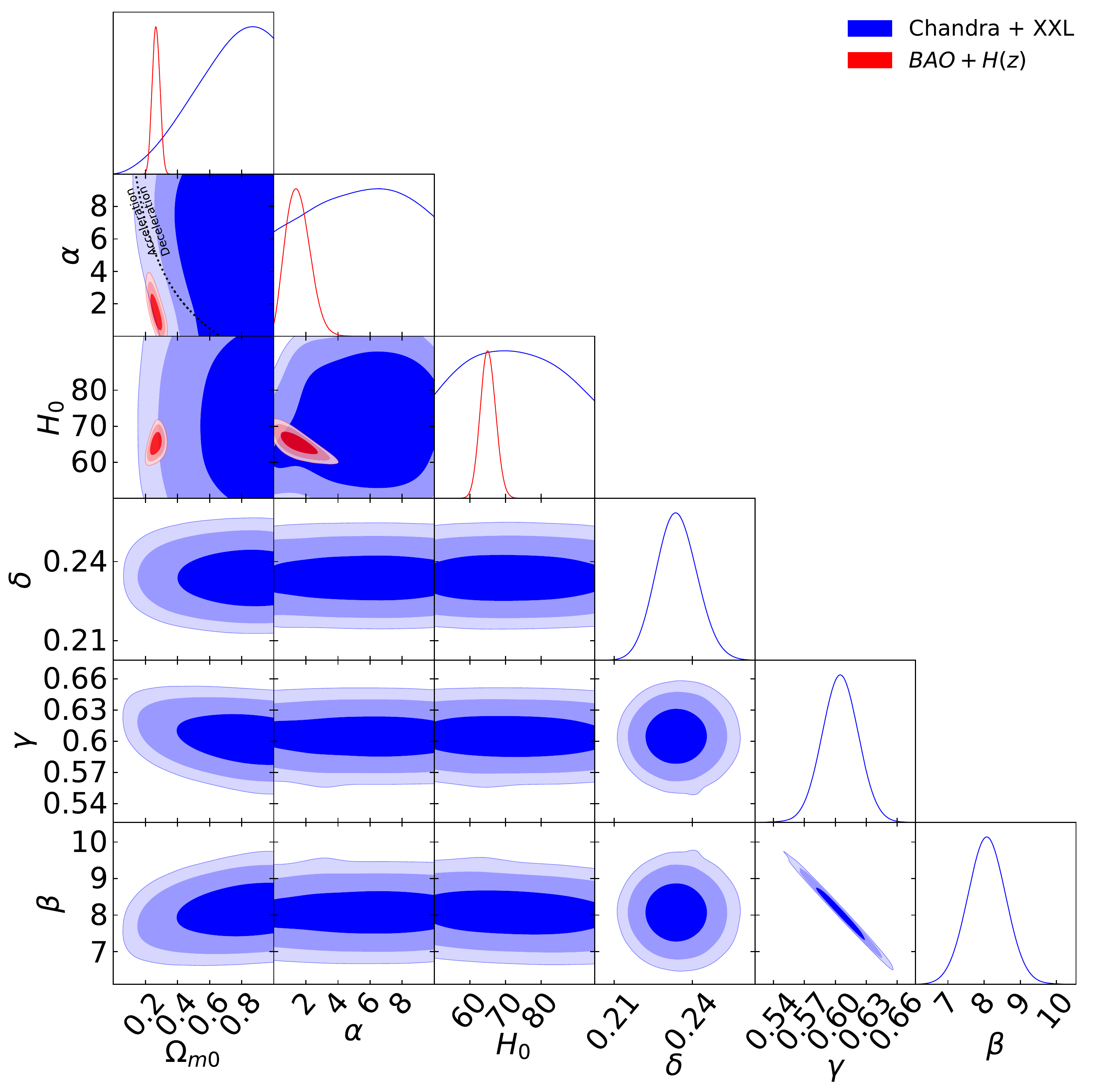}\par
    \includegraphics[width=\linewidth,height=7cm]{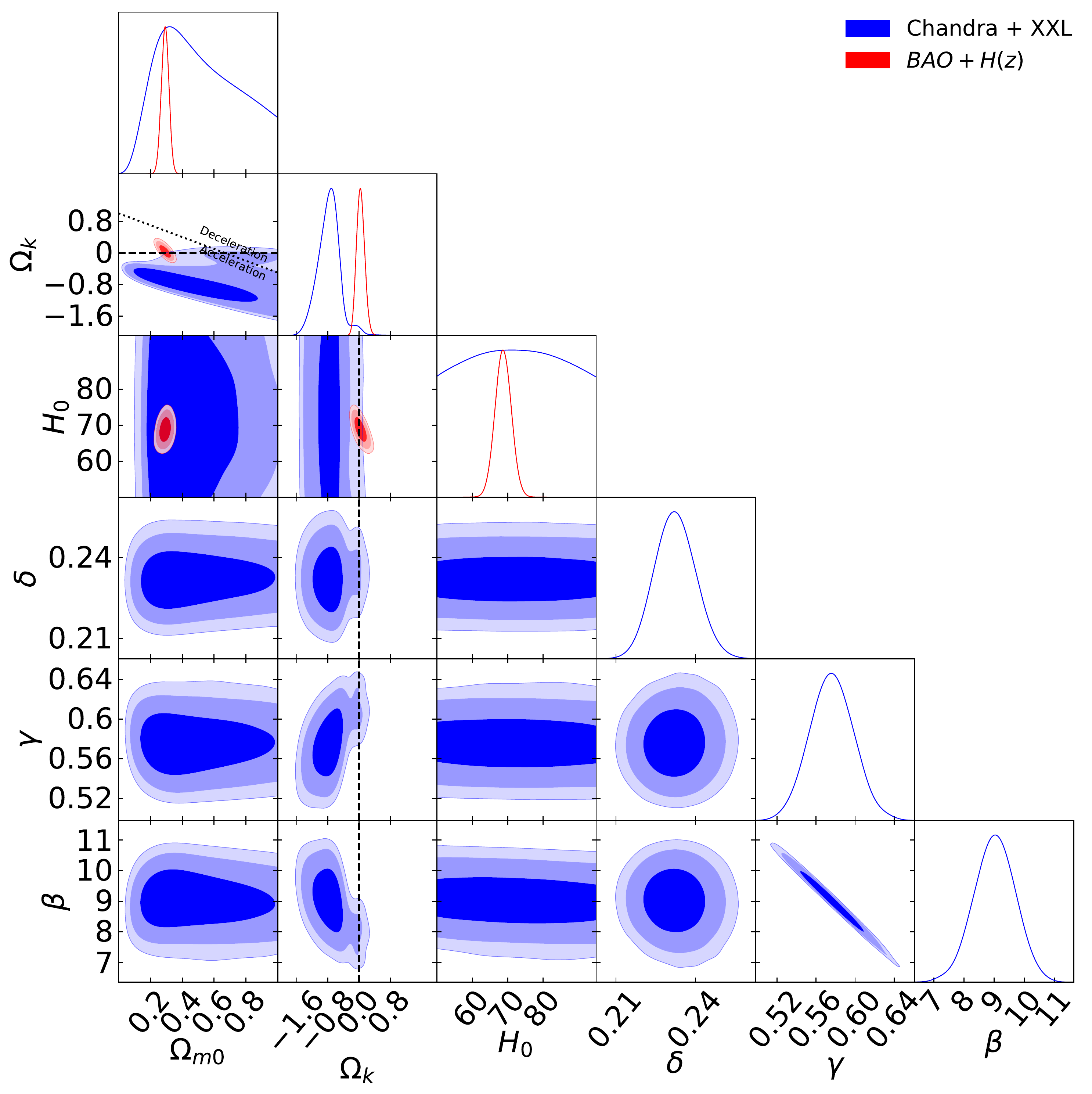}\par
    \includegraphics[width=\linewidth,height=7cm]{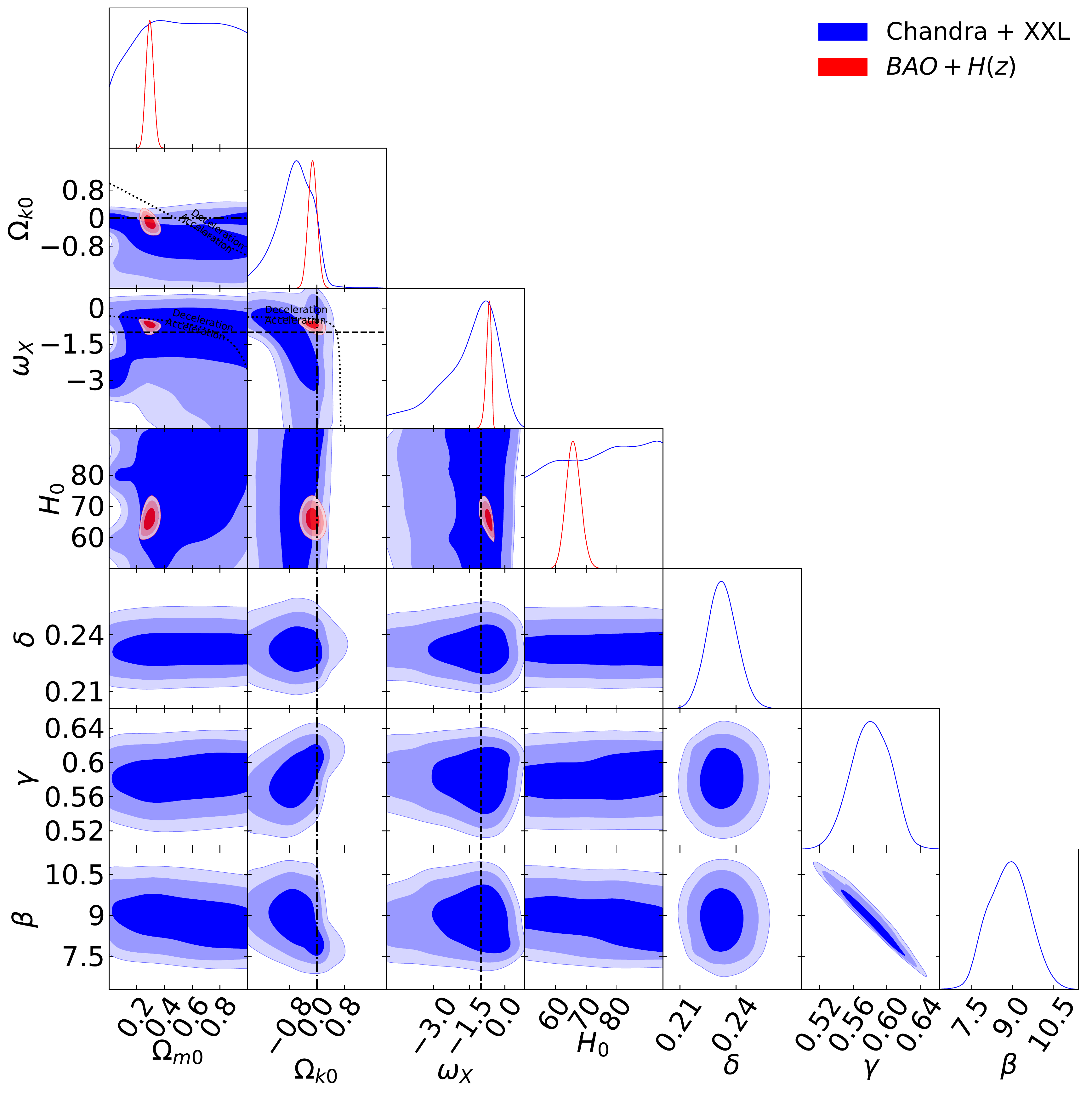}\par
    \includegraphics[width=\linewidth,height=7cm]{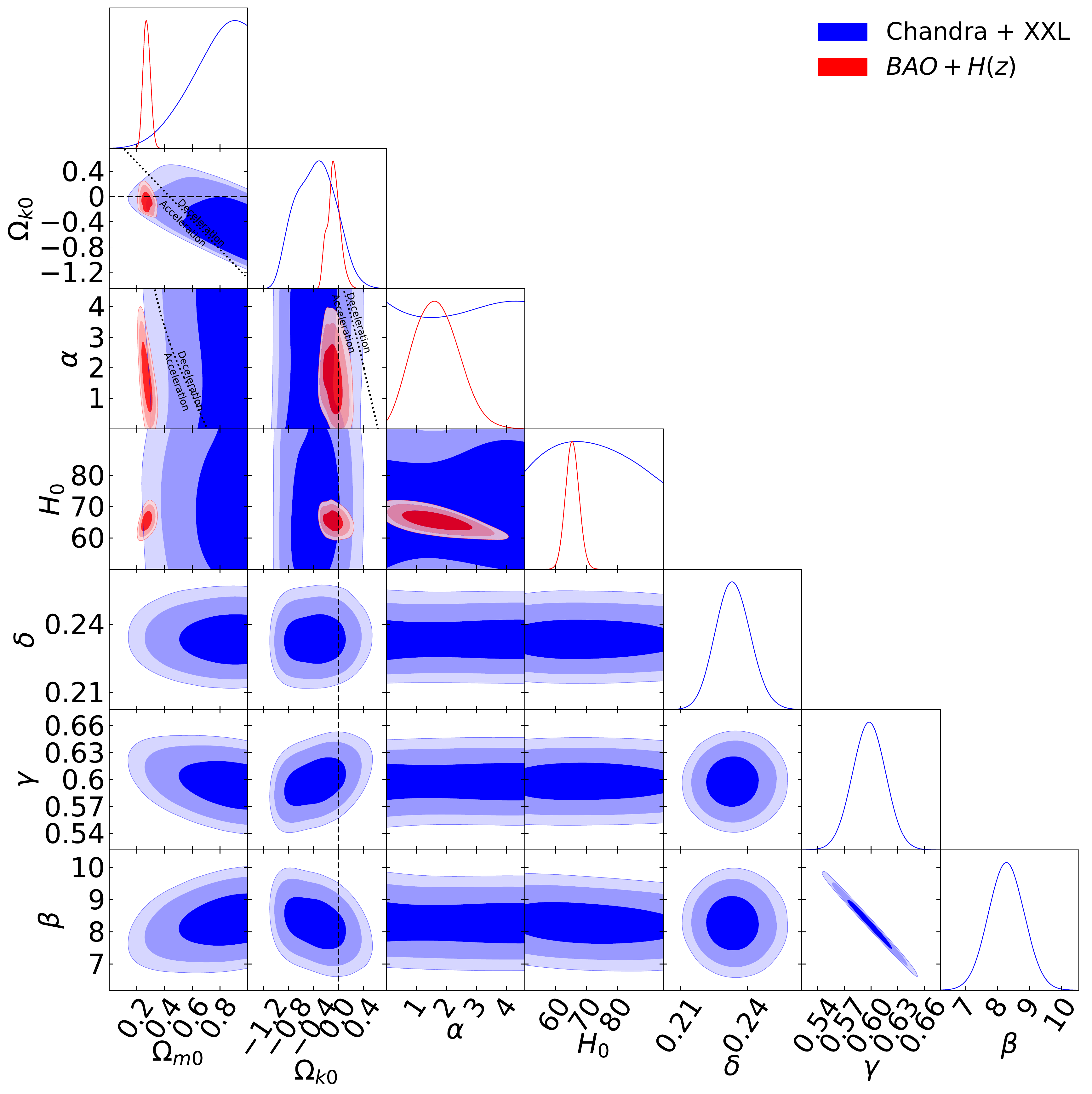}\par
\end{multicols}
\caption{One-dimensional likelihood distributions and two-dimensional likelihood contours at 1$\sigma$, 2$\sigma$, and 3$\sigma$ confidence levels using Chandra + XXL (blue) and BAO + $H(z)$ (red) data for all free parameters. Left column shows the flat $\Lambda$CDM model, flat XCDM parametrization, and flat $\phi$CDM model respectively. The black dotted lines in all plots are the zero acceleration lines. The black dashed lines in the flat XCDM parametrization plots are the $\omega_X=-1$ lines. Right column shows the non-flat $\Lambda$CDM model, non-flat XCDM parametrization, and non-flat $\phi$CDM model respectively. Black dotted lines in all plots are the zero acceleration lines. Black dashed lines in the non-flat $\Lambda$CDM and $\phi$CDM model plots and black dotted-dashed lines in the non-flat XCDM parametrization plots correspond to $\Omega_{k0} = 0$. The black dashed lines in the non-flat XCDM parametrization plots are the $\omega_X=-1$ lines.}
\label{fig:Eiso-Ep}
\end{figure*}

\begin{figure*}
\begin{multicols}{2}
    \includegraphics[width=\linewidth,height=7cm]{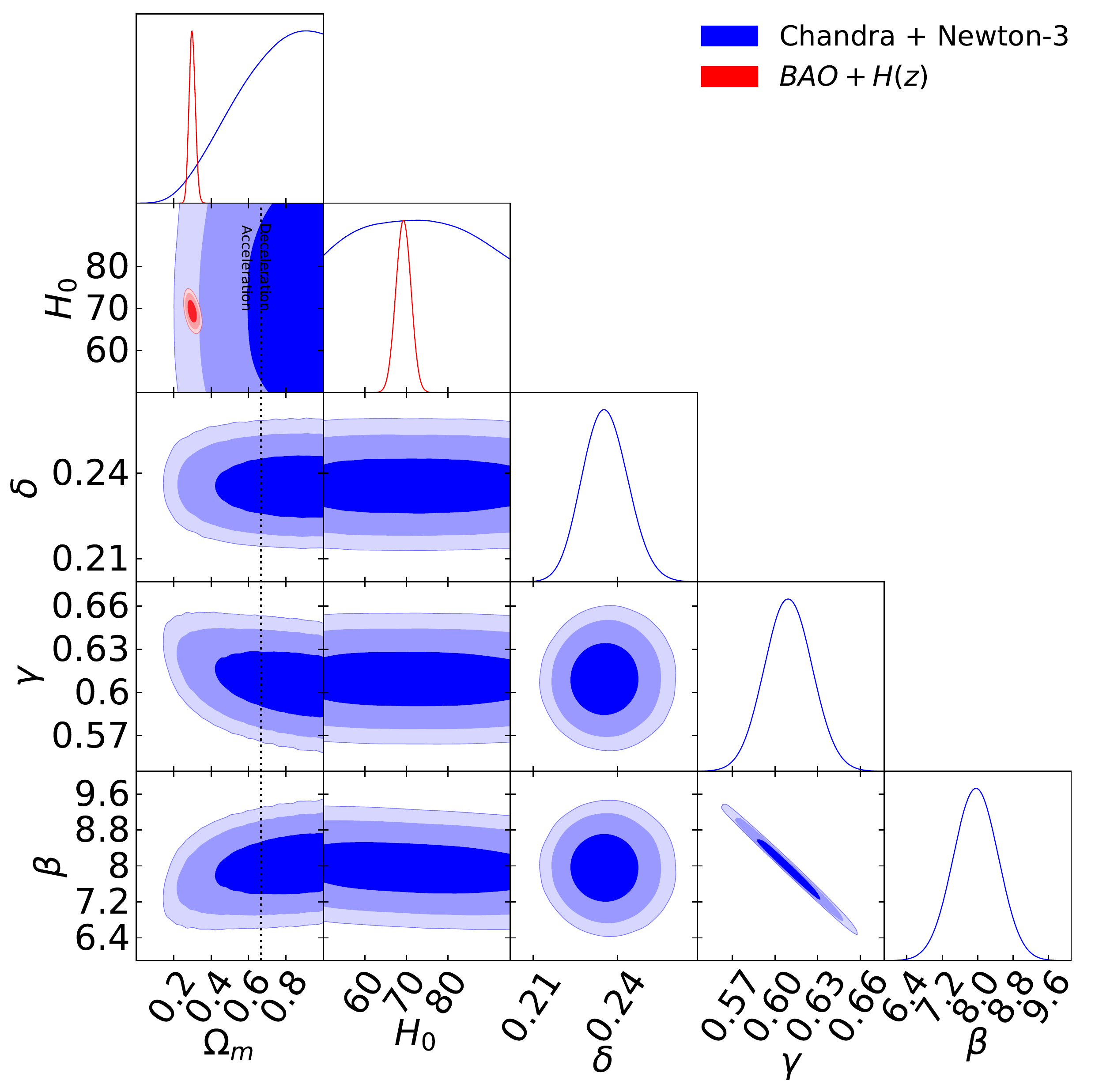}\par
    \includegraphics[width=\linewidth,height=7cm]{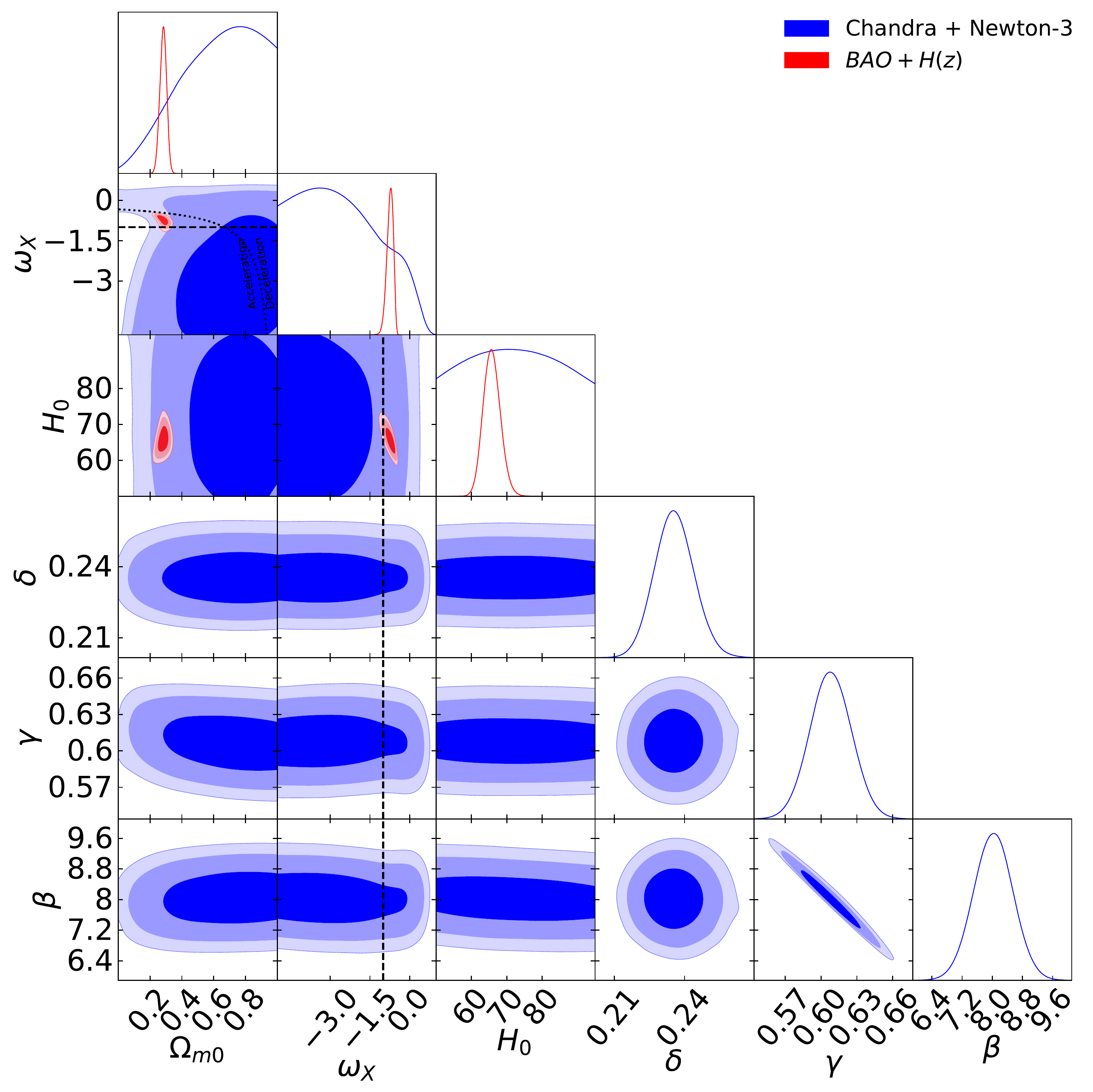}\par
    \includegraphics[width=\linewidth,height=7cm]{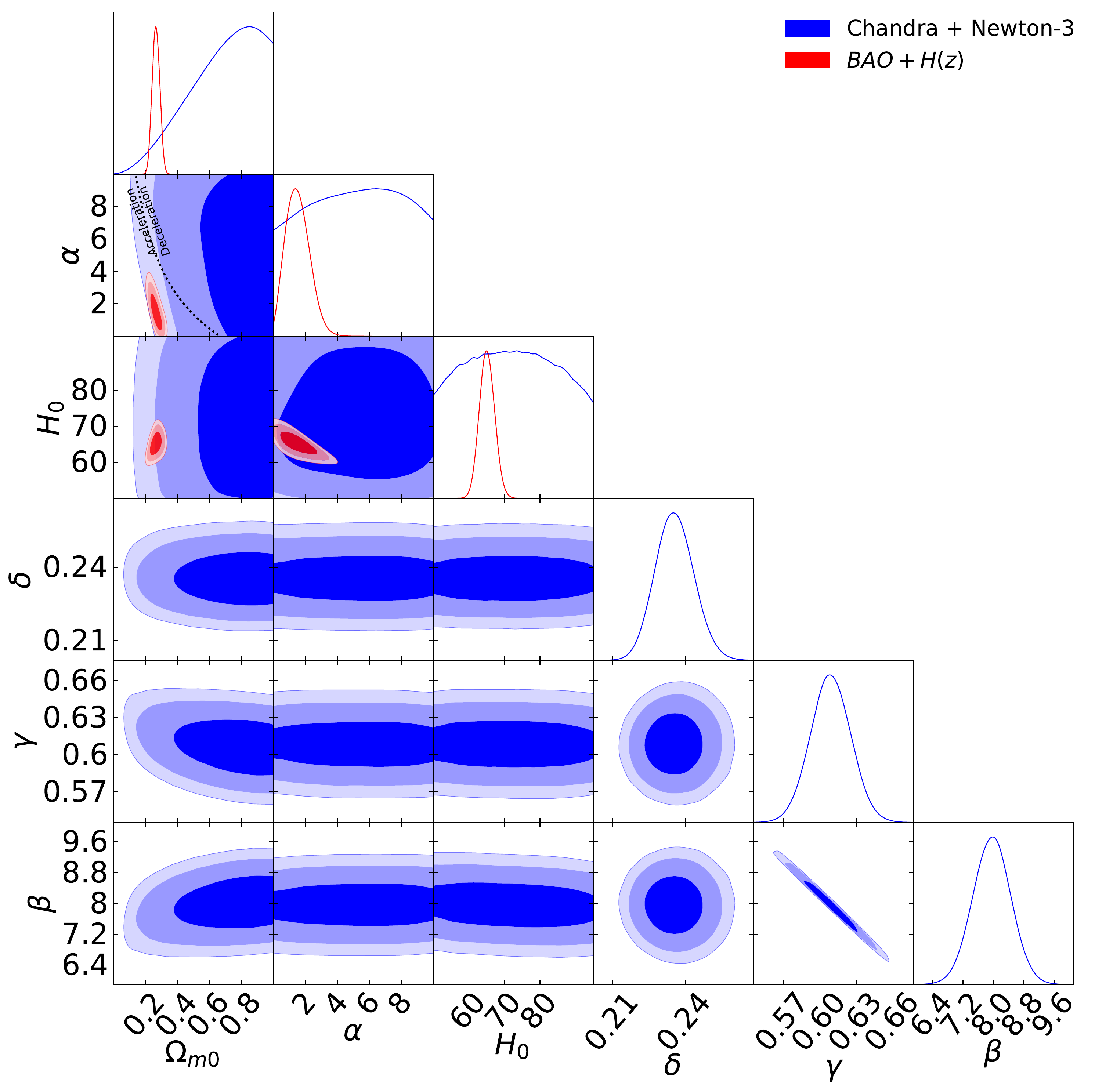}\par
    \includegraphics[width=\linewidth,height=7cm]{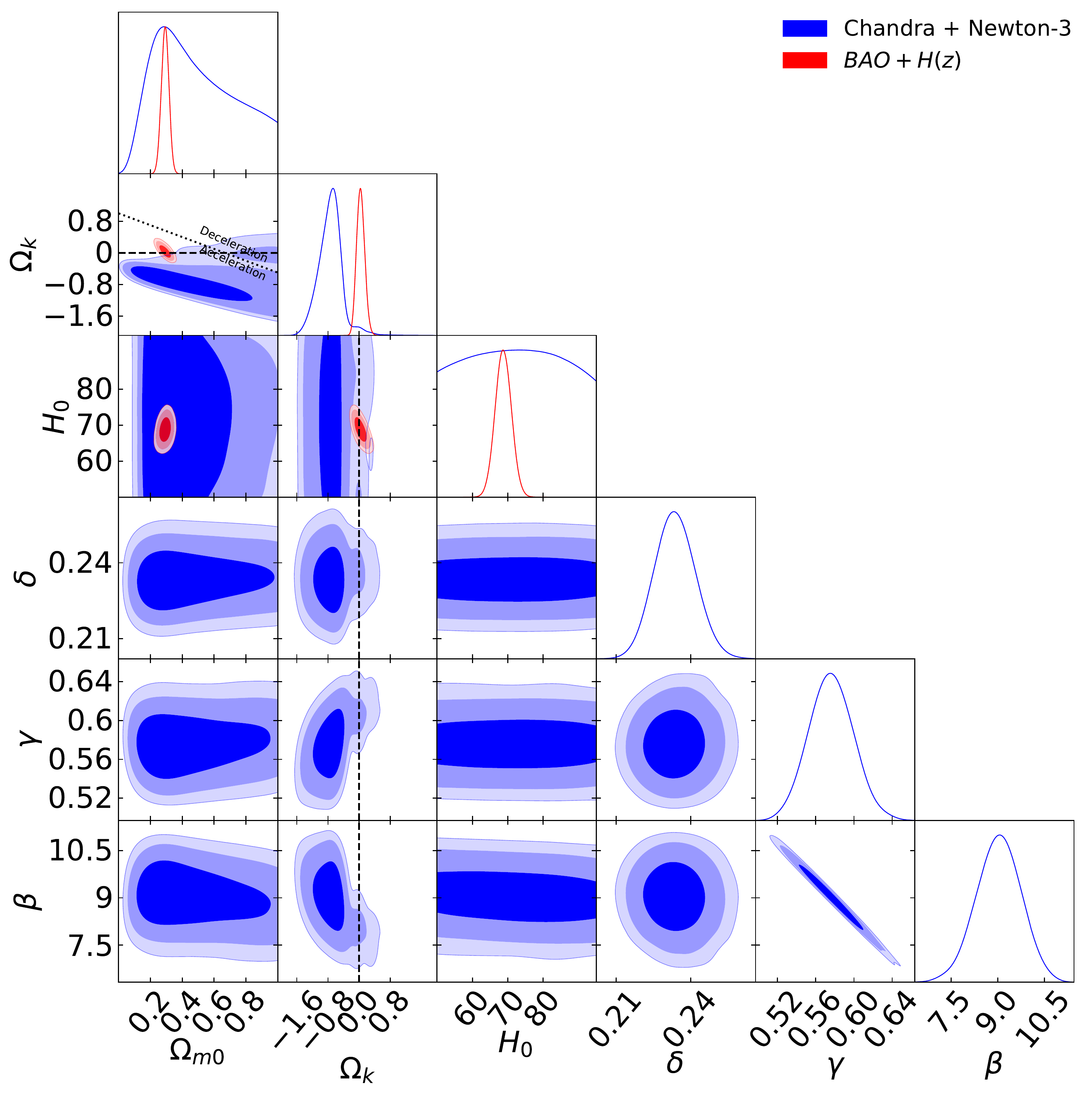}\par
    \includegraphics[width=\linewidth,height=7cm]{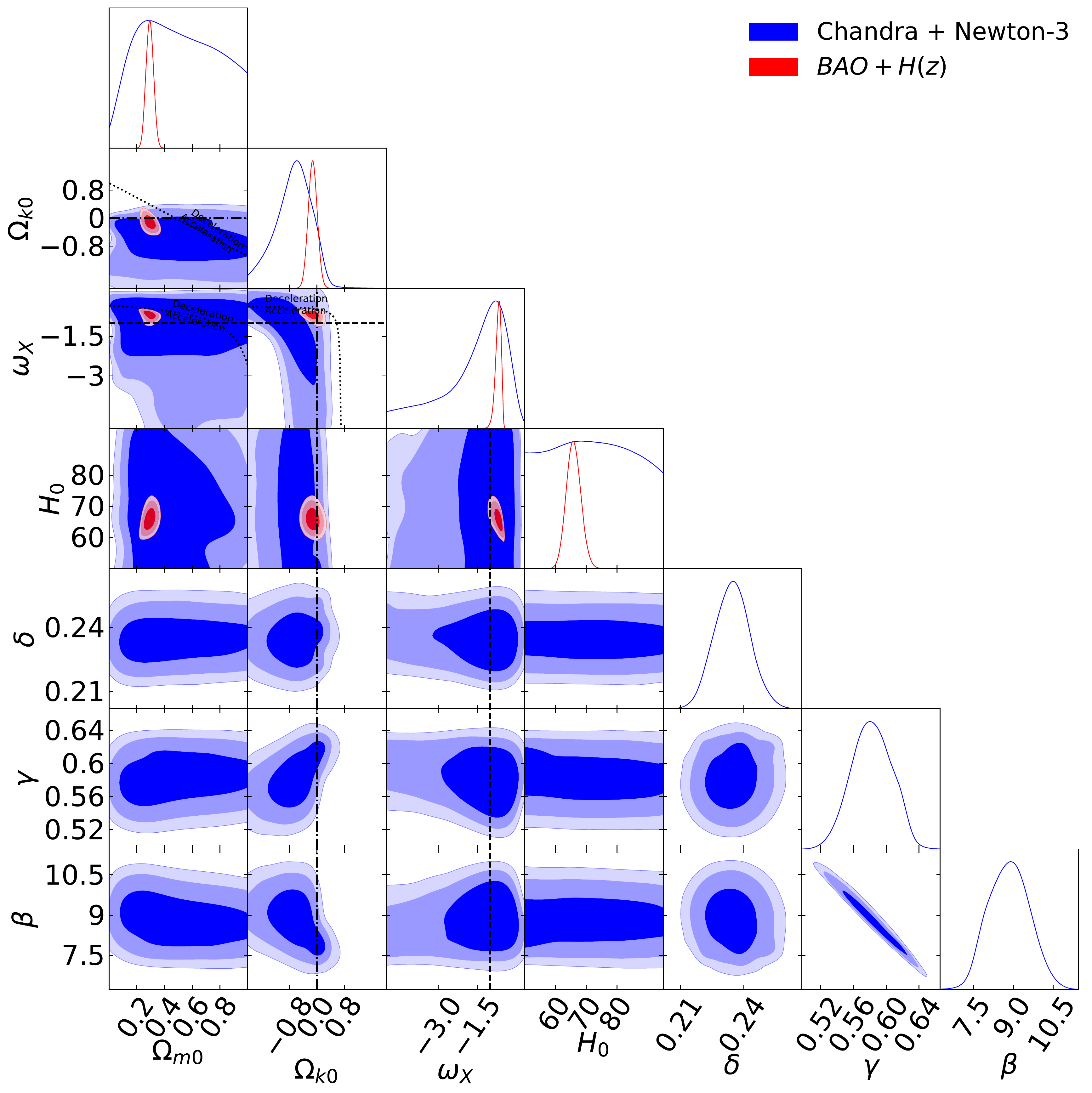}\par
    \includegraphics[width=\linewidth,height=7cm]{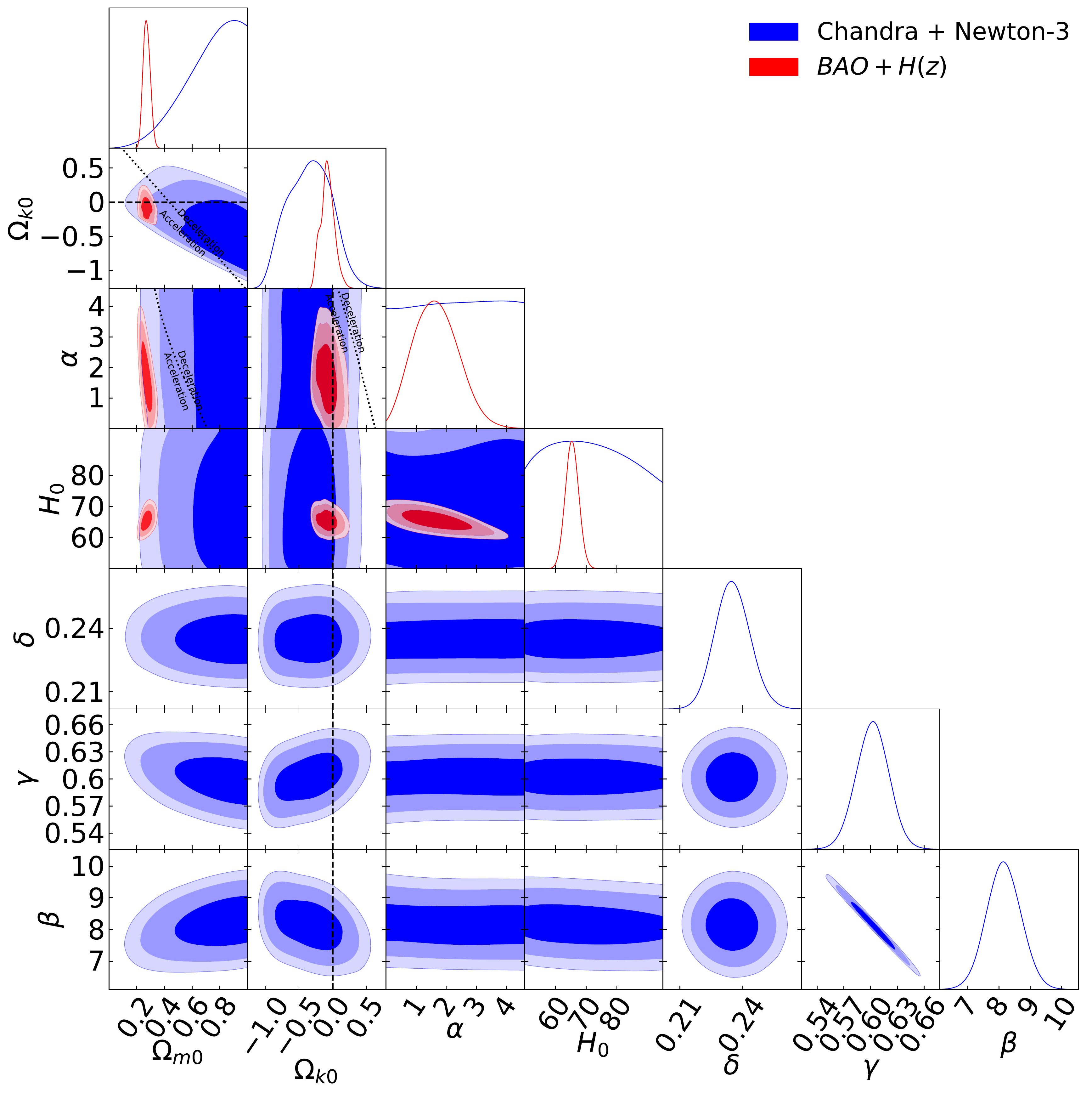}\par
\end{multicols}
\caption{One-dimensional likelihood distributions and two-dimensional likelihood contours at 1$\sigma$, 2$\sigma$, and 3$\sigma$ confidence levels using Chandra + Newton-3 (blue) and BAO + $H(z)$ (red) data for all free parameters. Left column shows the flat $\Lambda$CDM model, flat XCDM parametrization, and flat $\phi$CDM model respectively. The black dotted lines in all plots are the zero acceleration lines. The black dashed lines in the flat XCDM parametrization plots are the $\omega_X=-1$ lines. Right column shows the non-flat $\Lambda$CDM model, non-flat XCDM parametrization, and non-flat $\phi$CDM model respectively. Black dotted lines in all plots are the zero acceleration lines. Black dashed lines in the non-flat $\Lambda$CDM and $\phi$CDM model plots and black dotted-dashed lines in the non-flat XCDM parametrization plots correspond to $\Omega_{k0} = 0$. The black dashed lines in the non-flat XCDM parametrization plots are the $\omega_X=-1$ lines.}
\label{fig:Eiso-Ep}
\end{figure*}

\begin{figure*}
\begin{multicols}{2}
    \includegraphics[width=\linewidth,height=7cm]{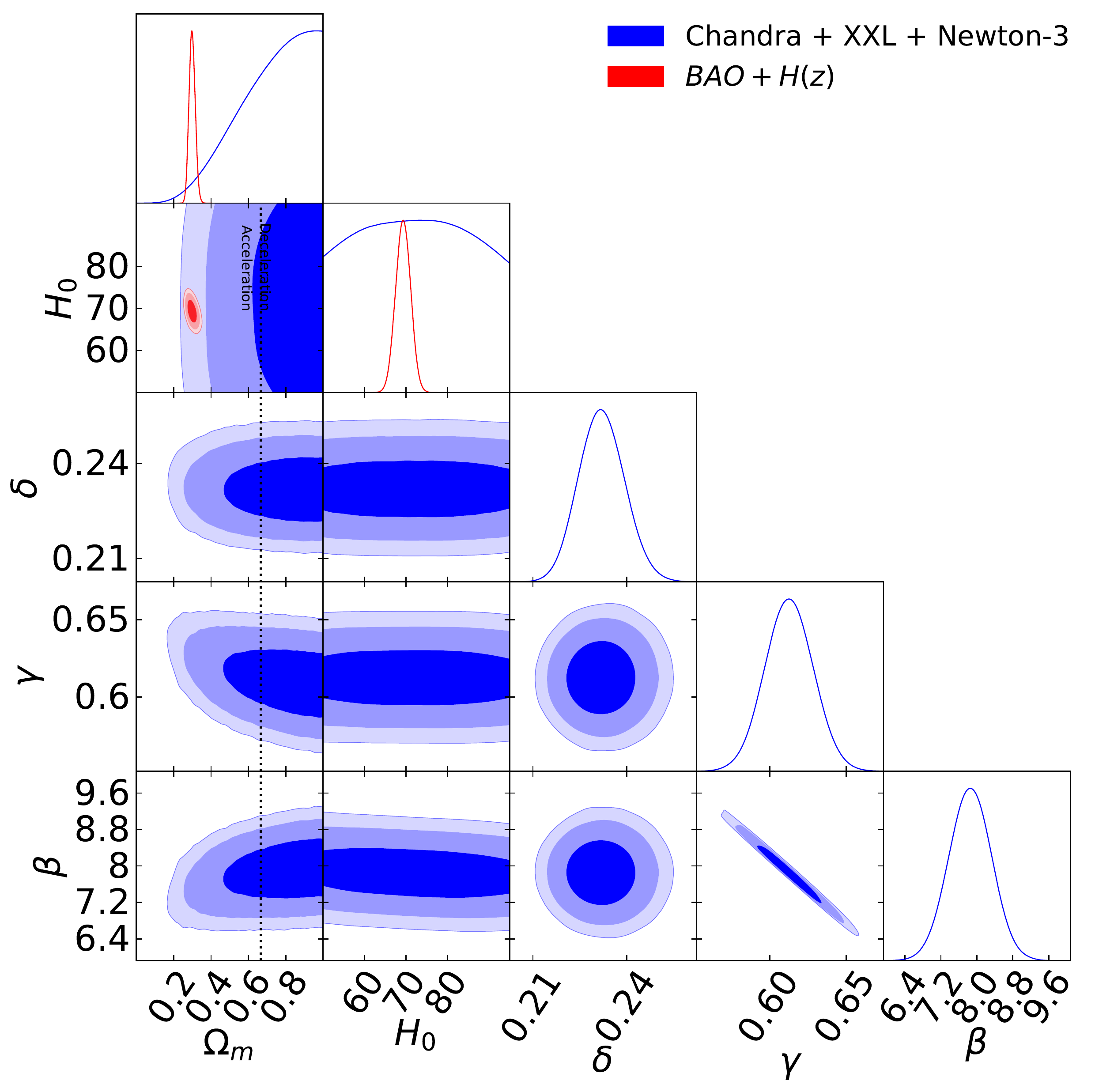}\par
    \includegraphics[width=\linewidth,height=7cm]{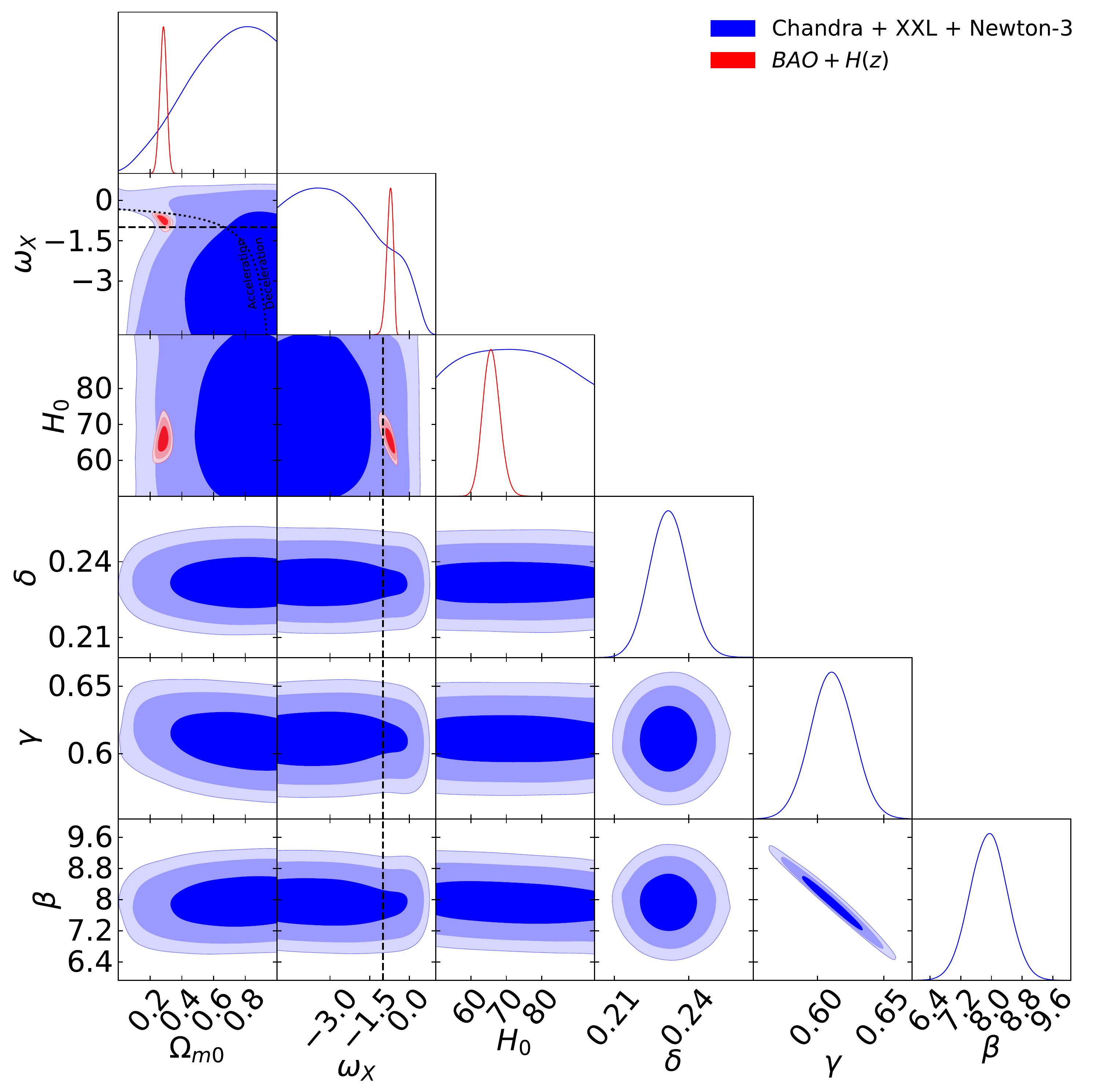}\par
    \includegraphics[width=\linewidth,height=7cm]{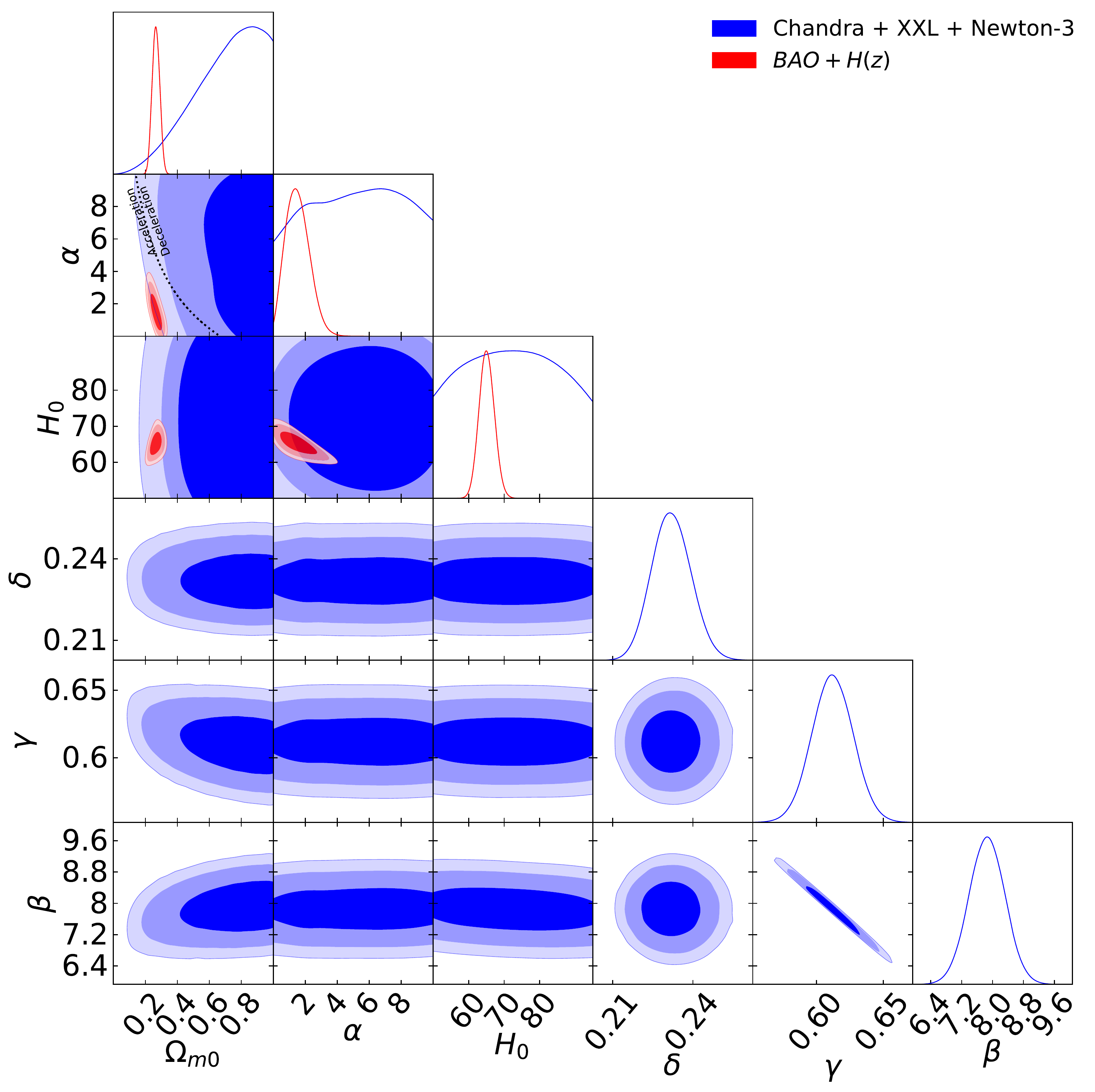}\par
    \includegraphics[width=\linewidth,height=7cm]{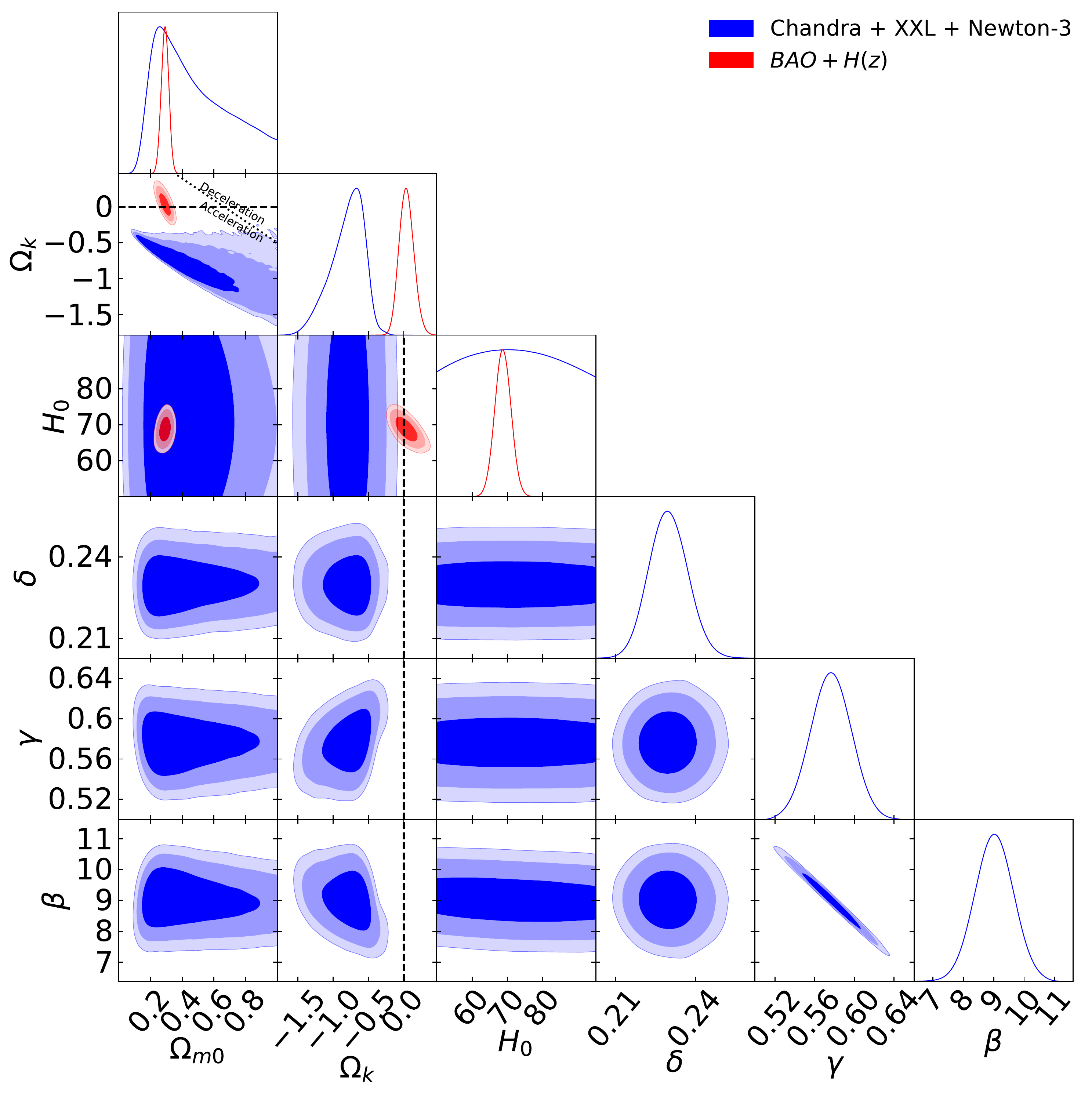}\par
    \includegraphics[width=\linewidth,height=7cm]{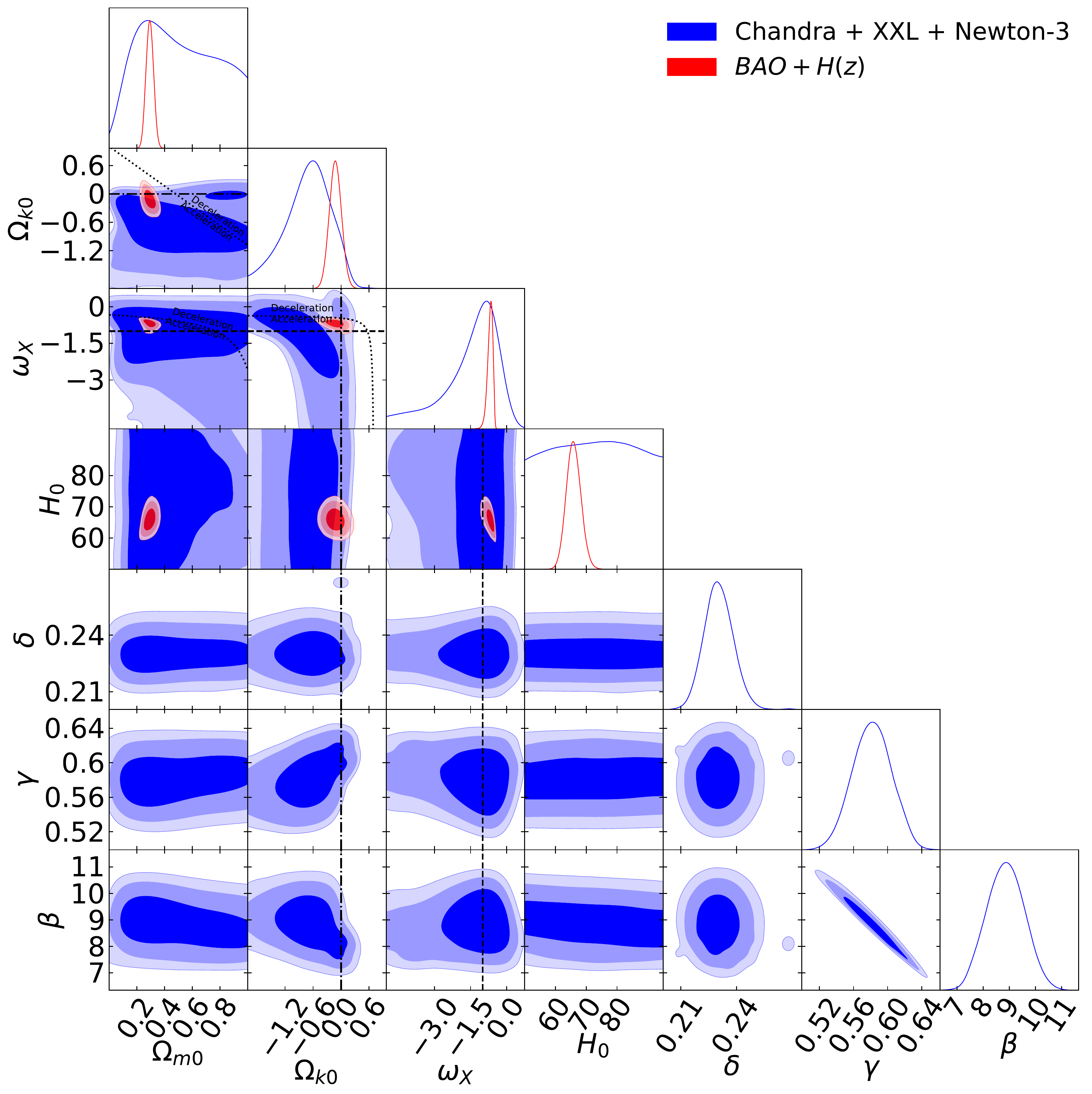}\par
    \includegraphics[width=\linewidth,height=7cm]{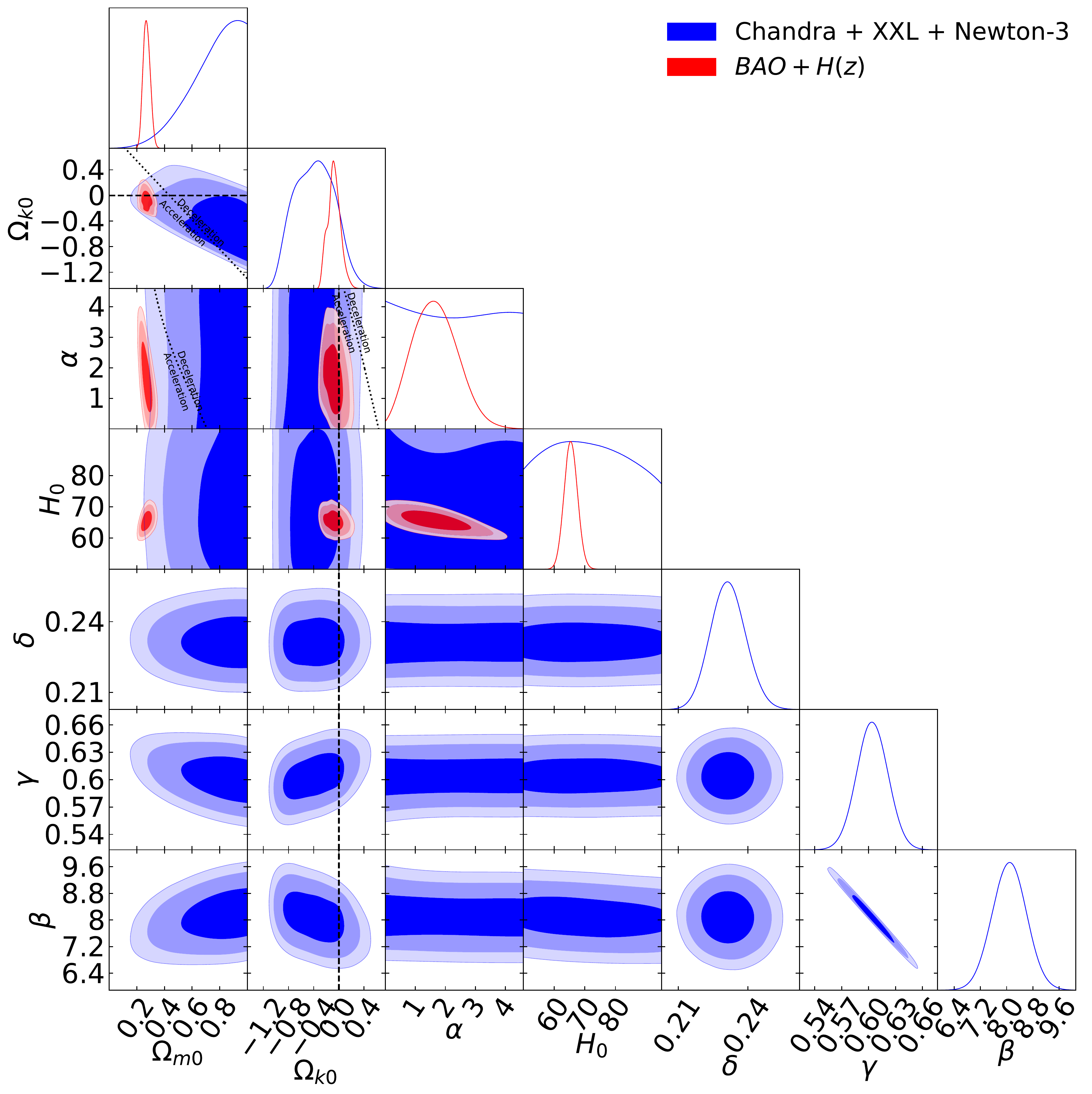}\par
\end{multicols}
\caption{One-dimensional likelihood distributions and two-dimensional likelihood contours at 1$\sigma$, 2$\sigma$, and 3$\sigma$ confidence levels using Chandra + XXL + Newton-3 (blue) and BAO + $H(z)$ (red) data for all free parameters. Left column shows the flat $\Lambda$CDM model, flat XCDM parametrization, and flat $\phi$CDM model respectively. The black dotted lines in all plots are the zero acceleration lines. The black dashed lines in the flat XCDM parametrization plots are the $\omega_X=-1$ lines. Right column shows the non-flat $\Lambda$CDM model, non-flat XCDM parametrization, and non-flat $\phi$CDM model respectively. Black dotted lines in all plots are the zero acceleration lines. Black dashed lines in the non-flat $\Lambda$CDM and $\phi$CDM model plots and black dotted-dashed lines in the non-flat XCDM parametrization plots correspond to $\Omega_{k0} = 0$. The black dashed lines in the non-flat XCDM parametrization plots are the $\omega_X=-1$ lines.}
\label{fig:Eiso-Ep}
\end{figure*}

\begin{figure*}
\begin{multicols}{2}
    \includegraphics[width=\linewidth,height=7cm]{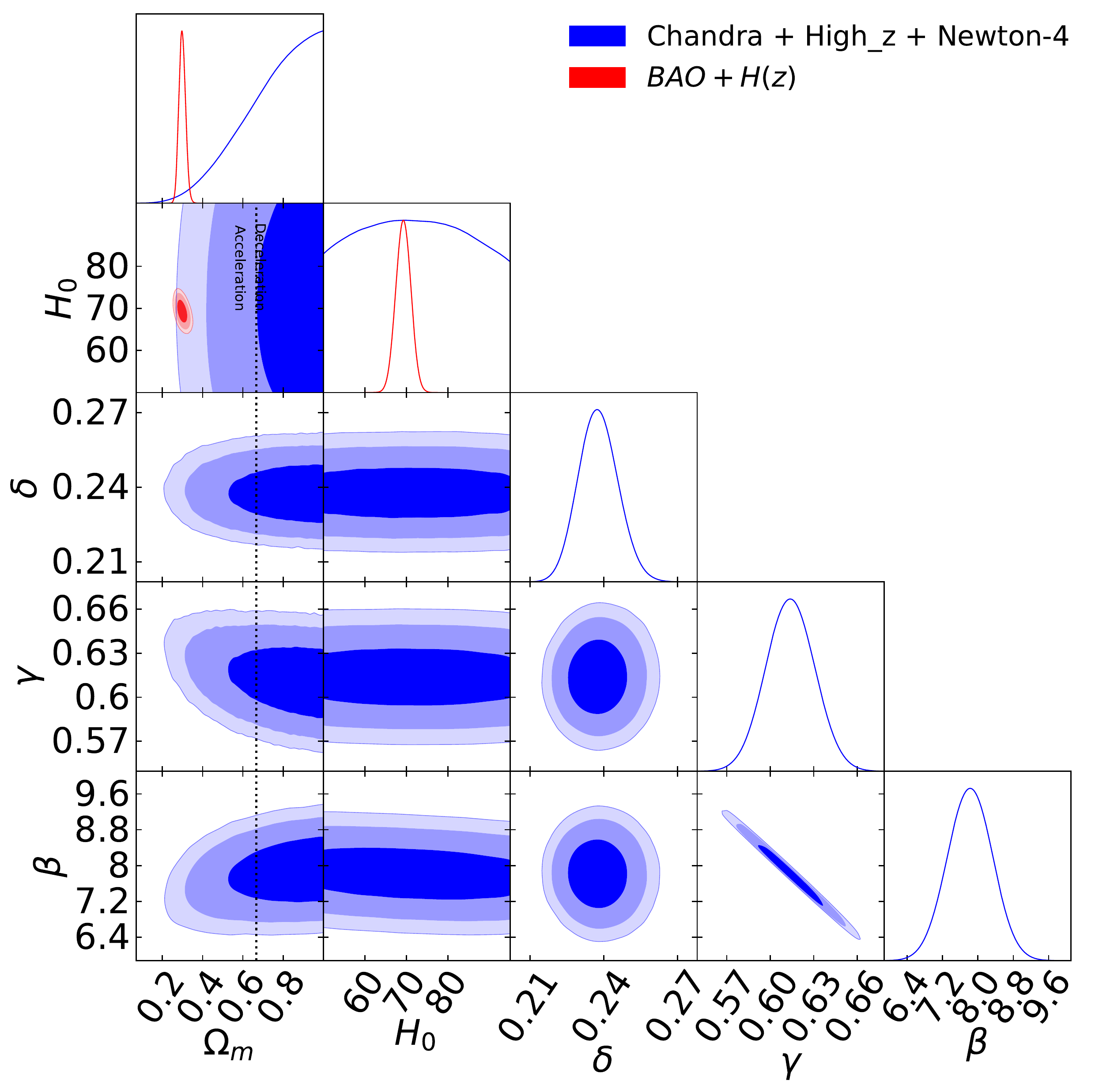}\par
    \includegraphics[width=\linewidth,height=7cm]{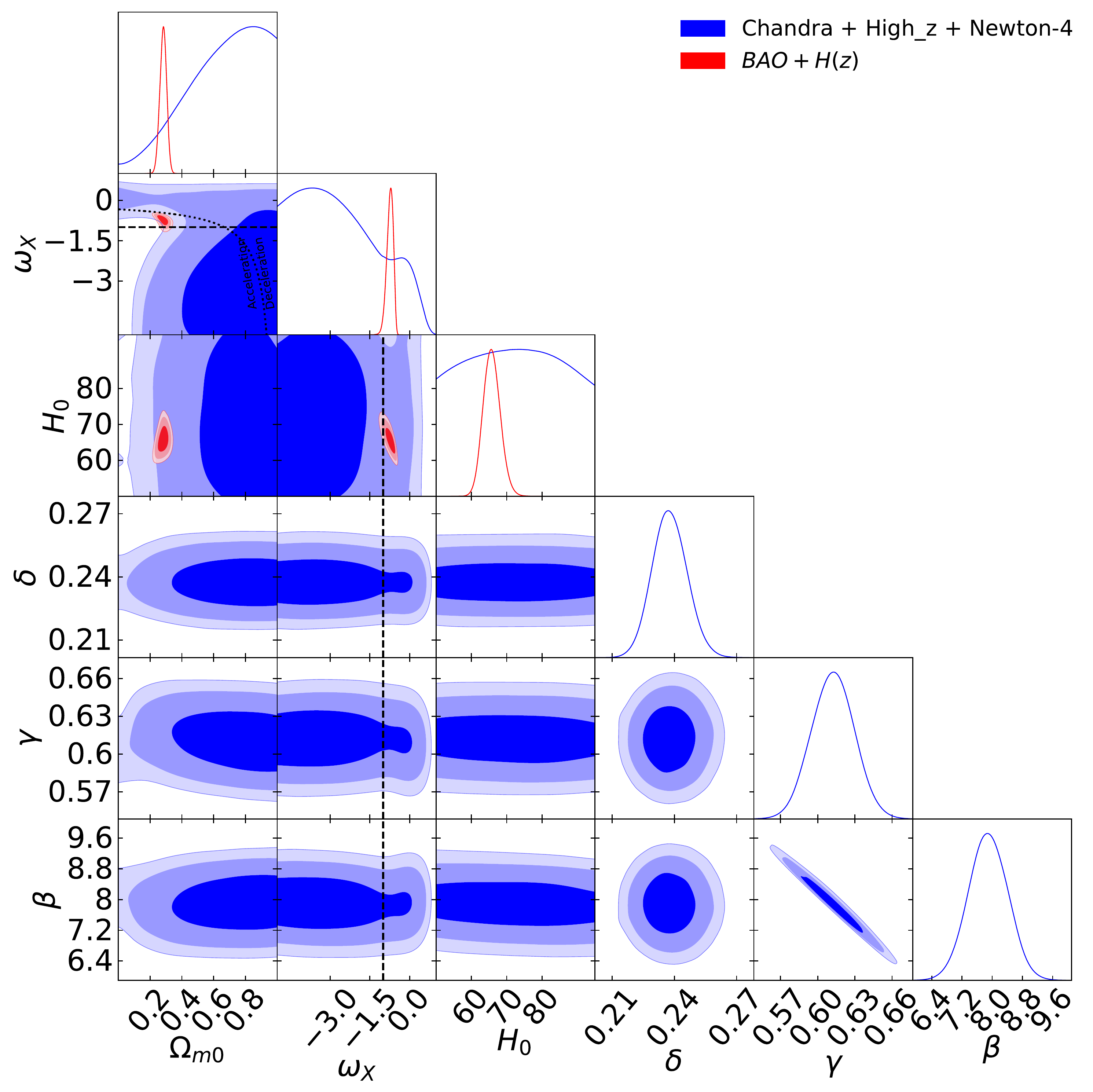}\par
    \includegraphics[width=\linewidth,height=7cm]{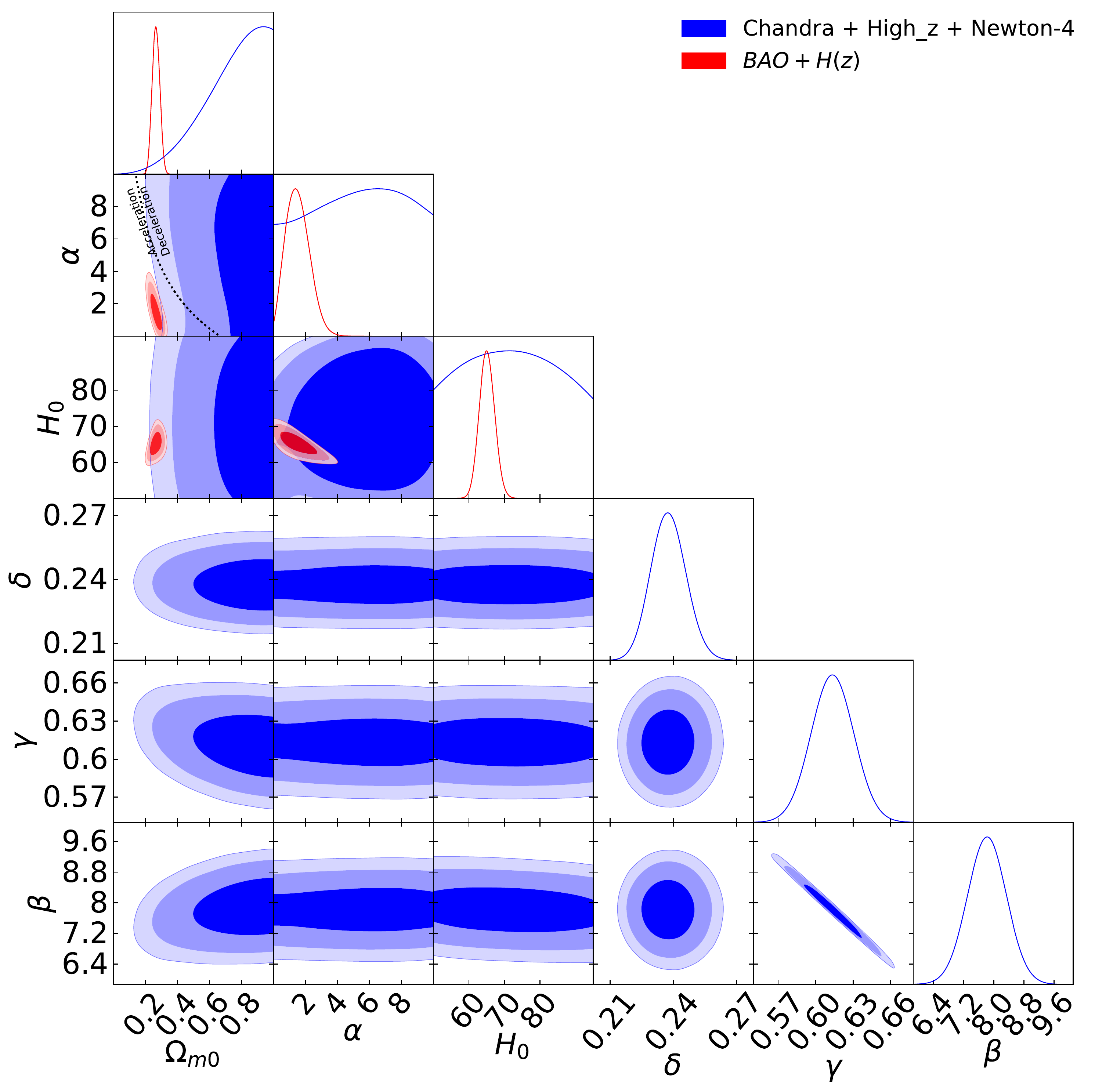}\par
    \includegraphics[width=\linewidth,height=7cm]{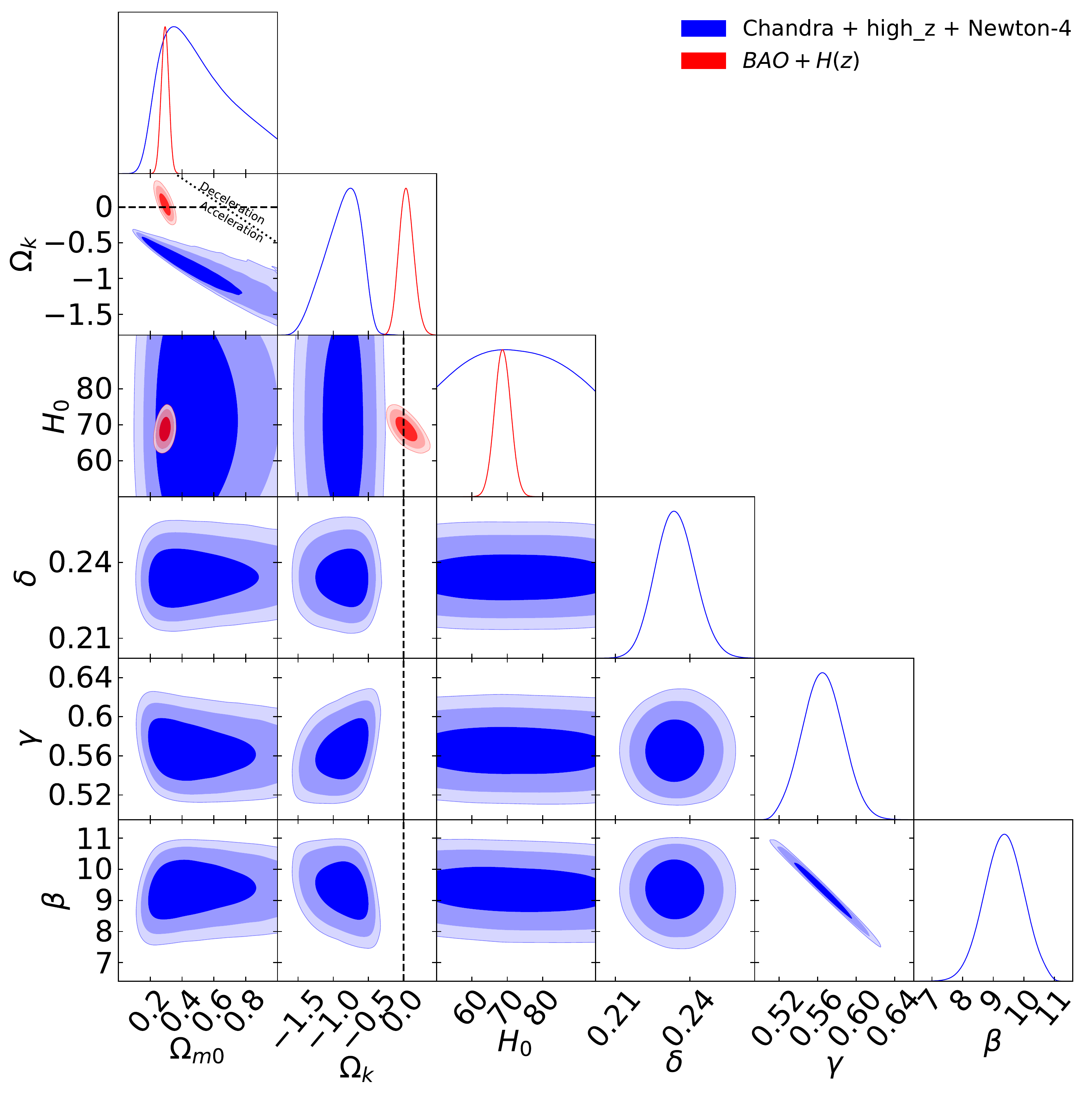}\par
    \includegraphics[width=\linewidth,height=7cm]{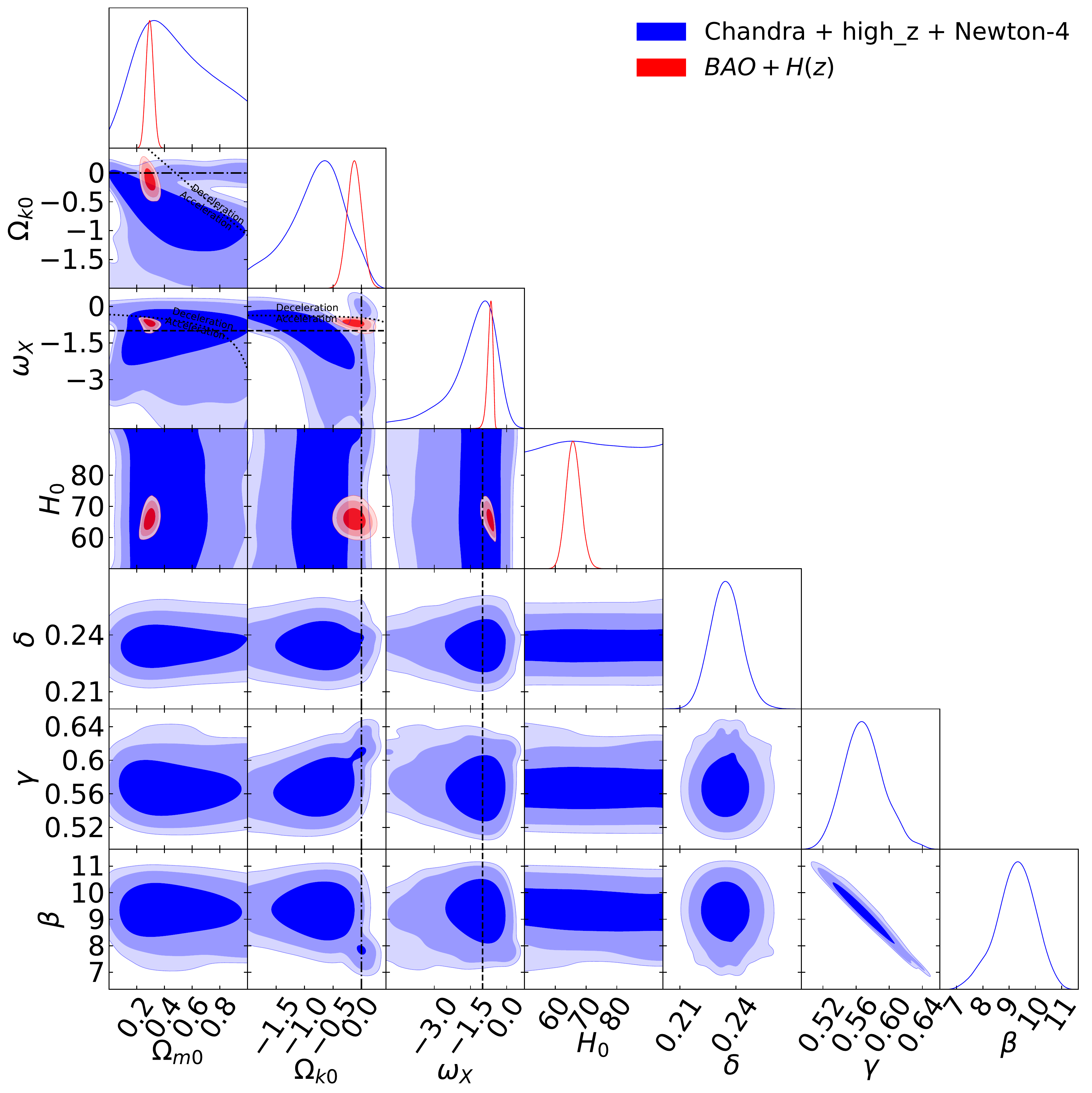}\par
    \includegraphics[width=\linewidth,height=7cm]{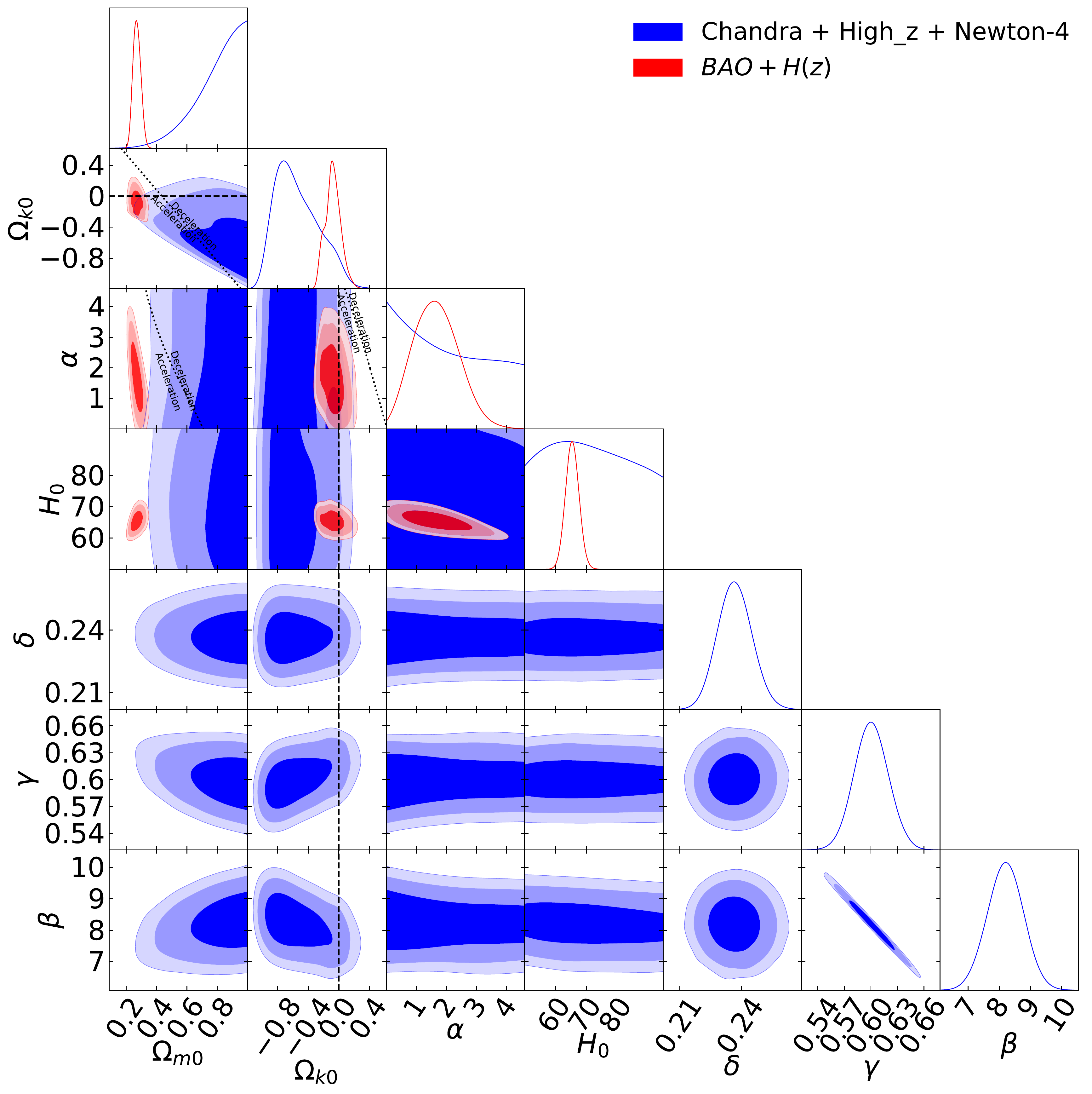}\par
\end{multicols}
\caption{One-dimensional likelihood distributions and two-dimensional likelihood contours at 1$\sigma$, 2$\sigma$, and 3$\sigma$ confidence levels using Chandra + High-$z$ + Newton-4 (blue) and BAO + $H(z)$ (red) data for all free parameters. Left column shows the flat $\Lambda$CDM model, flat XCDM parametrization, and flat $\phi$CDM model respectively. The black dotted lines in all plots are the zero acceleration lines. The black dashed lines in the flat XCDM parametrization plots are the $\omega_X=-1$ lines. Right column shows the non-flat $\Lambda$CDM model, non-flat XCDM parametrization, and non-flat $\phi$CDM model respectively. Black dotted lines in all plots are the zero acceleration lines. Black dashed lines in the non-flat $\Lambda$CDM and $\phi$CDM model plots and black dotted-dashed lines in the non-flat XCDM parametrization plots correspond to $\Omega_{k0} = 0$. The black dashed lines in the non-flat XCDM parametrization plots are the $\omega_X=-1$ lines.}
\label{fig:Eiso-Ep}
\end{figure*}

\begin{figure*}
\begin{multicols}{2}
    \includegraphics[width=\linewidth,height=7cm]{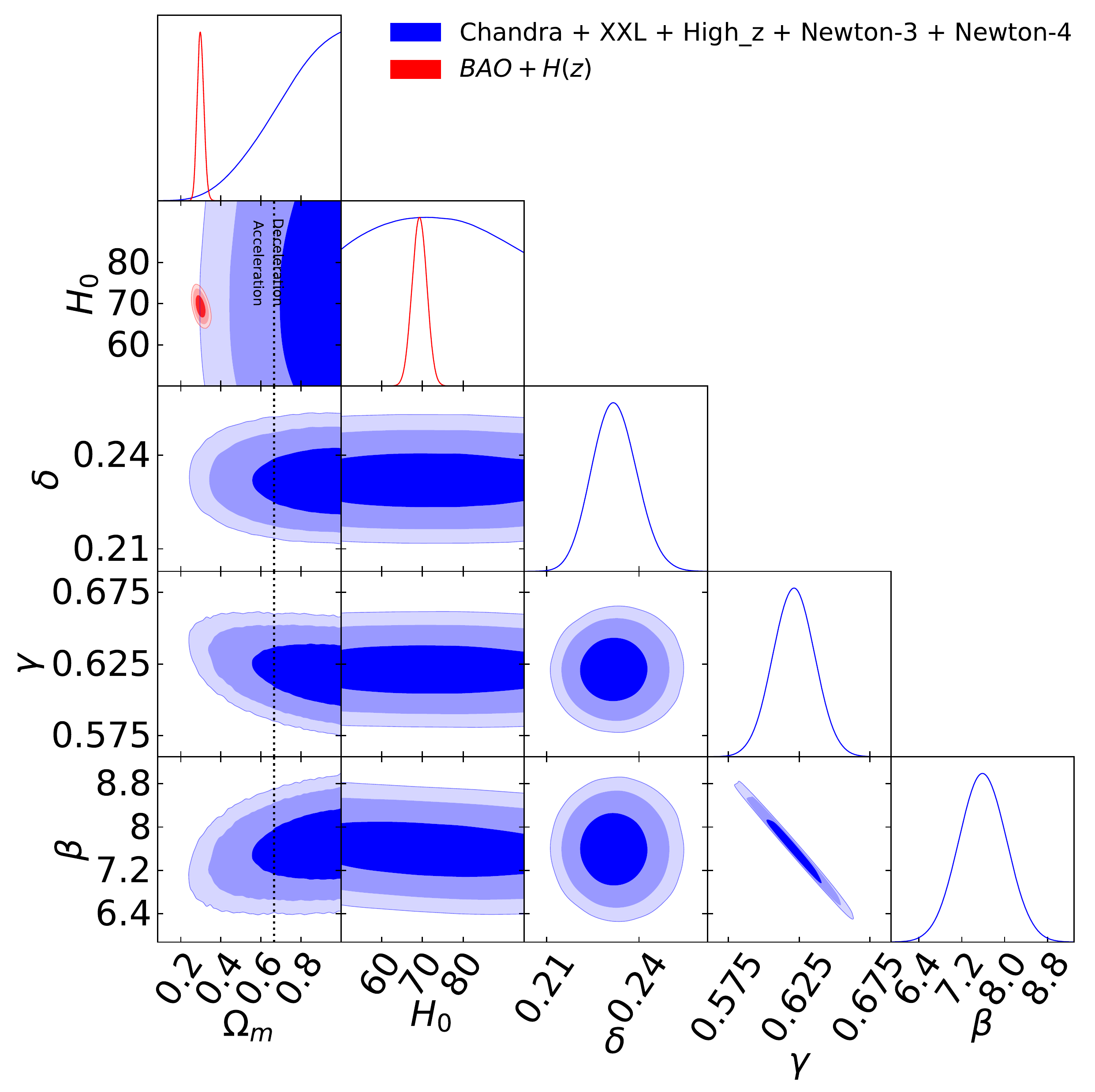}\par
    \includegraphics[width=\linewidth,height=7cm]{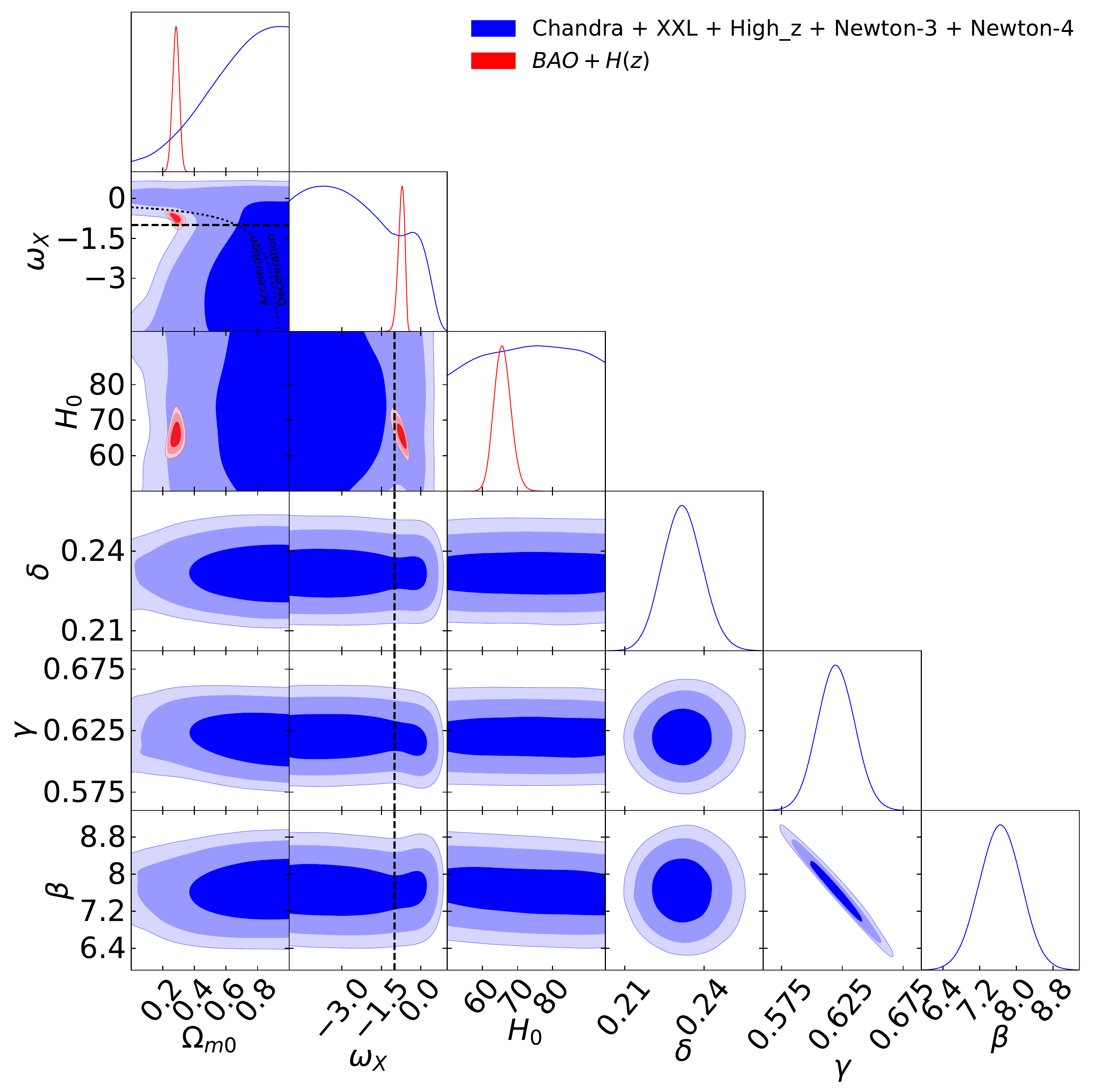}\par
    \includegraphics[width=\linewidth,height=7cm]{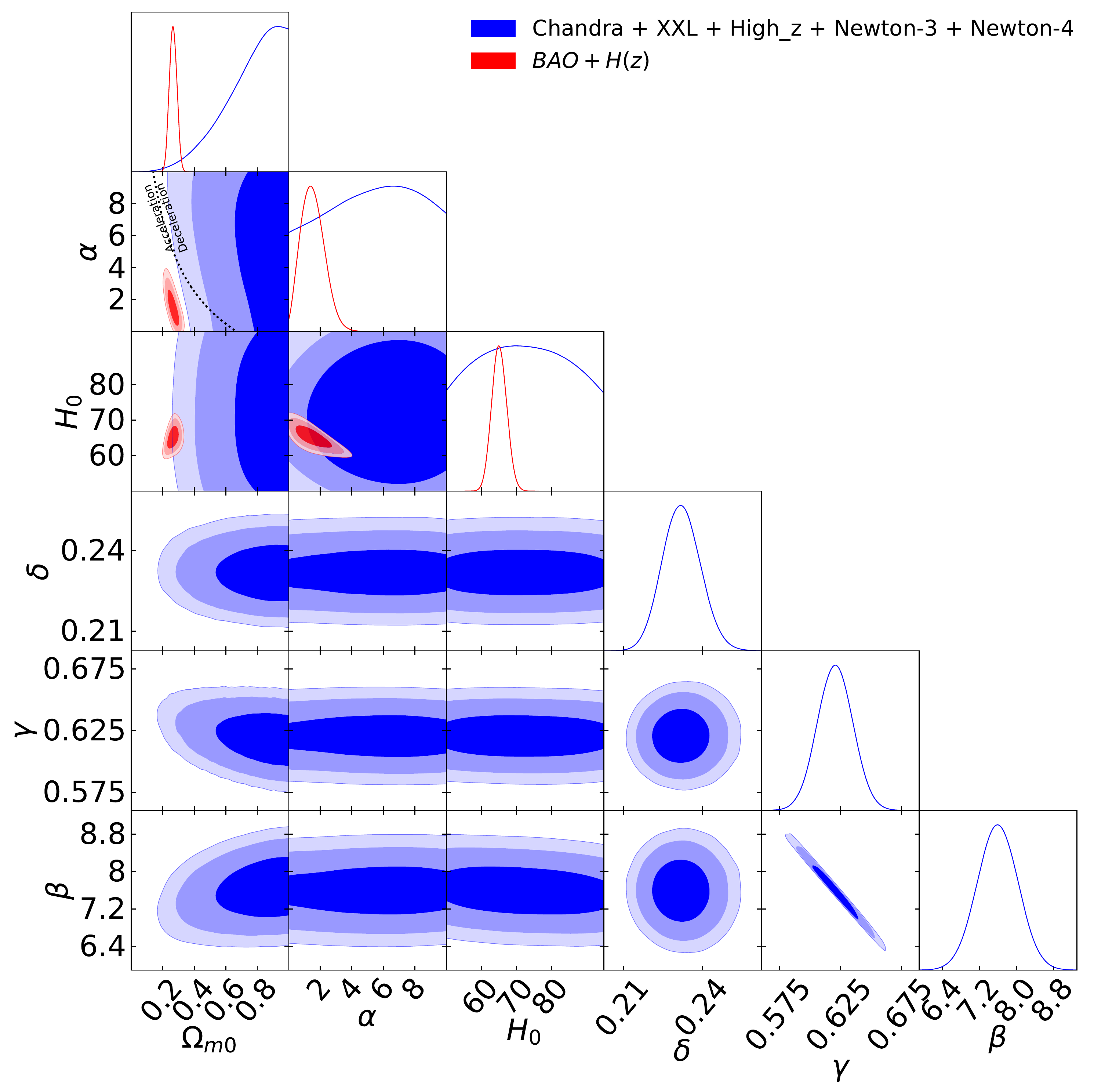}\par
    \includegraphics[width=\linewidth,height=7cm]{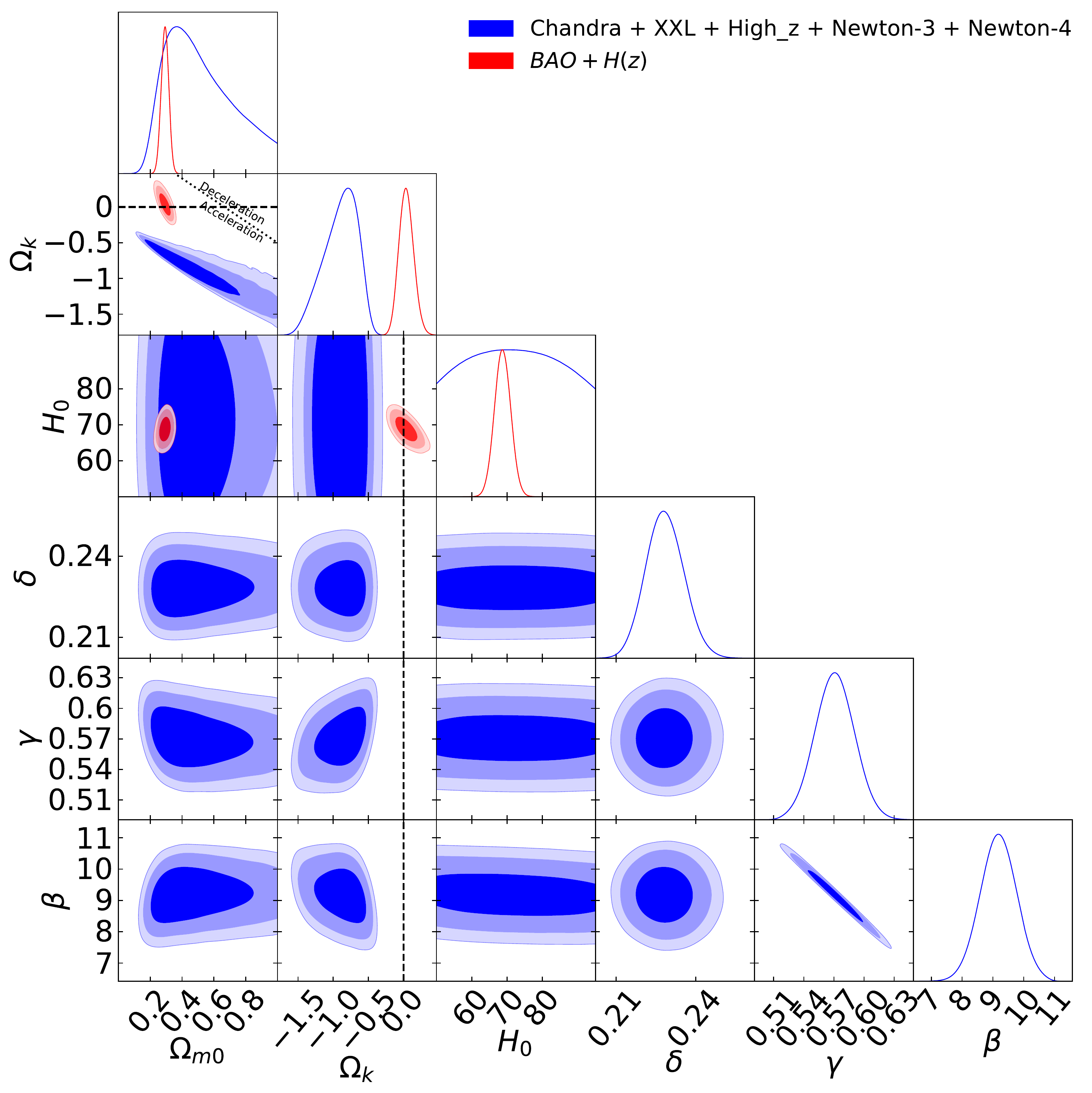}\par
    \includegraphics[width=\linewidth,height=7cm]{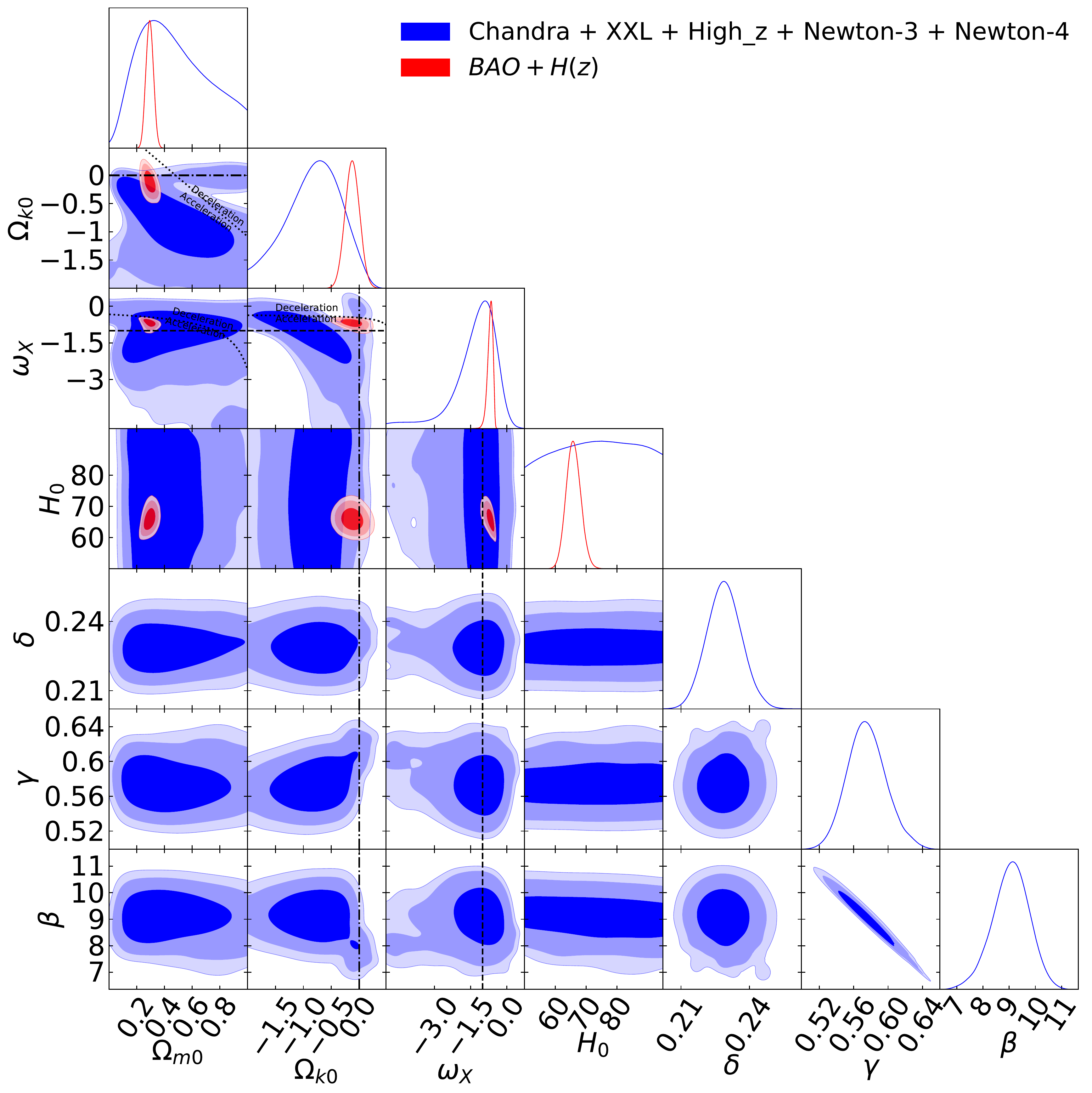}\par
    \includegraphics[width=\linewidth,height=7cm]{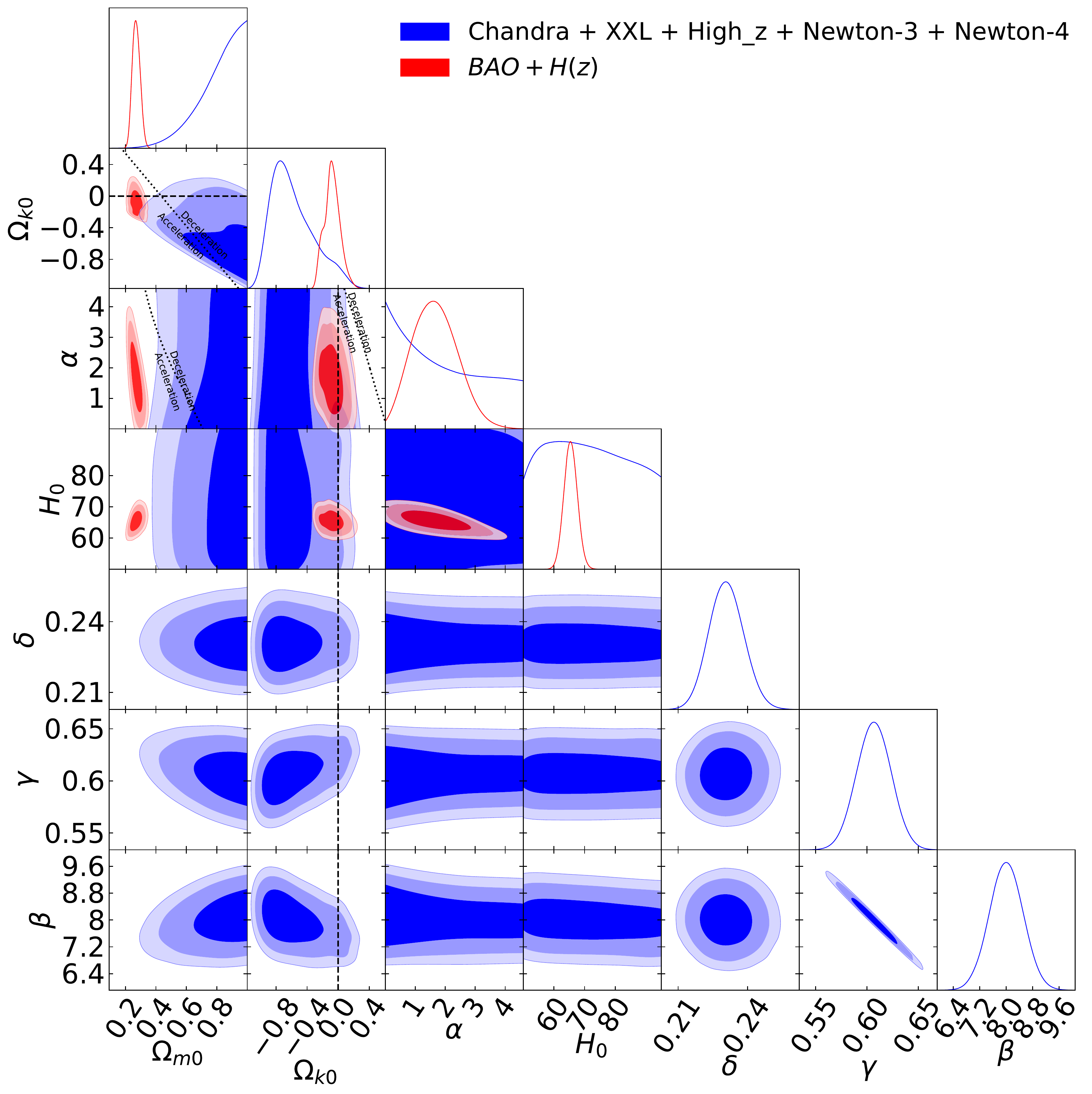}\par
\end{multicols}
\caption{One-dimensional likelihood distributions and two-dimensional likelihood contours at 1$\sigma$, 2$\sigma$, and 3$\sigma$ confidence levels using Chandra + XXL + High-$z$ + Newton-3 + Newton-4 (blue) and BAO + $H(z)$ (red) data for all free parameters. Left column shows the flat $\Lambda$CDM model, flat XCDM parametrization, and flat $\phi$CDM model respectively. The black dotted lines in all plots are the zero acceleration lines. The black dashed lines in the flat XCDM parametrization plots are the $\omega_X=-1$ lines. Right column shows the non-flat $\Lambda$CDM model, non-flat XCDM parametrization, and non-flat $\phi$CDM model respectively. Black dotted lines in all plots are the zero acceleration lines. Black dashed lines in the non-flat $\Lambda$CDM and $\phi$CDM model plots the black dotted-dashed lines in the non-flat XCDM parametrization plots correspond to $\Omega_{k0} = 0$. The black dashed lines in the non-flat XCDM parametrization plots are the $\omega_X=-1$ lines.}
\label{fig:Eiso-Ep}
\end{figure*}

For the SDSS-4XMM QSOs, beside the model- and redshift-dependent $\beta$ and $\gamma$ values, in the flat models especially, as shown in Fig.\ 1, these QSO data strongly favor currently decelerating cosmological expansion, contradicting constraints from most other cosmological data. Figure 1 also shows that even when the SDSS-4XMM constraints are consistent with currently accelerating cosmological expansion they pick out a different part of parameter space compared to what is favored by BAO + $H(z)$ data. Figure 2 shows that the lower and higher redshift halves of the SDSS-4XMM sub-sample result in somewhat different cosmological constraints.

For the SDSS-Chandra QSOs, which is the second largest sub-sample, comparing $\beta$ and $\gamma$ values for each of the six cosmological models, we see that these $\beta$ and $\gamma$ values are almost model-independent. For these QSOs, quantitative differences between $\beta$ and $\gamma$ values for different pairs of cosmological models are listed in Table 10. The difference between $\gamma$ values $(\Delta \gamma)$ from model to model ranges over $(0.04-0.96)\sigma$ while that between $\beta$ values $(\Delta \beta)$ ranges over $(0.01-0.99)\sigma$, both of which are almost statistically insignificant. This shows that SDSS-Chandra QSOs are standardizable via the $L_X-L_{UV}$ relation. For the XXL QSOs, which is the third largest sub-group, $\beta$ and $\gamma$ values are model-independent. For these QSOs, differences between $\beta$ and $\gamma$ values for different pair of cosmological model are listed in Table 11 and are almost negligible. This shows that the XXL sub-sample is standardizable through the $L_X-L_{UV}$ relation. Figure 3 shows that the SDSS-Chandra QSO constraints are largely consistent with those that follow from BAO + $H(z)$ data, while Fig.\ 4 shows that the much smaller XXL sub-group results in less-restrictive cosmological constraints that are consistent with those that follow from the BAO + $H(z)$ measurements.

We have also analysed some combinations of sub-groups to see whether these combinations have model-independent $L_X-L_{UV}$ relations and so can be used to measure cosmological model parameters. Results for $\beta$ and $\gamma$ for these combinations of QSOs are listed in Table 5. From this table and Table 12, for the High-$z$ + Newton-4 combination, $\beta$ and $\gamma$ values do not show significant model-dependency. This small sub-sample of 36 QSOs results in weak cosmological parameter constraints, see Fig.\ 5. For the larger SDSS-Chandra + XXL combination of 618 QSOs, Table 13, the difference between $\gamma$ values $(\Delta \gamma)$ from model to model range over $(0-1.04)\sigma$ and that between $\beta$ values $(\Delta \beta)$ range over $(0.01-1.14)\sigma$, which are almost statistically insignificant. The cosmological constraints from this larger sub-sample are more restrictive, and largely consistent with those that follow from BAO + $H(z)$ data, see Fig.\ 6. For the SDSS-Chandra + Newton-3 combination, Table 14, the difference between $\gamma$ values $(\Delta \gamma)$ from model to model range over $(0-1.21)\sigma$ and that between $\beta$ values $(\Delta \beta)$ range over $(0.00-1.26)\sigma$, which are almost statistically insignificant. Figure 7 shows that the cosmological constraints from this larger sub-sample of 556 QSOs are more restrictive and largely consistent with those that follow from BAO + $H(z)$ data. For the SDSS-Chandra + XXL + Newton-3 combination of 632 QSOs, Table 15, the difference between $\gamma$ values $(\Delta \gamma)$ from model to model range over $(0.05-1.44)\sigma$ and that between $\beta$ values $(\Delta \beta)$ range over $(0.00-1.49)\sigma$, which could both be statistically significant. This means that the SDSS-Chandra + XXL and Newton-3 sub-samples might not be mutually consistent. Figure 8 shows the resulting cosmological constraints. For the SDSS-Chandra + High-$z$ + Newton-4 combination of 578 QSOs, Table 16, the difference between $\gamma$ values $(\Delta \gamma)$ from model to model range over $(0.04-1.78)\sigma$ and that between $\beta$ values $(\Delta \beta)$ range over $(0.00-1.89)\sigma$, which could both be statistically significant. This means that the SDSS-Chandra and High-$z$ + Newton-4 sub-samples might not be mutually consistent. The resulting joint cosmological constraints are shown in Fig.\ 9. For the SDSS-Chandra + XXL + High-$z$ + Newton-3 + Newton-4 combination of 668 QSOs, Table 17, the difference between $\gamma$ values $(\Delta \gamma)$ from model to model range over $(0.00-2.07)\sigma$ and that between $\beta$ values $(\Delta \beta)$ range over $(0.00-2.16)\sigma$, which could both be statistically significant. This suggests that the SDSS-Chandra + XXL and High-$z$ + Newton-3 + Newton-4 sub-samples might not be mutually consistent. The resulting joint cosmological constraints are shown in Fig.\ 10.

While there are combinations of QSO sub-sets with model-independent $L_X-L_{UV}$ relations, perhaps the current best QSO $L_X-L_{UV}$ compilation for the purpose of constraining cosmological parameters is the $z \lesssim 1.5$ sub-set of about half the \citet{Lussoetal2020} QSOs used in \cite{KhadkaRatra2021} that provides relatively weak constraints which are consistent with those from better-established cosmological probes.  

\begin{table}
	\centering
	\small\addtolength{\tabcolsep}{-2.5pt}
	\small
	\caption{$L_X-L_{UV}$ relation parameters (and $\delta$) differences between different models for the SDSS-4XMM data.}
	\label{tab:BFP}
	\begin{threeparttable}
	\begin{tabular}{lccccccccccc} 
		\hline
		Model vs model & $\Delta \delta$ & $\Delta \gamma$ & $\Delta \beta$\\
		\hline
		Flat \lcdm\ vs non-flat \lcdm\ & $0.47\sigma$ & $3.44\sigma$ & $3.53\sigma$\\
		Flat \lcdm\ vs flat XCDM & $0.14\sigma$ & $0.37\sigma$ & $0.33\sigma$\\
		Flat \lcdm\ vs non-flat XCDM & $0.85\sigma$ & $3.54\sigma$ & $3.78\sigma$\\
		Flat \lcdm\ vs flat $\phi$CDM & $0.00\sigma$ & $0.00\sigma$ & $0.00\sigma$\\
		Flat \lcdm\ vs non-flat $\phi$CDM & $0.28\sigma$ & $0.94\sigma$ & $0.91\sigma$\\
		Non-flat \lcdm\ vs flat XCDM & $0.71\sigma$ & $3.01\sigma$ & $3.16\sigma$\\
		Non-flat \lcdm\ vs non-flat XCDM & $0.00\sigma$ & $0.09\sigma$ & $0.27\sigma$\\
		Non-flat \lcdm\ vs flat $\phi$CDM & $0.85\sigma$ & $3.44\sigma$ & $3.57\sigma$\\
		Non-flat \lcdm\ vs non-flat $\phi$CDM & $0.57\sigma$ & $2.49\sigma$ & $2.62\sigma$\\
		Flat XCDM vs non-flat XCDM & $0.71\sigma$ & $3.10\sigma$ & $3.40\sigma$\\
		Flat XCDM vs flat $\phi$CDM & $0.14\sigma$ & $0.37\sigma$ & $0.33\sigma$\\
		Flat XCDM vs non-flat $\phi$CDM & $0.14\sigma$ & $0.56\sigma$ & $0.57\sigma$\\
		Non-flat XCDM vs flat $\phi$CDM & $0.85\sigma$ & $3.54\sigma$ & $3.82\sigma$\\
		Non-flat XCDM vs non-flat $\phi$CDM & $0.57\sigma$ & $2.59\sigma$ & $2.87\sigma$\\
		Flat $\phi$CDM vs non-flat $\phi$CDM & $0.28\sigma$ & $0.94\sigma$ & $0.92\sigma$\\
		\hline
	\end{tabular}
    \end{threeparttable}
\end{table}

\begin{table}
\small\addtolength{\tabcolsep}{1.5pt}
\caption{$L_X-L_{UV}$ relation parameters (and $\delta$) differences between different models and lower and higher redshift halves of the SDSS-4XMM data set.}
\label{tab:BFP}
\begin{tabular}{cccc} 
\hline
Model & \hspace{8mm}$\Delta \delta$ \hspace{8mm} & \hspace{8mm}$\Delta \gamma$ \hspace{8mm} & $\Delta \beta$\\
\hline
\multicolumn{4}{c}{Between SDSS-4XMM-l and SDSS-4XMM-h} \\
\hline
Flat \lcdm\ & $2.82\sigma$ & $1.58\sigma$ & $1.69\sigma$\\
Non-flat \lcdm\ & $3.15\sigma$ & $2.5\sigma$ & $2.14\sigma$ \\
Flat XCDM & $2.63\sigma$ & $1.55\sigma$ & $1.62\sigma$\\
Non-flat XCDM & $3.25\sigma$ & $2.31\sigma$ & $2.12\sigma$\\
Flat $\phi$CDM & $2.82\sigma$ & $1.58\sigma$ & $1.67\sigma$\\
Non-flat $\phi$CDM & $2.93\sigma$ & $1.83\sigma$ & $1.87\sigma$\\
\hline
\end{tabular}
\end{table}

\section{Conclusion}
\label{con}

\cite{KhadkaRatra2021} discovered that \citet{Lussoetal2020} QSO data had $L_X-L_{UV}$ relation parameters that are cosmological-model as well as redshift dependent. These data are the compilation of seven different sub-samples. In this paper, we analysed these sub-groups, and some combination of sub-groups, to try to determine which QSO sub-groups are responsible for the issues pointed out in \cite{KhadkaRatra2021}.

From our sub-sample analyses here it is clear that $\beta$ and $\gamma$ values for the large SDSS-4XMM sub-sample are model as well as redshift dependent and that the SDSS-4XMM QSOs are responsible for the cosmological-model-dependent $L_X-L_{UV}$ relation found in \cite{KhadkaRatra2021} for the complete \citet{Lussoetal2020} QSO data. This finding indicates that current SDSS-4XMM QSOs are not standardizable candles and that this issue needs to be resolved if one is to use SDSS-4XMM QSOs for cosmological purposes. We are not sure why these current SDSS-4XMM QSOs are not standardizable but possibly a more careful examination of their X-ray and UV spectra might be able to provide an explanation for this.

Additionally, when High-$z$ + Newton-4 data are combined with other sub-samples, this results in model-dependent $\beta$ and $\gamma$ parameters (see Tables 16 and 17), which indicates that the higher-redshift High-$z$ and Newton-4 sub-sample might be inconsistent with the lower-redshift sub-samples. 

On the other hand, analyses of other big sub-samples, SDSS-Chandra and SDSS-Chandra + XXL, show that for these sub-samples the $L_X-L_{UV}$ relation parameters are almost independent of cosmological model. This indicates that current versions of these sub-samples can be standardized via the $L_X-L_{UV}$ relation and used for cosmological purposes. However, they provide only weak cosmological parameter constraints, constraints consistent with those determined using data obtained from better-established cosmological probes. 

Given that some of the sub-samples can be standardized through the $L_X-L_{UV}$ relation, it is not impossible that a more careful study of the SDSS-4XMM sample might find an underlying issue that when corrected makes the SDSS-4XMM QSOs a valuable cosmological probe.  

Additionally SDSS-Chandra + XXL QSOs, as well as reverberation-measured Mg II radius-luminosity relation QSOs \citep{Khadkaetal2021b}, are standardizable candles and useful cosmological probes, so future detection of more such quasars will help establish QSO data as a useful probe of the as yet largely unstudied  $1.5 \lesssim z \lesssim 4$ part of cosmological redshift space.

\begin{table}
	\centering
	\small\addtolength{\tabcolsep}{-2.5pt}
	\small
	\caption{$L_X-L_{UV}$ relation parameters (and $\delta$) differences between different models for the SDSS-4XMM-l data.}
	\label{tab:BFP}
	\begin{threeparttable}
	\begin{tabular}{lccccccccccc} 
		\hline
		Model vs model & $\Delta \delta$ & $\Delta \gamma$ & $\Delta \beta$\\
		\hline
		Flat \lcdm\ vs non-flat \lcdm\ & $0.10\sigma$ & $0.18\sigma$ & $0.18\sigma$\\
		Flat \lcdm\ vs flat XCDM & $0.00\sigma$ & $0.030\sigma$ & $0.04\sigma$\\
		Flat \lcdm\ vs non-flat XCDM & $0.00\sigma$ & $0.24\sigma$ & $0.25\sigma$\\
		Flat \lcdm\ vs flat $\phi$CDM & $0.00\sigma$ & $0.00\sigma$ & $0.01\sigma$\\
		Flat \lcdm\ vs non-flat $\phi$CDM & $0.00\sigma$ & $0.03\sigma$ & $0.02\sigma$\\
		Non-flat \lcdm\ vs flat XCDM & $0.10\sigma$ & $0.15\sigma$ & $0.14\sigma$\\
		Non-flat \lcdm\ vs non-flat XCDM & $0.10\sigma$ & $0.05\sigma$ & $0.07\sigma$\\
		Non-flat \lcdm\ vs flat $\phi$CDM & $0.10\sigma$ & $0.18\sigma$ & $0.19\sigma$\\
		Non-flat \lcdm\ vs non-flat $\phi$CDM & $0.10\sigma$ & $0.15\sigma$ & $0.16\sigma$\\
		Flat XCDM vs non-flat XCDM & $0.00\sigma$ & $0.22\sigma$ & $0.21\sigma$\\
		Flat XCDM vs flat $\phi$CDM & $0.00\sigma$ & $0.03\sigma$ & $0.05\sigma$\\
		Flat XCDM vs non-flat $\phi$CDM & $0.00\sigma$ & $0.00\sigma$ & $0.02\sigma$\\
		Non-flat XCDM vs flat $\phi$CDM & $0.00\sigma$ & $0.25\sigma$ & $0.26\sigma$\\
		Non-flat XCDM vs non-flat $\phi$CDM & $0.00\sigma$ & $0.22\sigma$ & $0.23\sigma$\\
		Flat $\phi$CDM vs non-flat $\phi$CDM & $0.00\sigma$ & $0.03\sigma$ & $0.03\sigma$\\
		\hline
	\end{tabular}
    \end{threeparttable}
\end{table}

\begin{table}
	\centering
	\small\addtolength{\tabcolsep}{-2.5pt}
	\small
	\caption{$L_X-L_{UV}$ relation parameters (and $\delta$) differences between different models for the SDSS-4XMM-h data.}
	\label{tab:BFP}
	\begin{threeparttable}
	\begin{tabular}{lccccccccccc} 
		\hline
		Model vs model & $\Delta \delta$ & $\Delta \gamma$ & $\Delta \beta$\\
		\hline
		Flat \lcdm\ vs non-flat \lcdm\ & $0.47\sigma$ & $1.12\sigma$ & $0.86\sigma$\\
		Flat \lcdm\ vs flat XCDM & $0.00\sigma$ & $0.034\sigma$ & $0.01\sigma$\\
		Flat \lcdm\ vs non-flat XCDM & $0.43\sigma$ & $1.02\sigma$ & $0.89\sigma$\\
		Flat \lcdm\ vs flat $\phi$CDM & $0.00\sigma$ & $0.04\sigma$ & $0.04\sigma$\\
		Flat \lcdm\ vs non-flat $\phi$CDM & $0.12\sigma$ & $0.23\sigma$ & $0.19\sigma$\\
		Non-flat \lcdm\ vs flat XCDM & $0.43\sigma$ & $1.03\sigma$ & $0.80\sigma$\\
		Non-flat \lcdm\ vs non-flat XCDM & $0.00\sigma$ & $0.05\sigma$ & $0.01\sigma$\\
		Non-flat \lcdm\ vs flat $\phi$CDM & $0.47\sigma$ & $1.16\sigma$ & $0.89\sigma$\\
		Non-flat \lcdm\ vs non-flat $\phi$CDM & $0.35\sigma$ & $0.87\sigma$ & $0.68\sigma$\\
		Flat XCDM vs non-flat XCDM & $0.43\sigma$ & $0.83\sigma$ & $0.04\sigma$\\
		Flat XCDM vs flat $\phi$CDM & $0.00\sigma$ & $0.00\sigma$ & $0.02\sigma$\\
		Flat XCDM vs non-flat $\phi$CDM & $0.11\sigma$ & $0.25\sigma$ & $0.18\sigma$\\
		Non-flat XCDM vs flat $\phi$CDM & $0.47\sigma$ & $1.07\sigma$ & $0.92\sigma$\\
		Non-flat XCDM vs non-flat $\phi$CDM & $0.35\sigma$ & $0.79\sigma$ & $0.72\sigma$\\
		Flat $\phi$CDM vs non-flat $\phi$CDM & $0.35\sigma$ & $0.28\sigma$ & $0.22\sigma$\\
		\hline
	\end{tabular}
    \end{threeparttable}
\end{table}

\begin{table}
	\centering
	\small\addtolength{\tabcolsep}{-2.5pt}
	\small
	\caption{$L_X-L_{UV}$ relation parameters (and $\delta$) differences between different models for the SDSS-Chandra data.}
	\label{tab:BFP}
	\begin{threeparttable}
	\begin{tabular}{lccccccccccc} 
		\hline
		Model vs model & $\Delta \delta$ & $\Delta \gamma$ & $\Delta \beta$\\
		\hline
		Flat \lcdm\ vs non-flat \lcdm\ & $0.18\sigma$ & $0.96\sigma$ & $0.99\sigma$\\
		Flat \lcdm\ vs flat XCDM & $0.00\sigma$ & $0.08\sigma$ & $0.09\sigma$\\
		Flat \lcdm\ vs non-flat XCDM & $0.09\sigma$ & $0.79\sigma$ & $0.8\sigma$\\
		Flat \lcdm\ vs flat $\phi$CDM & $0.00\sigma$ & $0.04\sigma$ & $0.01\sigma$\\
		Flat \lcdm\ vs non-flat $\phi$CDM & $0.00\sigma$ & $0.27\sigma$ & $0.23\sigma$\\
		Non-flat \lcdm\ vs flat XCDM & $0.18\sigma$ & $0.89\sigma$ & $0.92\sigma$\\
		Non-flat \lcdm\ vs non-flat XCDM & $0.09\sigma$ & $0.15\sigma$ & $0.16\sigma$\\
		Non-flat \lcdm\ vs flat $\phi$CDM & $0.18\sigma$ & $0.92\sigma$ & $0.98\sigma$\\
		Non-flat \lcdm\ vs non-flat $\phi$CDM & $0.18\sigma$ & $0.70\sigma$ & $0.78\sigma$\\
		Flat XCDM vs non-flat XCDM & $0.09\sigma$ & $0.72\sigma$ & $0.72\sigma$\\
		Flat XCDM vs flat $\phi$CDM & $0.00\sigma$ & $0.04\sigma$ & $0.08\sigma$\\
		Flat XCDM vs non-flat $\phi$CDM & $0.00\sigma$ & $0.19\sigma$ & $0.14\sigma$\\
		Non-flat XCDM vs flat $\phi$CDM & $0.09\sigma$ & $0.75\sigma$ & $0.79\sigma$\\
		Non-flat XCDM vs non-flat $\phi$CDM & $0.09\sigma$ & $0.54\sigma$ & $0.59\sigma$\\
		Flat $\phi$CDM vs non-flat $\phi$CDM & $0.00\sigma$ & $0.23\sigma$ & $0.22\sigma$\\
		\hline
	\end{tabular}
    \end{threeparttable}
\end{table}

\begin{table}
	\centering
	\small\addtolength{\tabcolsep}{-2.5pt}
	\small
	\caption{$L_X-L_{UV}$ relation parameters (and $\delta$) differences between different models for the XXL data.}
	\label{tab:BFP}
	\begin{threeparttable}
	\begin{tabular}{lccccccccccc} 
		\hline
		Model vs model & $\Delta \delta$ & $\Delta \gamma$ & $\Delta \beta$\\
		\hline
		Flat \lcdm\ vs non-flat \lcdm\  & $0.04\sigma$ & $0.07\sigma$ & $0.05\sigma$\\
		Flat \lcdm\ vs flat XCDM & $0.04\sigma$ & $0.06\sigma$ & $0.05\sigma$\\
		Flat \lcdm\ vs non-flat XCDM & $0.04\sigma$ & $0.03\sigma$ & $0.00\sigma$\\
		Flat \lcdm\ vs flat $\phi$CDM & $0.00\sigma$ & $0.07\sigma$ & $0.05\sigma$\\
		Flat \lcdm\ vs non-flat $\phi$CDM & $0.00\sigma$ & $0.07\sigma$ & $0.05\sigma$\\
		Non-flat \lcdm\ vs flat XCDM & $0.00\sigma$ & $0.01\sigma$ & $0.00\sigma$\\
		Non-flat \lcdm\ vs non-flat XCDM & $0.00\sigma$ & $0.04\sigma$ & $0.05\sigma$\\
		Non-flat \lcdm\ vs flat $\phi$CDM & $0.04\sigma$ & $0.00\sigma$ & $0.00\sigma$\\
		Non-flat \lcdm\ vs non-flat $\phi$CDM & $0.04\sigma$ & $0.01\sigma$ & $0.00\sigma$\\
		Flat XCDM vs non-flat XCDM & $0.00\sigma$ & $0.03\sigma$ & $0.05\sigma$\\
		Flat XCDM vs flat $\phi$CDM & $0.04\sigma$ & $0.01\sigma$ & $0.00\sigma$\\
		Flat XCDM vs non-flat $\phi$CDM & $0.04\sigma$ & $0.00\sigma$ & $0.00\sigma$\\
		Non-flat XCDM vs flat $\phi$CDM & $0.00\sigma$ & $0.04\sigma$ & $0.05\sigma$\\
		Non-flat XCDM vs non-flat $\phi$CDM & $0.04\sigma$ & $0.03\sigma$ & $0.05\sigma$\\
		Flat $\phi$CDM vs non-flat $\phi$CDM & $0.00\sigma$ & $0.01\sigma$ & $0.00\sigma$\\
		\hline
	\end{tabular}
    \end{threeparttable}
\end{table}

\begin{table}
	\centering
	\small\addtolength{\tabcolsep}{-2.5pt}
	\small
	\caption{$L_X-L_{UV}$ relation parameters (and $\delta$) differences between different models for the High-$z$ + Newton-4 data.}
	\label{tab:BFP}
	\begin{threeparttable}
	\begin{tabular}{lccccccccccc} 
		\hline
		Model vs model & $\Delta \delta$ & $\Delta \gamma$ & $\Delta \beta$\\
		\hline
		Flat \lcdm\ vs non-flat \lcdm\  & $0.03\sigma$ & $0.02\sigma$ & $0.03\sigma$\\
		Flat \lcdm\ vs flat XCDM & $0.00\sigma$ & $0.00\sigma$ & $0.01\sigma$\\
		Flat \lcdm\ vs non-flat XCDM & $0.03\sigma$ & $0.03\sigma$ & $0.02\sigma$\\
		Flat \lcdm\ vs flat $\phi$CDM & $0.02\sigma$ & $0.03\sigma$ & $0.02\sigma$\\
		Flat \lcdm\ vs non-flat $\phi$CDM & $0.02\sigma$ & $0.05\sigma$ & $0.02\sigma$\\
		Non-flat \lcdm\ vs flat XCDM & $0.03\sigma$ & $0.02\sigma$ & $0.03\sigma$\\
		Non-flat \lcdm\ vs non-flat XCDM & $0.00\sigma$ & $0.02\sigma$ & $0.01\sigma$\\
		Non-flat \lcdm\ vs flat $\phi$CDM & $0.02\sigma$ & $0.02\sigma$ & $0.04\sigma$\\
		Non-flat \lcdm\ vs non-flat $\phi$CDM & $0.02\sigma$ & $0.03\sigma$ & $0.05\sigma$\\
		Flat XCDM vs non-flat XCDM & $0.03\sigma$ & $0.03\sigma$ & $0.02\sigma$\\
		Flat XCDM vs flat $\phi$CDM & $0.02\sigma$ & $0.03\sigma$ & $0.01\sigma$\\
		Flat XCDM vs non-flat $\phi$CDM & $0.02\sigma$ & $0.05\sigma$ & $0.02\sigma$\\
		Non-flat XCDM vs flat $\phi$CDM & $0.02\sigma$ & $0.00\sigma$ & $0.03\sigma$\\
		Non-flat XCDM vs non-flat $\phi$CDM & $0.02\sigma$ & $0.02\sigma$ & $0.04\sigma$\\
		Flat $\phi$CDM vs non-flat $\phi$CDM & $0.00\sigma$ & $0.02\sigma$ & $0.01\sigma$\\
		\hline
	\end{tabular}
    \end{threeparttable}
\end{table}

\begin{table}
	\centering
	\small\addtolength{\tabcolsep}{-2.5pt}
	\small
	\caption{$L_X-L_{UV}$ relation parameters (and $\delta$) differences between different models for the SDSS-Chandra + XXL data.}
	\label{tab:BFP}
	\begin{threeparttable}
	\begin{tabular}{lccccccccccc} 
		\hline
		Model vs model & $\Delta \delta$ & $\Delta \gamma$ & $\Delta \beta$\\
		\hline
		Flat \lcdm\ vs non-flat \lcdm\ & $0.18\sigma$ & $1.04\sigma$ & $1.14\sigma$\\
		Flat \lcdm\ vs flat XCDM & $0.00\sigma$ & $0.04\sigma$ & $0.08\sigma$\\
		Flat \lcdm\ vs non-flat XCDM & $0.09\sigma$ & $0.79\sigma$ & $0.92\sigma$\\
		Flat \lcdm\ vs flat $\phi$CDM & $0.00\sigma$ & $0.00\sigma$ & $0.01\sigma$\\
		Flat \lcdm\ vs non-flat $\phi$CDM & $0.00\sigma$ & $0.28\sigma$ & $0.27\sigma$\\
		Non-flat \lcdm\ vs flat XCDM & $0.18\sigma$ & $1.01\sigma$ & $1.07\sigma$\\
		Non-flat \lcdm\ vs non-flat XCDM & $0.09\sigma$ & $0.15\sigma$ & $0.15\sigma$\\
		Non-flat \lcdm\ vs flat $\phi$CDM & $0.18\sigma$ & $1.04\sigma$ & $1.13\sigma$\\
		Non-flat \lcdm\ vs non-flat $\phi$CDM & $0.18\sigma$ & $0.77\sigma$ & $0.88\sigma$\\
		Flat XCDM vs non-flat XCDM & $0.09\sigma$ & $0.76\sigma$ & $0.85\sigma$\\
		Flat XCDM vs flat $\phi$CDM & $0.00\sigma$ & $0.04\sigma$ & $0.07\sigma$\\
		Flat XCDM vs non-flat $\phi$CDM & $0.00\sigma$ & $0.24\sigma$ & $0.19\sigma$\\
		Non-flat XCDM vs flat $\phi$CDM & $0.09\sigma$ & $0.79\sigma$ & $0.90\sigma$\\
		Non-flat XCDM vs non-flat $\phi$CDM & $0.09\sigma$ & $0.55\sigma$ & $0.67\sigma$\\
		Flat $\phi$CDM vs non-flat $\phi$CDM & $0.00\sigma$ & $0.28\sigma$ & $0.26\sigma$\\
		\hline
	\end{tabular}
    \end{threeparttable}
\end{table}

\begin{table}
	\centering
	\small\addtolength{\tabcolsep}{-2.5pt}
	\small
	\caption{$L_X-L_{UV}$ relation parameters (and $\delta$) differences between different models for the Chandra + Newton-3 data.}
	\label{tab:BFP}
	\begin{threeparttable}
	\begin{tabular}{lccccccccccc} 
		\hline
		Model vs model & $\Delta \delta$ & $\Delta \gamma$ & $\Delta \beta$\\
		\hline
		Flat \lcdm\ vs non-flat \lcdm\  & $0.18\sigma$ & $1.19\sigma$ & $1.26\sigma$\\
		Flat \lcdm\ vs flat XCDM & $0.00\sigma$ & $0.04\sigma$ & $0.08\sigma$\\
		Flat \lcdm\ vs non-flat XCDM & $0.09\sigma$ & $0.92\sigma$ & $1.00\sigma$\\
		Flat \lcdm\ vs flat $\phi$CDM & $0.00\sigma$ & $0.00\sigma$ & $0.00\sigma$\\
		Flat \lcdm\ vs non-flat $\phi$CDM & $0.09\sigma$ & $0.23\sigma$ & $0.26\sigma$\\
		Non-flat \lcdm\ vs flat XCDM & $0.18\sigma$ & $1.15\sigma$ & $1.19\sigma$\\
		Non-flat \lcdm\ vs non-flat XCDM & $0.09\sigma$ & $0.18\sigma$ & $0.19\sigma$\\
		Non-flat \lcdm\ vs flat $\phi$CDM & $0.18\sigma$ & $1.21\sigma$ & $1.26\sigma$\\
		Non-flat \lcdm\ vs non-flat $\phi$CDM & $0.09\sigma$ & $0.91\sigma$ & $1.01\sigma$\\
		Flat XCDM vs non-flat XCDM & $0.09\sigma$ & $0.88\sigma$ & $0.94\sigma$\\
		Flat XCDM vs flat $\phi$CDM & $0.00\sigma$ & $0.26\sigma$ & $0.08\sigma$\\
		Flat XCDM vs non-flat $\phi$CDM & $0.09\sigma$ & $0.24\sigma$ & $0.18\sigma$\\
		Non-flat XCDM vs flat $\phi$CDM & $0.09\sigma$ & $0.94\sigma$ & $1.00\sigma$\\
		Non-flat XCDM vs non-flat $\phi$CDM & $0.00\sigma$ & $0.67\sigma$ & $0.77\sigma$\\
		Flat $\phi$CDM vs non-flat $\phi$CDM & $0.09\sigma$ & $0.29\sigma$ & $0.26\sigma$\\
		\hline
	\end{tabular}
    \end{threeparttable}
\end{table}

\begin{table}
	\centering
	\small\addtolength{\tabcolsep}{-2.5pt}
	\small
	\caption{$L_X-L_{UV}$ relation parameters (and $\delta$) differences between different models for the SDSS-Chandra + XXL + Newton-3 data.}
	\label{tab:BFP}
	\begin{threeparttable}
	\begin{tabular}{lccccccccccc} 
		\hline
		Model vs model & $\Delta \delta$ & $\Delta \gamma$ & $\Delta \beta$\\
		\hline
		Flat \lcdm\ vs non-flat \lcdm\ & $0.20\sigma$ & $1.44\sigma$ & $1.49\sigma$\\
		Flat \lcdm\ vs flat XCDM & $0.00\sigma$ & $0.09\sigma$ & $0.12\sigma$\\
		Flat \lcdm\ vs non-flat XCDM & $0.09\sigma$ & $1.14\sigma$ & $1.21\sigma$\\
		Flat \lcdm\ vs flat $\phi$CDM & $0.00\sigma$ & $0.05\sigma$ & $0.00\sigma$\\
		Flat \lcdm\ vs non-flat $\phi$CDM & $0.39\sigma$ & $0.34\sigma$ & $0.00\sigma$\\
		Non-flat \lcdm\ vs flat XCDM & $0.19\sigma$ & $1.40\sigma$ & $1.39\sigma$\\
		Non-flat \lcdm\ vs non-flat XCDM & $0.09\sigma$ & $0.20\sigma$ & $0.17\sigma$\\
		Non-flat \lcdm\ vs flat $\phi$CDM & $0.20\sigma$ & $1.44\sigma$ & $1.49\sigma$\\
		Non-flat \lcdm\ vs non-flat $\phi$CDM & $0.20\sigma$ & $1.07\sigma$ & $1.17\sigma$\\
		Flat XCDM vs non-flat XCDM & $0.09\sigma$ & $1.09\sigma$ & $1.11\sigma$\\
		Flat XCDM vs flat $\phi$CDM & $0.00\sigma$ & $0.05\sigma$ & $0.12\sigma$\\
		Flat XCDM vs non-flat $\phi$CDM & $0.00\sigma$ & $0.31\sigma$ & $0.22\sigma$\\
		Non-flat XCDM vs flat $\phi$CDM & $0.09\sigma$ & $1.13\sigma$ & $1.21\sigma$\\
		Non-flat XCDM vs non-flat $\phi$CDM & $0.09\sigma$ & $0.79\sigma$ & $0.92\sigma$\\
		Flat $\phi$CDM vs non-flat $\phi$CDM & $0.00\sigma$ & $0.35\sigma$ & $0.34\sigma$\\
		\hline
	\end{tabular}
    \end{threeparttable}
\end{table}

\begin{table}
	\centering
	\small\addtolength{\tabcolsep}{-2.5pt}
	\small
	\caption{$L_X-L_{UV}$ relation parameters (and $\delta$) differences between different models for the SDSS-Chandra + High-$z$ + Newton-4 data.}
	\label{tab:BFP}
	\begin{threeparttable}
	\begin{tabular}{lccccccccccc} 
		\hline
		Model vs model & $\Delta \delta$ & $\Delta \gamma$ & $\Delta \beta$\\
		\hline
		Flat \lcdm\ vs non-flat \lcdm\ & $0.38\sigma$ & $1.78\sigma$ & $1.89\sigma$\\
		Flat \lcdm\ vs flat XCDM & $0.00\sigma$ & $0.08\sigma$ & $0.10\sigma$\\
		Flat \lcdm\ vs non-flat XCDM & $0.025\sigma$ & $1.67\sigma$ & $1.74\sigma$\\
		Flat \lcdm\ vs flat $\phi$CDM & $0.09\sigma$ & $0.04\sigma$ & $0.00\sigma$\\
		Flat \lcdm\ vs non-flat $\phi$CDM & $0.09\sigma$ & $0.57\sigma$ & $0.51\sigma$\\
		Non-flat \lcdm\ vs flat XCDM & $0.35\sigma$ & $1.70\sigma$ & $1.80\sigma$\\
		Non-flat \lcdm\ vs non-flat XCDM & $0.08\sigma$ & $0.10\sigma$ & $0.10\sigma$\\
		Non-flat \lcdm\ vs flat $\phi$CDM & $0.44\sigma$ & $1.78\sigma$ & $1.89\sigma$\\
		Non-flat \lcdm\ vs non-flat $\phi$CDM & $0.27\sigma$ & $1.23\sigma$ & $1.35\sigma$\\
		Flat XCDM vs non-flat XCDM & $0.25\sigma$ & $1.59\sigma$ & $1.66\sigma$\\
		Flat XCDM vs flat $\phi$CDM & $0.09\sigma$ & $0.04\sigma$ & $0.10\sigma$\\
		Flat XCDM vs non-flat $\phi$CDM & $0.09\sigma$ & $0.48\sigma$ & $0.41\sigma$\\
		Non-flat XCDM vs flat $\phi$CDM & $0.33\sigma$ & $1.67\sigma$ & $1.74\sigma$\\
		Non-flat XCDM vs non-flat $\phi$CDM & $0.17\sigma$ & $1.12\sigma$ & $1.22\sigma$\\
		Flat $\phi$CDM vs non-flat $\phi$CDM & $0.18\sigma$ & $0.54\sigma$ & $0.51\sigma$\\
		\hline
	\end{tabular}
    \end{threeparttable}
\end{table}
\begin{table}
	\centering
	\small\addtolength{\tabcolsep}{-2.5pt}
	\small
	\caption{$L_X-L_{UV}$ relation parameters (and $\delta$) differences between different models for the SDSS-Chandra + XXL + High-$z$ + Newton-3 + Newton-4 data.}
	\label{tab:BFP}
	\begin{threeparttable}
	\begin{tabular}{lccccccccccc} 
		\hline
		Model vs model & $\Delta \delta$ & $\Delta \gamma$ & $\Delta \beta$\\
		\hline
		Flat \lcdm\ vs non-flat \lcdm\  & $0.40\sigma$ & $2.07\sigma$ & $2.16\sigma$\\
		Flat \lcdm\ vs flat XCDM & $0.00\sigma$ & $0.05\sigma$ & $0.10\sigma$\\
		Flat \lcdm\ vs non-flat XCDM & $0.28\sigma$ & $1.78\sigma$ & $1.89\sigma$\\
		Flat \lcdm\ vs flat $\phi$CDM & $0.00\sigma$ & $0.00\sigma$ & $0.00\sigma$\\
		Flat \lcdm\ vs non-flat $\phi$CDM & $0.10\sigma$ & $0.64\sigma$ & $0.62\sigma$\\
		Non-flat \lcdm\ vs flat XCDM & $0.38\sigma$ & $2.02\sigma$ & $2.08\sigma$\\
		Non-flat \lcdm\ vs non-flat XCDM & $0.09\sigma$ & $0.13\sigma$ & $0.13\sigma$\\
		Non-flat \lcdm\ vs flat $\phi$CDM & $0.40\sigma$ & $2.12\sigma$ & $2.16\sigma$\\
		Non-flat \lcdm\ vs non-flat $\phi$CDM & $0.30\sigma$ & $1.45\sigma$ & $1.54\sigma$\\
		Flat XCDM vs non-flat XCDM & $0.27\sigma$ & $1.74\sigma$ & $1.82\sigma$\\
		Flat XCDM vs flat $\phi$CDM & $0.00\sigma$ & $0.05\sigma$ & $0.10\sigma$\\
		Flat XCDM vs non-flat $\phi$CDM & $0.09\sigma$ & $0.59\sigma$ & $0.53\sigma$\\
		Non-flat XCDM vs flat $\phi$CDM & $0.28\sigma$ & $1.82\sigma$ & $1.93\sigma$\\
		Non-flat XCDM vs non-flat $\phi$CDM & $0.19\sigma$ & $1.21\sigma$ & $1.35\sigma$\\
		Flat $\phi$CDM vs non-flat $\phi$CDM & $0.10\sigma$ & $0.66\sigma$ & $0.62\sigma$\\
		\hline
	\end{tabular}
    \end{threeparttable}
\end{table}

\section{ACKNOWLEDGEMENTS}
We thank Guido Risaliti and Salvatore Capozziello for useful discussions. This research was supported in part by DOE grant DE- SC0011840. Part of the computation for this project was performed on the Beocat Research Cluster at Kansas State University.

\section*{Data availability}
The data underlying this article are publicly available in \cite{Lussoetal2020}.



\bibliographystyle{mnras}
\bibliography{mybibfile}

\bsp	
\label{lastpage}
\end{document}